\documentclass[letterpaper,11pt,oneside]{book}
\usepackage{amssymb,amsmath,amsfonts}
\usepackage[english]{babel}
\usepackage[utf8]{inputenc}
\usepackage{multicol}
\usepackage{graphicx}
\usepackage{float}
\usepackage{caption}
\usepackage{subcaption}
\usepackage{wrapfig}
\usepackage{color}
\usepackage{cancel}
\usepackage{fancyhdr}
\usepackage{mathrsfs}
\usepackage{amsfonts}
\usepackage{geometry}
\usepackage{setspace}
\usepackage{afterpage}
\usepackage{cite} %hace las referencias [1-5], en lugar de [1,2,3,4,5]
\usepackage{tikz}

\usepackage[colorlinks=true,urlcolor=blue,anchorcolor=blue,citecolor=blue,filecolor=blue,linkcolor=blue,menucolor=blue,pagecolor=blue,linktocpage=true,pdfproducer=medialab,pdfa=true]{hyperref}

\begin{document}

%\newgeometry{left=4cm,right=2.5cm,top=2	cm,bottom=2.5cm}

\thispagestyle{empty}

\begin{titlepage}
	\centering
	{\includegraphics[height=4cm]{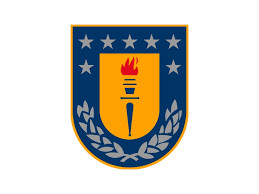}\\
	\textbf{UNIVERSIDAD DE CONCEPCI\'ON} \\ 
	\textbf{FACULTAD DE CIENCIAS F\'ISICAS Y MATEM\'ATICAS} \\
	\textbf{DOCTORADO EN CIENCIAS F\'ISICAS} \par}
	\vspace{2cm}
	{\LARGE \textbf{INTEGRABLE SYSTEMS \\ AND THE BOUNDARY DYNAMICS OF \\ (HIGHER SPIN) GRAVITY ON AdS$_3$}  \par}	
	\vspace{1.7 cm}
	{\itshape \it Thesis presented to the Facultad de Ciencias F\'isicas y Matem\'aticas of the\\ Universidad de Concepci\'on to qualify for the academic degree of\\ Doctor en Ciencias F\'isicas \par}
    \vspace{1.7cm}
	%{\Large BY \par}
	{\Large \textbf{EMILIO GERM\'AN OJEDA LEYTON} \par}
	\vspace{1.7cm}
	\begin{flushright}
		\begin{tabular}[c]{lll}
			\large{Thesis Advisor} & : & \large{Dr. Alfredo P\'erez Donoso} \\
			& & \large{\textit{Centro de Estudios Cient\'ificos (CECs)}} \\
		\end{tabular} 
	\end{flushright} 
	\vspace{2cm}
	{\large January, 2021 \par}
	{\large Concepci\'on, Chile \par}
\end{titlepage}

\newpage
%Recupera geometría del texto completo
\restoregeometry

%\thispagestyle{empty}
%\ \\
%\vspace{17cm}

%\begin{flushleft}
%The total or partial reproduction is authorized, for academic purposes, by any means or procedure, including the bibliographic citation of the document.
%\end{flushleft}

\newpage
\frontmatter
\setcounter{page}{1}
\thispagestyle{empty}
\ \\
\vspace{10cm}

\begin{large}
\hspace{1.5cm}\begin{tabular}[c]{lll}
Thesis Advisor & : & Dr. Alfredo P\'erez Donoso \\
& & {\normalsize \textit{Laboratorio de F\'isica Te\'orica,}} \\ 
& & {\normalsize \textit{Centro de Estudios Cient\'ificos (CECs),}} \\
& & \textit{Valdivia, Chile.} \\ \\
Assessment Committee & : & Dr. Ricardo Troncoso Pérez \\
& & {\normalsize \textit{Laboratorio de F\'isica Te\'orica,}} \\ 
& & {\normalsize \textit{Centro de Estudios Cient\'ificos (CECs),}} \\ 
& & \textit{Valdivia, Chile.} \\ \\
& & Dr. Fernando Izaurieta Aranda \\
& & {\normalsize \textit{Departamento de F\'isica,}} \\ 
& & {\normalsize \textit{Universidad de Concepci\'on (UDEC),}} \\
& & \textit{Concepci\'on, Chile.} 
\end{tabular}
\end{large}

\newpage

\newpage

\thispagestyle{empty}
\ \\
\vspace{13cm}

\begin{flushright}
	{\Large {\it Dedicated to my parents. }}
\end{flushright}

\afterpage{\null\newpage}

\doublespacing
\thispagestyle{plain}
\chapter*{Acknowledgments}
\addcontentsline{toc}{chapter}{Acknowledgments}
\begin{quotation}

Throughout the writing of this thesis I have received a great deal of support and assistance.

First of all, my parents, who supported me with love and understanding. All my achievements would not have been possible without them.

I would also like to thank to my thesis advisor Alfredo Pérez, for providing guidance and feedback throughout this project and the whole doctoral process.

I am also very grateful with the members of the Centro de Estudios Científicos, specially with professors Ricardo Troncoso, Jorge Zanelli and David Tempo, my colleagues in the attic, where I spent much of my time in Valdivia, and the administrative staff, mainly to Patricia Fernandoy and Angélica Treimún.

Finally, to Camila, who has been at my side from the very beginning, and whose friendship and company has been undoubtedly the greatest gift.

\end{quotation}

%\newpage
%\thispagestyle{plain}

\afterpage{\null\newpage}
%\newpage
%\thispagestyle{plain}
%\ \\

%\thispagestyle{plain}
%\ \\
%\vspace{3cm}
%\singlespacing
%\hspace{7cm}
%\begin{flushright}
%\begin{Large}
%\textit{``If the doors of perception\\
%were cleansed,\\
%every thing would appear to man\\
%as it is: Infinite.''}
%\vspace{1cm}\\
%Johnny Herrera
%\end{Large}
%\end{flushright}

\ \\

\newpage
\doublespacing
\chapter*{Abstract}
\addcontentsline{toc}{chapter}{Abstract}
\begin{quotation}		
This thesis extends a previously found relation between the integrable KdV hierarchy and the boundary dynamics of pure gravity on AdS$_3$ described in the highest weight gauge, to a more general class of integrable systems associated to three-dimensional gravity on AdS$_3$ and higher spin gravity with
gauge group $SL(N,\mathbb{R})\times SL(N,\mathbb{R})$ in the diagonal gauge. We present new sets of boundary conditions for the (higher spin) gravitational theories on AdS$_3$, where the dynamics of the boundary degrees of freedom is described by two independent left and right members of a hierarchy of integrable equations. For the pure gravity case, the associated hierarchy corresponds to the  Gardner hierarchy, also known as the ``mixed KdV-mKdV'' one, while for the case of higher spin gravity, they are identified with the ``modified Gelfand-Dickey'' hierarchies. The complete integrable structure of the hierarchies, i.e., the phase space, the Poisson brackets and the infinite number of commuting conserved charges, are directly obtained from the asymptotic structure and the conserved surface integrals in the gravitational theories. Consequently, the corresponding Miura transformation is recovered from a purely geometric construction in the bulk. Black hole solutions that fit within our boundary conditions, the Hamiltonian reduction at the boundary and more general thermodynamic ensembles called ``Generalized Gibbs Ensemble'' (GGE) are also discussed.

\end{quotation}

\afterpage{\null\newpage}

\newpage
\thispagestyle{plain}
\doublespacing
\tableofcontents

%\newpage
%\thispagestyle{plain}
%\singlespacing
%\listoffigures

%\newpage

%\thispagestyle{plain}
%\doublespacing
%\chapter*{Resumen}
%\addcontentsline{toc}{chapter}{Resumen}
%\begin{quotation}

\newpage
\thispagestyle{plain}

\newpage

\mainmatter
%\pagenumbering{arabic}
\setcounter{page}{1}
\thispagestyle{plain}

\afterpage{\null\newpage}

\chapter*{Introduction}
\addcontentsline{toc}{chapter}{Introduction}

%\begin{quotation}
%\emph{....}
%\end{quotation}

The asymptotic structure of spacetime plays a fundamental role in
the description of General Relativity in three dimensions. Since this is a topological theory, it does not possess local propagating degrees of freedom, and consequently, its dynamics is completely dictated by the choice of boundary conditions. In the case of a negative cosmological constant, in 1986, J. Brown and M. Henneaux presented a set of boundary conditions \cite{Brown:1986nw}, whose asymptotic symmetries are spanned by two copies of the Virasoro algebra with central charge $c=3l/2G$. Interestingly, this choice of boundary conditions is not unique. There are other possible sensible choices that can be consistently used, with different physical consequences \cite{Compere:2013bya,Troessaert:2013fma,Avery:2013dja,Afshar:2016wfy,Perez:2016vqo,Grumiller:2016pqb,Ojeda:2019xih,Perez:2020klz}. In particular, as was first shown in ref. \cite{Perez:2016vqo} and later extended in refs. \cite{Fuentealba:2017omf, Ojeda:2019xih, Ojeda:2020bgz}, the dynamics of the gravitational excitations at the boundary are related with two-dimensional integrable systems, and therefore, the asymptotic symmetries are spanned by an infinite set of commuting charges.
 
The key of the relationship between three-dimensional gravity and
two-dimensional integrable systems is the precise way in which the
length and time scales are fixed in the asymptotic region of spacetime
in the gravitational theory. Indeed, the asymptotic values of the
lapse and shift functions in an ADM decomposition of the metric, can
be chosen to depend explicitly on the dynamical fields in a very tight
form, compatible with the action principle.

In the case of the KdV hierarchy analyzed in ref. \cite{Perez:2016vqo},
the fall-off of the metric coincides with the one of Brown and Henneaux
when one includes the most general form of the Lagrange multipliers
\cite{Henneaux:2013dra,Bunster:2014mua}. However, the lapse and shift
functions, and consequently the boundary metric of AdS$_{3}$, are
no longer kept fixed at infinity, because they are now allowed to
depend explicitly on the Virasoro currents. As a consequence of this
particular choice of chemical potentials, Einstein equations in the
asymptotic region precisely reduce to two independent left and right
copies of the $k$-th element of the KdV hierarchy. Furthermore, the
complete integrable structure of the KdV system, i.e., the phase space,
the Poisson brackets and the infinite number of commuting conserved
charges, are directly obtained from the asymptotic analysis and the
conserved surface integrals in the gravitational theory. The standard
Brown--Henneaux boundary conditions \cite{Brown:1986nw} are contained
as a particular case ($k=0$) of the KdV hierarchy.

It is worth pointing out that there exists a deep relation between
the KdV hierarchy and two-dimensional conformal field theories (CFT$_{2}$),
indeed, the infinite (quantum) commuting KdV charges can be expressed
as composite operators in terms of the stress tensor of a CFT$_{2}$
\cite{Sasaki:1987mm,Eguchi:1989hs,Bazhanov:1994ft}. This fact has
been recently used to describe Generalized Gibbs ensembles in these
theories \cite{Calabrese:2011vdk,Sotiriadis:2014uza,PhysRevLett.115.157201,Vidmar_2016,deBoer:2016bov,Perez:2016vqo,Pozsgay_2017,Dymarsky:2018lhf,Maloney:2018hdg,Maloney:2018yrz,Dymarsky:2018iwx,Brehm:2019fyy,Dymarsky:2019etq},
as well as constructing black hole solutions carrying non-trivial
KdV charges \cite{Dymarsky:2020tjh}.

Different extensions of this new duality between a three-dimensional gravitational
theory in the bulk with certain specific boundary conditions, and a two-dimensional integrable system at the boundary, have been reported in the literature. In ref. \cite{Fuentealba:2017omf}, the case of General Relativity without cosmological constant was studied. The associated integrable system corresponds to a generalization of
the Hirota-Satsuma coupled KdV system \cite{Hirota:1981wb}, which
possesses a BMS$_{3}$ Poisson structure. Another interesting generalization was proposed in ref. \cite{Melnikov:2018fhb}, where it was shown that the integrable system associated to gravity on AdS$_{3}$ coupled to two $U(1)$ Chern-Simons fields is the Benjamin-Ono$_{2}$ integrable system. 

The main objective of this thesis is to exhibit two new generalizations of this duality that were reported in refs.\cite{Ojeda:2019xih} and \cite{Ojeda:2020bgz}. In particular, in ref. \cite{Ojeda:2019xih} a new set of boundary conditions for pure gravity on AdS$_3$ was proposed, such that boundary dynamics is described by the called Gardner (or mixed KdV-mKdV) hierarchy. These boundary conditions are closely related with the soft hairy ones in ref. \cite{Afshar:2016wfy}, but where now the chemical potentials are chosen to depend on the dynamical fields in a very precise way. This nontrivial way of fixing the chemical potentials is the key to make contact with the integrable system. In ref. \cite{Ojeda:2020bgz} these results were further extended to the case of higher spin gravity on AdS$_3$ with gauge group $SL\left(N,\mathbb{R}\right)\times SL\left(N,\mathbb{R}\right)$, where the corresponding integrable system was shown to be the N-th modified Gelfand-Dickey hierarchy.

\section*{Thesis plan}
%\addcontentsline{toc}{chapter}{Thesis plan}

In chapter \ref{chapter:Integrable systems}, a brief review of some general properties of the integrable systems is presented. In particular, we focus on the integrable systems which are relevant for the main goal of this thesis, i.e., the (modified) KdV and the family of (modified) Gelfand-Dickey hierarchies. In chapter \ref{chapter:Gardner Part} we discuss the new boundary conditions for pure gravity on AdS$_3$ reported in \cite{Ojeda:2019xih} which were shown to be related with the Gardner (or mixed KdV-mKdV) hierarchy. The higher spin generalization proposed in ref. \cite{Ojeda:2020bgz} is described in chapter \ref{chapter:Gelfand-Dickey hierarchy Part}. Finally we include a section devoted to the conclusions. Seven appendices with some technical details are also included.

\newpage{}

\chapter{An introduction to integrable systems\label{chapter:Integrable systems}}

The integrability of a differential equation is determined by the
existence of integrals of motion or conserved quantities. Formally,
Liouville's theorem in classical mechanics states the conditions under
which the equations of motion of a dynamical system can always be
solved by a mathematical procedure. For a system with a finite number
of degrees of freedom, the Liouville theorem says:

``\textit{A Hamiltonian system with a $2N$-dimensional phase space
	is integrable by the method of quadratures if and only if there exist
}\textbf{\textit{$N$}}\textit{, functionally }\textbf{\textit{independent}}\textit{
	conserved quantities which are in }\textbf{\textit{involution}}\textit{.}''

For a system with an infinite number of degrees of freedom, it is necessary to consider an infinite number of independent conserved quantities
in involution.

In this first chapter, the most relevant aspects of non-linear integrable systems are reviewed. It will be addressed using two equivalent formulations. The first corresponds to the bi-Hamiltonian formulation needed to make contact with the asymptotic dynamics of gravitational theories in $2+1$ dimensions, and the second one is the Lax pair formulation, which clarifies the role played by the Miura transformation and allows us to generalize the results in ref. \cite{Perez:2016vqo, Ojeda:2019xih} to the $3$ dimensional gravitation theory endowed with higher spin fields \cite{Ojeda:2020bgz}.

\section{Korteweg-de Vries hierarchy\label{sec:KdV hierarchy}}

The Korteweg-de Vries equation is one of the most important non-linear
partial differential equations, which historically helped to understand
in detail the problem of integrability. This is a nonlinear dispersive
wave equation which allows to describe \textquotedblleft solitary
waves,\textquotedblright{} observed by first time in 1834 by the Scottish engineer John Scott Russell (1808--1882). Remarkably, the KdV equation appears in many different branches of mathematics and physics.

\subsection{KdV equation as a Hamiltonian system}

The Korteweg-de Vries equation is given by

\begin{equation}
	\dot{\mathcal{L}}=\frac{3}{2}\mathcal{L}\mathcal{L^{\prime}}+\frac{1}{4}\mathcal{L^{\prime\prime\prime}},\label{eq:KdVEquation1}
\end{equation}
where dots and primes denote derivatives with respect to the temporal
and spatial coordinates respectively.

The dynamics of eq. \eqref{eq:KdVEquation1} can be described
using the Hamiltonian formalism, where the Poisson bracket of two
arbitrary functional $F$ and $G$ is given by

\begin{equation}
	\left\{ F,G\right\} _{1}=\int d\phi\left(\frac{\delta F}{\delta\mathcal{L}}\mathcal{D}_{\left(1\right)}\frac{\delta G}{\delta\mathcal{L}}\right),\label{eq:PoissonBracketKdV1}
\end{equation}
where $\mathcal{D}_{\left(1\right)}:=2\partial_{\phi}$. 

The Poisson bracket \eqref{eq:PoissonBracketKdV1}, together with the Hamiltonian

\begin{equation}
	H_{\left(1\right)}^{\text{KdV}}=\oint d\phi\left(\frac{1}{8}\mathcal{L}^{3}-\frac{1}{16}\mathcal{L}^{\prime2}\right),\label{eq:HKdV1}
\end{equation}
precisely reproduce the KdV equation \eqref{eq:KdVEquation1}, which can be rewritten as

\begin{equation}
	\dot{\mathcal{L}}=\left\{ \mathcal{L},H_{\left(1\right)}^{\text{KdV}}\right\} _{1}=\mathcal{D}_{\left(1\right)}\left(\frac{\delta H_{\left(1\right)}^{\text{KdV}}}{\delta\mathcal{L}}\right).\label{eq:KdVEq1}
\end{equation}

The integrability of the KdV equation and the existence of
a hierarchy of integrable equations, rely on the fact that this system is actually
bi-Hamiltonian. Indeed, there exists an alternative symplectic structure
characterized by the following operator

\begin{equation}
	\mathcal{D}_{\left(2\right)}:=\partial_{\phi}\mathcal{L}+2\mathcal{L}\partial_{\phi}+\frac{1}{2}\partial_{\phi}^{3}.\label{eq:D2KdV}
\end{equation}
Consequently, the Poisson bracket of two arbitrary functionals $F$
and $G$ associated with the operator $\mathcal{D}_{\left(2\right)}$
is
\begin{equation}
	\left\{ F,G\right\} _{2}=\int d\phi\left(\frac{\delta F}{\delta\mathcal{L}}\mathcal{D}_{\left(2\right)}\frac{\delta G}{\delta\mathcal{L}}\right).\label{eq:PoissonBracketKdV2}
\end{equation}
Using the second Poisson bracket \eqref{eq:PoissonBracketKdV2} and the Hamiltonian
\begin{equation}
	H_{\left(0\right)}^{\text{KdV}}=\oint d\phi\left(\frac{1}{4}\mathcal{L}^{2}\right),\label{eq:HKdV0}
\end{equation}
we can alternatively rewrite the KdV equation as
\begin{equation}
	\dot{\mathcal{L}}=\left\{ \mathcal{L},H_{\left(0\right)}^{\text{KdV}}\right\} _{2}=\mathcal{D}_{\left(2\right)}\left(\frac{\delta H_{\left(0\right)}^{\text{KdV}}}{\delta\mathcal{L}}\right).\label{eq:KdVEq2}
\end{equation}

\subsection{Recursion relation and conserved charges of the KdV hierarchy}

From eqs. \eqref{eq:KdVEq1} and \eqref{eq:KdVEq2}, the following relation is obtained
\[
\dot{\mathcal{L}}=\mathcal{D}_{\left(1\right)}\left(\frac{\delta H_{\left(1\right)}^{\text{KdV}}}{\delta\mathcal{L}}\right)=\mathcal{D}_{\left(2\right)}\left(\frac{\delta H_{\left(0\right)}^{\text{KdV}}}{\delta\mathcal{L}}\right),
\]
which relates the gradient of the Hamiltonian $H_{\left(1\right)}^{\text{KdV}}$
with the gradient of the Hamiltonian $H_{\left(0\right)}^{\text{KdV}}$
as 

\begin{equation}
	\left(\frac{\delta H_{\left(1\right)}^{\text{KdV}}}{\delta\mathcal{L}}\right)=\mathcal{D}_{\left(1\right)}^{-1}\mathcal{D}_{\left(2\right)}\left(\frac{\delta H_{\left(0\right)}^{\text{KdV}}}{\delta\mathcal{L}}\right).\label{eq:Recursion Relation KdV 0 and 1}
\end{equation}

This can be used to define a recursion relation that associates the gradient of the $k$-th Hamiltonian $H_{\left(k\right)}^{\text{KdV}}$ with the gradient of the $\left(k-1\right)$-th Hamiltonian $H_{\left(k-1\right)}^{\text{KdV}}$ through

\begin{equation}
	R_{\left(k\right)}=\mathcal{D}_{\left(1\right)}^{-1}\mathcal{D}_{\left(2\right)}R_{\left(k-1\right)},\label{eq:Recursion Relation KdV}
\end{equation}
with 
\begin{equation}
	R_{\left(k\right)}:=\frac{\delta H_{\left(k\right)}^{\text{KdV}}}{\delta\mathcal{L}},\label{eq:Gelfand-Dickey KdV}
\end{equation}
which are the called Gelfand-Dickey polynomials associated to the
$k$-th Hamiltonian of the hierarchy.

Using the recursion relation \eqref{eq:Recursion Relation KdV}, one can construct an infinite number of Hamiltonians $H_{\left(k\right)}^{\text{KdV}}$. Each of these Hamiltonians defines an independent conserved charge of the KdV equation \eqref{eq:KdVEquation1}.

Finally, one can prove that the $H_{\left(k\right)}^{\text{KdV}}$
are in involution with respect to both Poisson brackets. Let us consider $k>m$,
the Poisson bracket associated to the operator $\mathcal{D}_{\left(2\right)}$
of two conserved charges can be written as

\begin{eqnarray*}
	\left\{ H_{\left(k\right)}^{\text{KdV}},H_{\left(m\right)}^{\text{KdV}}\right\} _{2} & = & \int d\phi\left(R_{\left(k\right)}\mathcal{D}_{\left(2\right)}R_{\left(m\right)}\right),\\
	& = & \int d\phi\left(R_{\left(k\right)}\mathcal{D}_{\left(1\right)}R_{\left(m+1\right)}\right),\\
	& = & -\int d\phi\left(\mathcal{D}_{\left(1\right)}R_{\left(k\right)}R_{\left(m+1\right)}\right),\\
	& = & -\int d\phi\left(\mathcal{D}_{\left(2\right)}R_{\left(k-1\right)}R_{\left(m+1\right)}\right),\\
	& = & -\left\{ H_{\left(m+1\right)}^{\text{KdV}},H_{\left(k-1\right)}^{\text{KdV}}\right\} _{2},\\
	& = & \left\{ H_{\left(k-1\right)}^{\text{KdV}},H_{\left(m+1\right)}^{\text{KdV}}\right\} _{2}.
\end{eqnarray*}
Applying this procedure $m-k$ times, one finally gets
\[
\left\{ H_{\left(k\right)}^{\text{KdV}},H_{\left(m\right)}^{\text{KdV}}\right\} _{2}=\left\{ H_{\left(m\right)}^{\text{KdV}},H_{\left(k\right)}^{\text{KdV}}\right\} _{2}=0.
\]
This proves that all the conserved quantities are in involution. Furthermore, we can use these conserved quantities $H_{\left(k\right)}^{\text{KdV}}$ as Hamiltonians of new differential equations. Therefore, it is then possible to define a complete hierarchy of integrable equations. The $k$-th equation has the form

\[
\dot{\mathcal{L}}_{\left(k\right)}=\left\{ \mathcal{L},H_{\left(k\right)}^{\text{KdV}}\right\} _{1}=\left\{ \mathcal{L},H_{\left(k-1\right)}^{\text{KdV}}\right\} _{2}.
\]
For example, the second member of the hierarchy can be obtained with
the operator $\mathcal{D}_{\left(2\right)}$ and the Hamiltonian $H_{\left(1\right)}^{\text{KdV}}$. It is given by

\begin{equation}
	\dot{\mathcal{L}}=\frac{15}{8}\mathcal{L}^{2}\mathcal{L}^{\prime}+\frac{5}{4}\mathcal{L}^{\prime}\mathcal{L}^{\prime\prime}+\frac{5}{8}\mathcal{L}\mathcal{L}^{\prime\prime\prime}+\frac{1}{16}\mathcal{L}^{\prime\prime\prime\prime\prime}.\label{eq:KdVEquation2}
\end{equation}

\section{Lax pairs\label{sec:Lax pairs}}

In 1967, Gardner, Greene, Kruskal, and Miura introduced in ref. \cite{Gardner1967} the inverse scattering transformation\emph{ }in the study
of exact solutions of the initial-value problem for the KdV equation. This led to a new method for solving non-linear partial differential equations called the inverse scattering method, which is the analogue of the Fourier transformation method for linear partial differential equations. One year after that, based on this method, Peter Lax showed in ref. \cite{Lax:1968} that the following pair of differential operators

\begin{eqnarray}
	L & = & \partial_{\phi}^{2}+\mathcal{L},\label{eq:Lax L - KdV}\\
	P & = & \partial_{\phi}^{3}+\frac{3}{2}\mathcal{L}\partial_{\phi}+\frac{3}{4}\mathcal{L}^{\prime},\label{eq:Lax P - KdV}
\end{eqnarray}
together with the equation

\begin{equation}
	\partial_{t}L=\left[P,L\right],\label{eq:Lax equation}
\end{equation}
reproduce the KdV equation.

Equation \eqref{eq:Lax equation} is called the isospectral Lax equation,
and relates an integrable non-linear partial differential equation
like \eqref{eq:KdVEquation1} with the following pair of equations

\begin{eqnarray}
	L\psi & = & \lambda\psi,\label{eq:Spectral Lax}\\
	\dot{\psi} & = & P\psi.\label{eq:Time evolution eigenfunction}
\end{eqnarray}
The first equation is called the spectral equation of $L$, and the second
one describes the time evolution for the eigenfunction $\psi$. Eq. \eqref{eq:Lax equation}
guarantees that the eigenvalue $\lambda$ is time independent, as
can be seen by taking the time derivative of  eq. \eqref{eq:Spectral Lax}
\[
\partial_{t}L\psi+L\partial_{t}\psi=\partial_{t}\lambda\psi+\lambda\partial_{t}\psi.
\]
If one replaces eqs. \eqref{eq:Time evolution eigenfunction} and \eqref{eq:Spectral Lax} the following condition is obtained

\[
\left(\partial_{t}L-\left[P,L\right]\right)\psi=\partial_{t}\lambda\psi.
\]
Therefore, if eq. \eqref{eq:Lax equation} is satisfied, then $\partial_{t}\lambda=0$,
i.e., the KdV equation preserves the spectrum
of the operator $L$. Furthermore, all the members of the KdV hierarchy fulfill this property.

The second operator $P$ is given by

\[
P=\left(L^{\frac{3}{2}}\right)_{\geq0}=\partial_{\phi}^{3}+\frac{3}{2}\mathcal{L}\partial_{\phi}+\frac{3}{4}\mathcal{L}^{\prime},
\]
where $\left(\ldots\right)_{\geq0}$ correspond to the purely differential
part of the operator (See Appendix \ref{Appendix 1: Op P}).

As was discussed above, the entire KdV hierarchy preserves the spectrum
of the operator $L$. Thus, the $k$-th equation of the hierarchy
is obtained by using the following commutator

\[
\dot{\mathcal{L}}=\left[\left(L^{\frac{2k+1}{2}}\right)_{\geq0},L\right].
\]

\subsection{Gelfand-Dickey hierarchies}

In order to generalize the previous construction to higher order operators, one can consider the following $n$-th order differential operator\footnote{More general cases can be found in ref. \cite{Blaszak1998}}

\begin{equation}
	L=\partial_{\phi}^{n}+\mathcal{U}_{n-2}\partial_{\phi}^{n-2}+\ldots+\mathcal{U}_{1}\partial_{\phi}+\mathcal{U}_{0},\label{eq:n-th order Lax}
\end{equation}
where the coefficients $\mathcal{U}_{n-2},\ldots,\mathcal{U}_{0}$
are the $n-1$ real or complex fields of the theory under discussion.

The operator \eqref{eq:n-th order Lax}, together with the operator

\[
P_{m}=\left(L^{\frac{m}{n}}\right)_{\geq0},
\]
are called a Lax pair $\left(L,P_{m}\right)$. The isospectral Lax equation then takes the form

\begin{equation}
	\frac{\partial L}{\partial t_{\left(m\right)}}=\left[\left(L^{\frac{m}{n}}\right)_{\geq0},L\right],\label{eq:Lax equation final}
\end{equation}
defining the equations associated to the Gelfand-Dickey hierarchy. In particular, for the choice of integer $\left(n=2,\:m=3\right)$ one recovers the KdV equation.

\subsection*{Boussinesq equation $\left(n=3,\:m=2\right)$}

As an example, let us consider the following pair of operators $L=\partial_{\phi}^{3}+\mathcal{L}\partial_{\phi}+\mathcal{W}$
and $P_{2}=\left(L^{\frac{2}{3}}\right)_{\geq0}=\partial_{\phi}^{2}+\frac{2}{3}\mathcal{L}$. Equation \eqref{eq:Lax equation final} then implies

\begin{eqnarray*}
\dot{\mathcal{L}} & = & -\mathcal{L}^{\prime\prime}+2\mathcal{W}^{\prime},\\
\dot{\mathcal{W}} & = & \mathcal{W}^{\prime\prime}-\frac{2}{3}\mathcal{L}\mathcal{L}^{\prime}-\frac{2}{3}\mathcal{L}^{\prime\prime\prime}.
\end{eqnarray*}
Combining both equations and eliminating the field $\mathcal{W}$, one finds the following equation that must be satisfied by $\mathcal{L}$

\begin{equation}
	\ddot{\mathcal{L}}=-\frac{4}{3}\left(\mathcal{L}\mathcal{L}^{\prime}\right)^{\prime}-\frac{1}{3}\mathcal{L}^{\prime\prime\prime\prime}.\label{eq:Lax - Boussinesq equation}
\end{equation}
This is known as the ``Good'' Boussinesq equation.

\subsection{Poisson structures}

The Hamiltonian formulation presented in sec. \ref{sec:KdV hierarchy},
can be recovered by using the Lax pair to define the first and second Poisson structures. Following \cite{Blaszak1998}, the first Poisson structure associated to the $n$-th order differential operator $L$ in eq. \eqref{eq:n-th order Lax} is given by the commutator

\begin{equation}
	\dot{L}=\Theta_{1}\left(L\right)\nabla H=\left[L,\nabla H\right]_{\geq0},\label{eq:First Hamiltonian Op Lax}
\end{equation}
where $\nabla H$ is a pseudo-differential operator defined by\footnote{See Appendix \ref{Appendix 1: Op P} for some properties of the pseudo-differential
	operators.}

\[
\nabla H=\sum_{i=0}^{n-2}\partial_{\phi}^{-i-1}\frac{\delta H}{\delta\mathcal{U}_{i}}.
\]
The second Poisson structure is given by
\begin{equation}
	\dot{L}=\Theta_{2}\left(L\right)\nabla H=\left(L\,\nabla H\right)_{\geq0}L-L\left(\nabla H\,L\right)_{\geq0}+\frac{1}{n}\left[\partial^{-1}\text{res}\left(\left[\nabla H,L\right]\right),L\right], \label{eq:Second Hamiltonian Op Lax}
\end{equation}
where $\text{res}\left(\ldots\right)$ means the residue of the operator,
i.e., the coefficient of the order $\partial^{-1}$. Equations \eqref{eq:First Hamiltonian Op Lax} and \eqref{eq:Second Hamiltonian Op Lax} can be rewritten in the following form

\begin{equation}
\dot{\mathcal{U}}_{\left(k\right)} = \mathscr{O}_{1}\frac{\delta H_{\left(k+1\right)}}{\delta\mathcal{U}}=\mathscr{O}_{2}\frac{\delta H_{\left(k\right)}}{\delta\mathcal{U}},\label{eq: Bi-Hamiltonian equations GD}
\end{equation}
where $\mathcal{U}$ denotes the set of fields $\left(\mathcal{U}_{n-2},\ldots,\mathcal{U}_{0}\right)$
and the operators $\mathscr{O}_{1}$ and $\mathscr{O}_{2}$ are certain differential $\left(n-1\right)\times\left(n-1\right)$
matrix operators.

\subsection{Modified Gelfand-Dickey hierarchies}

The $n$-th order differential operator in eq. \eqref{eq:n-th order Lax}
can be factorized as

\begin{equation}
	L=\left(\partial_{\phi}-\mathcal{V}_{n-1}\right)\left(\partial_{\phi}-\mathcal{V}_{n-2}\right)\cdots\left(\partial_{\phi}-\mathcal{V}_{0}\right),\label{eq:n-th order Lax - Miura}
\end{equation}
If we compare this expression with \eqref{eq:n-th order Lax}, we find an equation which allow us to write the field $\mathcal{V}_{n-1}$ in terms of the remaining $\left(n-1\right)$-$\mathcal{V}$ fields. This can be traced back to the absence of a term in the order $\partial^{n-1}$ in eq. \eqref{eq:n-th order Lax}. Now we can write a set of $\left(n-1\right)$-differential equations which relate the fields $\mathcal{U}$ with the fields $\mathcal{V}$ and its spatial derivatives.

\begin{equation}
	\mathcal{U}_{i}=\mathcal{U}_{i}\left(\mathcal{V}_{0},\ldots,\mathcal{V}_{n-2}\right),\qquad i=0,\ldots,n-2.\label{eq: Lax - Miura Transformation}
\end{equation}
These equations define the Miura transformation. The first example
was discovered in 1968 by Robert Miura in the study of Korteweg-de Vries equation \cite{Miura:1968}.

The fields $\mathcal{V}$ satisfy a system of non-linear partial differential
equations called modified Gelfand-Dickey equations. These equations
are also bi-Hamiltonian, i.e., they are endowed with their first and
second Poisson structures.

To obtain the Poisson structures, let us take the time derivative of the Miura transformation in eq. \eqref{eq: Lax - Miura Transformation}

\begin{equation}
	\dot{\mathcal{U}}_{\left(k\right)}=\left(\frac{\partial\mathcal{U}}{\partial\mathcal{V}}\right)\dot{\mathcal{V}}_{\left(k\right)}=M\dot{\mathcal{V}}_{\left(k\right)},\label{eq: Modified GD u dot =00003D M v dot}
\end{equation}
where $M=\left(\frac{\partial\mathcal{U}}{\partial\mathcal{V}}\right)$
is a $\left(n-1\right)\times\left(n-1\right)$-matrix differential
operator called the Fréchet derivative, whose formal adjoint $M^{\dagger}$
relates the Gelfand-Dickey polynomials of both systems

\begin{equation}
	\frac{\delta H_{\left(k\right)}}{\delta\mathcal{V}}=M^{\dagger}\frac{\delta H_{\left(k\right)}}{\delta\mathcal{U}}.\label{eq:Modified GD dH/dv =00003D Mdaga dH/du}
\end{equation}
By means of eqs. \eqref{eq: Bi-Hamiltonian equations GD}, \eqref{eq: Modified GD u dot =00003D M v dot}
and \eqref{eq:Modified GD dH/dv =00003D Mdaga dH/du} we can obtain
the equations of the modified Gelfand-Dickey hierarchy
\begin{equation}
	\dot{\mathcal{V}}_{\left(k\right)}=\mathscr{D}_{1}\frac{\delta H_{\left(k\right)}}{\delta\mathcal{V}}=\mathscr{D}_{2}\frac{\delta H_{\left(k-1\right)}}{\delta\mathcal{V}},\label{eq:Bi-Hamiltonian equations mGD}
\end{equation}
where the operators $\mathscr{D}_{1}$ and $\mathscr{D}_{2}$ are
$\left(n-1\right)\times\left(n-1\right)$ matrix differential operators
given by

\begin{eqnarray}
	\mathscr{D}_{1} & = & M^{-1}\mathscr{O}_{2}\left(M^{\dagger}\right)^{-1},\label{eq:modified First Hamiltonian Op}\\
	\mathscr{D}_{2} & = & \mathscr{D}_{1}M^{\dagger}\mathscr{O}_{1}^{-1}M\mathscr{D}_{1}.\label{eq:modified Second Hamiltonian Op}
\end{eqnarray}

\subsection*{Modified KdV hierarchy }

The KdV Lax operator \eqref{eq:Lax L - KdV}, rewritten as in eq. \eqref{eq:n-th order Lax - Miura}, takes the form

\[
L=\partial_{\phi}^{2}+\mathcal{L}=\left(\partial_{\phi}-\mathcal{J}\right)\left(\partial_{\phi}+\mathcal{J}\right).
\]
From here one can read the Miura transformation

\begin{equation}
	\mathcal{L}=\mathcal{J}^{\prime}-\mathcal{J}^{2}.\label{eq:Miura Transformation mKdV}
\end{equation}
Besides, the Fréchet derivative and its formal ajdoint are given by

\begin{eqnarray*}
	M & = & \partial_{\phi}-2\mathcal{J},\\
	M^{\dagger} & = & -\partial_{\phi}-2\mathcal{J}.
\end{eqnarray*}
The modified KdV equation can then be obtained by replacing the
Miura transformation \eqref{eq:Miura Transformation mKdV} in the
KdV equation \eqref{eq:KdVEquation1}

\begin{equation}
	\dot{\mathcal{J}}=-\frac{3}{2}\mathcal{J}^{2}\mathcal{J}^{\prime}+\frac{1}{4}\mathcal{J}^{\prime\prime\prime}.\label{eq:mKdVEquation1}
\end{equation}
Finally, the entire modified KdV hierarchy can be obtained using the recursion relation and the following two Poisson structures

\begin{eqnarray}
	\mathscr{D}_{1} & = & -\frac{1}{2}\partial_{\phi},\label{eq:D1mKdV}\\
	\mathscr{D}_{2} & = & -2\partial_{\phi}\left(\mathcal{J}\partial_{\phi}^{-1}\left(\mathcal{J}\partial_{\phi}\right)\right)+\frac{1}{2}\partial_{\phi}^{3}.\label{eq:D2mKdV}
\end{eqnarray}

Remarkably, there is another integrable system called the Gardner hierarchy, which combines both, the KdV and mKdV hierarchies. The first member of the Gardner hierarchy is given by

\begin{equation}
	\dot{\mathcal{J}}= \frac{3}{4}\mathcal{J}\mathcal{J}^{\prime}-\frac{3}{4}\mathcal{J}^{2}\mathcal{J}^{\prime}+\frac{1}{4}\mathcal{J}^{\prime\prime\prime},\label{eq:GardnerEquation1}
\end{equation}
which is known as the Gardner equation. \\

It is worth mention that there is an alternative formulation of the integrable systems discussed in this chapter, called the ``zero curvature formulation.'' In this framework the Lax pair is described by a two-dimensional gauge connection with vanishing field strength which is valued on a certain semisimple Lie algebra. In the case of the (modified) KdV hierarchy the corresponding Lie algebra is $SL(2,\mathbb{R})$, while in the case of the N-th (modified) Gelfand-Dickey hierarchy is $SL(N,\mathbb{R})$. The zero curvature formulation will be widely used in the next chapters in connection with the gravitation theories.

\chapter{Boundary conditions for gravity on AdS$_{3}$ and the Gardner hierarchy \label{chapter:Gardner Part}}

In this chapter, we explore the consequences of extending the ``soft
hairy'' boundary conditions on AdS$_{3}$ introduced in refs. \cite{Afshar:2016wfy,Afshar:2016kjj}
by choosing the chemical potentials as appropriate \emph{local} functions
of the fields, and its relation with integrable systems. For this
class of boundary conditions, the fall-off of the metric near infinity
differs from the one of Brown and Henneaux, and they have the particular
property that, by virtue of the topological nature of three-dimensional
General Relativity, they can also be interpreted as being defined
in the near horizon region of spacetimes with event horizons. 

The corresponding integrable system is shown to be the Gardner hierarchy
of non-linear partial differential equations, also known as the ``mixed
KdV-mKdV'' hierarchy (see e.g. \cite{Ames1992nonlinear}), whose
first member is given by \footnote{Eq. \eqref{eq:GardnerEquation1} is recovered by applying the following rescaling $\mathcal{J} \rightarrow -\frac{a}{b}\mathcal{J}$, $t \rightarrow -\sqrt{2b}\frac{b}{4a^3} t $,  $\phi  \rightarrow \frac{\sqrt{2b}}{a} \phi  $.}
\begin{equation}
	\dot{\mathcal{J}}=3a\mathcal{J}\mathcal{J^{\prime}}+3b\mathcal{J}^{2}\mathcal{J^{\prime}}-2\mathcal{J^{\prime\prime\prime}}.\label{eq:Gardner Equation}
\end{equation}
Here, $a,$ $b$ are arbitrary constants. Equation \eqref{eq:Gardner Equation}
has the very special property that simultaneously combines both, KdV
and modified KdV (mKdV) equations. Remarkably, in spite of the deformations
generated by the arbitrary parameters $a$ and $b$, the integrability
of this equation and of the complete hierarchy associated to it, is
maintained. Indeed, considering an appropriated scaling, when $b=0$
and $a\neq0$, eq. \eqref{eq:Gardner Equation} precisely reduces
to the KdV equation, while when $a=0$ and
$b\neq0$ it coincides with the mKdV equation.

From the point of view of the gravitational theory, there exists a
precise choice of chemical potentials, that extends the boundary conditions
in \cite{Afshar:2016wfy,Afshar:2016kjj}, such that Einstein equations
reduce to two (left and right) copies of eq. \eqref{eq:Gardner Equation}.
Moreover, the infinite number of abelian conserved quantities of the
Gardner hierarchy are obtained from the asymptotic symmetries and
the canonical generators in General Relativity using the Regge-Teitelboim
approach \cite{Regge:1974zd}. These results are generalized to the
whole hierarchy, and can also be applied to the case with a vanishing
cosmological constant. We also show that if the hierarchy is ``extended
backwards'' one recovers the soft hairy boundary conditions of refs.
\cite{Afshar:2016wfy,Afshar:2016kjj} as a particular case.

For a negative cosmological constant, black hole solutions are naturally
accommodated within our boundary conditions, and are described by
static configurations associated to the corresponding member of the
Gardner hierarchy. In particular, BTZ black holes \cite{Banados:1992wn,Banados:1992gq}
are described by constant values of the left and right fields $\mathcal{J}_{\pm}$.
Static solutions with nonconstants $\mathcal{J}_{\pm}$ can also be
found, and describe black holes carrying nontrivial conserved charges.
The thermodynamic properties of the black holes in the ensembles defined
by our boundary conditions will be also shown.

\newpage{}

\section{Asymptotic behavior of the gravitational field\label{sec:G-Asymp}}

\subsection{Chern-Simons formulation of gravity on AdS$_{3}$\label{subsec:Chern-Simons pure gravity AdS3}}

In this section we describe the asymptotic behavior (fall-off) of
the gravitational field without specifying yet what is fixed at the
boundary of spacetime, i.e., without imposing at this step of the
analysis a precise boundary condition. For the purpose of simplicity
and clarity in the presentation, we mostly work in the Chern--Simons
formulation of three-dimensional Einstein gravity with a negative
cosmological constant \cite{Achucarro:1987vz,Witten:1988hc}.

The action for General Relativity on AdS$_{3}$ can be written as
a Chern--Simons action for the gauge group $SL(2,\mathbb{R})\times SL(2,\mathbb{R})$

\[
I=I_{CS}\left[A^{+}\right]-I_{CS}\left[A^{-}\right],
\]
where

\begin{equation}
	I_{CS}\left[A\right]=\frac{\kappa}{4\pi}\int\left\langle AdA+\frac{2}{3}A^{3}\right\rangle .\label{eq:G-Ics}
\end{equation}
Here, the gauge connections $A^{\pm}$ are 1-forms valued on the $sl(2,\mathbb{R})$
algebra, and are related to the vielbein $e$ and spin connection
$\omega$ through $A^{\pm}=\omega\pm e\ell^{-1}$. The level in \eqref{eq:G-Ics}
is given by $\kappa=\ell/4G$, where $\ell$ is the AdS radius and
$G$ is the Newton constant, while the bilinear form $\left\langle ,\right\rangle $
is defined by the trace in the fundamental representation of $sl(2,\mathbb{R})$.
The $sl(2,\mathbb{R})$ generators $L_{n}$ with $n=0,\pm1$, obey
the commutation relations $\left[L_{n},L_{m}\right]=\left(n-m\right)L_{n+m}$,
with non-vanishing components of the bilinear form given by $\left\langle L_{1}L_{-1}\right\rangle =-1$,
and $\left\langle L_{0}^{2}\right\rangle =1/2$.

\newpage

\subsection{Asymptotic form of the gauge field\label{subsec:Asymptotic-form-of1}}

In order to describe the fall-off of the gauge connection, we closely
follow the analysis in refs. \cite{Afshar:2016wfy,Afshar:2016kjj}.
We will assume that the
gauge fields in the asymptotic region take the form

\begin{equation}
	A^{\pm}=b_{\pm}^{-1}\left(d+\mathfrak{a}^{\pm}\right)b_{\pm},\label{eq:G-gaugetransform}
\end{equation}
where $b_{\pm}$ are gauge group elements depending only on the radial
coordinate, and $\mathfrak{a}^{\pm}=\mathfrak{a}_{t}^{\pm}dt+\mathfrak{a}_{\phi}^{\pm}d\phi$
correspond to auxiliary connections that only depend on the temporal
and angular coordinates $t$ and $\phi$, respectively \cite{Coussaert:1995zp}.
It is worth pointing out that, by virtue of the lack of local propagating
degrees of freedom of three-dimensional Einstein gravity, all the
relevant physical information necessary for the asymptotic analysis
is completely encoded in the auxiliary connections $\mathfrak{a}^{\pm}$,
independently of the precise choice of $b_{\pm}$. We will not specify
any particular $b_{\pm}$ until section \ref{sec:G-metric}, where
the metric formulation will be discussed.

The auxiliary connections $\mathfrak{a}^{\pm}$ are chosen to be given
by

\begin{equation}
	\mathfrak{a}^{\pm}=L_{0}\left(\zeta^{\pm}dt\pm\mathcal{J}^{\pm}d\phi\right).\label{eq:G-diagonalg}
\end{equation}
Here, $\mathfrak{a}^{\pm}$ are diagonal matrices in the fundamental
representation of $sl(2,\mathbb{R})$, and by that reason we say that
the connections $\mathfrak{a}^{\pm}$ in eq. \eqref{eq:G-diagonalg}
are written in the ``diagonal gauge.'' The fields $\zeta^{\pm}$
are defined along the temporal components of $\mathfrak{a}^{\pm}$,
and consequently they correspond to Lagrange multipliers (chemical
potentials). On the other hand, $\mathcal{J}^{\pm}$ belong to the
spatial components of the gauge connections and therefore they are
identified as the dynamical fields.

The field equations are determined by the vanishing of the field strength
$F^{\pm}=dA^{\pm}+A^{\pm2}=0$, and take the form

\begin{equation}
	\dot{\mathcal{J}_{\pm}}=\pm\zeta_{\pm}^{\prime}.\label{eq:G-fieldeq}
\end{equation}
Note that there are no time derivatives associated to the fields $\zeta_{\pm}$,
which is consistent with the fact that they are chemical potentials.

It is important to emphasize that, as was pointed out in refs. \cite{Afshar:2016wfy,Afshar:2016kjj},
one can also interpret the connections in \eqref{eq:G-gaugetransform}
and \eqref{eq:G-diagonalg} as describing the behavior of the fields
in a region near a horizon. Indeed, the reconstructed metric can always
be expressed (in a co-rotating frame) as the direct product of the
two-dimensional Rindler metric times $S^{1}$, together with appropriate
deviations from it. Furthermore, the charges are independent of the
gauge group element $b_{\pm}\left(r\right)$, and consequently they
do not depend on the precise value of the radial coordinate where
they are evaluated. In this sense, one can also interpret the present
analysis as describing ``near horizon boundary conditions.'' This
near horizon interpretation can also be extended to higher spacetime
dimensions \cite{Grumiller2020ssng}.

\subsection{Consistency with the action principle and canonical generators}

A fundamental requirement that must be guaranteed is the consistency
of the asymptotic form of the gauge connections in eqs. \eqref{eq:G-gaugetransform}
and \eqref{eq:G-diagonalg} with the action principle. In order to
analyze the physical consequences of this requirement, we will use
the Regge-Teitelboim approach \cite{Regge:1974zd} in the canonical
formulation of Chern--Simons theory.

The canonical action can be written as

\[
I_{can}\left[A^{\pm}\right]=-\frac{\kappa}{4\pi}\int dtd^{2}x\varepsilon^{ij}\left\langle A_{i}^{\pm}\dot{A_{j}^{\pm}}-A_{t}^{\pm}F_{ij}^{\pm}\right\rangle +B_{\infty}^{\pm},
\]
where $B_{\infty}^{\pm}$ are boundary terms needed to ensure that
the action principle attains an extremum. Their variations are then
given by

\[
\delta B_{\infty}^{\pm}=-\frac{\kappa}{2\pi}\int dtd\phi\left\langle A_{t}^{\pm}\delta A_{\phi}^{\pm}\right\rangle ,
\]
so that, when one evaluates them for the asymptotic form of the connections
\eqref{eq:G-gaugetransform} and \eqref{eq:G-diagonalg} yields 
\begin{equation}
	\delta B_{\infty}^{\pm}=\mp\frac{\kappa}{4\pi}\int dtd\phi\zeta_{\pm}\delta\mathcal{J_{\pm}}.\label{eq:G-deltaB}
\end{equation}

Consistency of the analysis requires that one must be able to ``take
the delta outside'' in the variation of the boundary terms \eqref{eq:G-deltaB},
i.e., they must be integrable in the functional space\footnote{The requirement of integrability of the boundary terms can relaxed
	in the presence of ingoing or outgoing radiation, where their lack
	of integrability precisely gives the rate of change of the charges
	in time \cite{Barnich:2011mi,Bunster:2018yjr,Bunster:2019mup}. In
	the present case this possibility is not at hand, because General
	Relativity in 2+1-dimensions does not have local propagating degrees
	of freedom.}. This can only be achieved provided one specifies a precise boundary
condition, which turns out to be equivalent to specify what fields
are fixed, i.e., without functional variation, at the boundary of
spacetime. Following ref. \cite{Perez:2016vqo}, a generic possible
choice of boundary conditions is to assume that the chemical potentials
$\zeta_{\pm}$ depend on the dynamical fields $\mathcal{J_{\pm}}$
through

\begin{equation}
	\zeta_{\pm}=\frac{4\pi}{\kappa}\frac{\delta H^{\pm}}{\delta\mathcal{J_{\pm}}},\label{eq:G-zeta}
\end{equation}
where $H^{\pm}=\int d\phi\mathcal{H}^{\pm}[\mathcal{J_{\pm}},\mathcal{J}_{\pm}^{\prime},\mathcal{J_{\pm}}^{\prime\prime},\dots]$
are functionals depending \emph{locally} on the fields $\mathcal{J_{\pm}}$
and their spatial derivatives. Here we also assume that the left and
right sectors are decoupled. With the particular choice of boundary
conditions given by \eqref{eq:G-zeta}, the delta in \eqref{eq:G-deltaB}
can be immediately taken outside, and consequently the boundary terms
necessary to improve the canonical action take the form

\[
B^{\pm}=\mp\int dtH^{\pm}.
\]
An immediate consequence of this choice is that the total energy of
the system, defined as the on-shell value of the generator of translations
in time, can be directly written in terms of the ``Hamiltonians''
$H^{\pm}$ as
\begin{equation}
	E=H^{+}+H^{-}.\label{eq:G-Energia}
\end{equation}

The conserved charges associated to the asymptotic symmetries are
also sensitive to the choice of boundary conditions, but in a more
subtle way. The form of the connections $\mathfrak{a}^{\pm}$ in eq.
\eqref{eq:G-diagonalg} is preserved under gauge transformations $\delta\mathfrak{a}^{\pm}=d\lambda^{\pm}+\left[\mathfrak{a}^{\pm},\lambda^{\pm}\right]$,
with gauge parameters $\lambda^{\pm}$ given by\footnote{Extra terms along $L_{1}$ and $L_{-1}$ might also be added, however
	they are pure gauge.} 
\[
\lambda^{\pm}=\eta_{\pm}L_{0},
\]
provided the fields $\mathcal{J}^{\pm}$ and the chemical potentials
$\zeta^{\pm}$ transform as

\begin{equation}
	\delta\mathcal{J_{\pm}}=\pm\eta_{\pm}^{\prime},\label{eq:G-deltaJ}
\end{equation}
\begin{equation}
	\delta\zeta_{\pm}=\dot{\eta}_{\pm}.\label{eq:G-deltaz}
\end{equation}

Following the Regge-Teitelboim approach \cite{Regge:1974zd}, the
variation of the conserved charges reads

\begin{equation}
	\delta Q^{\pm}\left[\eta_{\pm}\right]=\pm\frac{\kappa}{4\pi}\oint d\phi\eta_{\pm}\delta\mathcal{J_{\pm}}.\label{eq:G-deltaQ}
\end{equation}
Here, the parameters $\eta^{\pm}$ are not arbitrary, because now
the chemical potentials $\zeta^{\pm}$ depend on the fields $\mathcal{J}^{\pm}$
and their spatial derivatives. Hence, one must use the chain rule
in the variation at the left hand-side of eq. \eqref{eq:G-deltaz},
giving the following first order differential equations in time for
$\eta^{\pm}$

\begin{equation}
	\dot{\eta}_{\pm}=\pm\frac{4\pi}{\kappa}\frac{\delta}{\delta\mathcal{J_{\pm}}}\int d\phi\frac{\delta H^{\pm}}{\delta\mathcal{J_{\pm}}}\partial_{\phi}\eta_{\pm}.\label{eq:G-etapunto}
\end{equation}
Generically, these differential equations will depend explicitly on
$\mathcal{J_{\pm}}$, and consequently finding explicit solutions
to them becomes a very hard task. However, this can be achieved for
certain special choices of boundary conditions, which are related
to the integrable hierarchy associated to the Gardner equation. This
will be discussed in detail in the next section.

\section{Different choices of boundary conditions in the diagonal gauge\label{sec:G-Different-choices-of}}

\subsection{Soft hairy boundary conditions\label{sec:G-softhairy}}

A simple choice of boundary conditions corresponds to fix the chemical
potentials $\zeta_{\pm}$ at boundary, such that they are arbitrary
functions without variation, i.e., $\delta\zeta_{\pm}=0$. This possibility
was analyzed in detail in refs. \cite{Afshar:2016wfy,Afshar:2016kjj},
and it was termed ``soft hairy boundary conditions.'' The asymptotic
symmetry algebra is then spanned by the generators

\[
Q_{\text{soft hairy}}^{\pm}\left[\eta_{\pm}\right]=\pm\frac{\kappa}{4\pi}\oint d\phi\eta_{\pm}\mathcal{J_{\pm}},
\]
where, by the consistency with time evolution, the parameters $\eta_{\pm}$
are arbitrary functions without variation ($\delta\eta_{\pm}=0$)
. The global charges are then characterized by $\mathcal{J_{\pm}}$,
which obey the following Poisson brackets

\begin{equation}
	\left\{ \mathcal{J_{\pm}}\left(\phi\right),\mathcal{J_{\pm}}\left(\bar{\phi}\right)\right\} =\pm\frac{4\pi}{\kappa}\partial_{\phi}\delta\left(\phi-\bar{\phi}\right).\label{eq:G-Poissonbracket}
\end{equation}
The asymptotic symmetry algebra is then given by two copies of $\hat{u}\left(1\right)$
current algebras.

In a co-rotating frame, $\zeta_{+}=\zeta_{-}=const.$, the generator
of time evolution is identified with the sum of the left and right
zero modes of $\mathcal{J_{\pm}}$, that commute with all the members
of the algebra. In this sense, one can say that the higher modes describe
``soft hair excitations'' in the sense of Hawking, Perry and Strominger
\cite{Hawking:2016msc,Hawking:2016sgy}, because they do not change
the energy of the gravitational configuration. Besides, in this frame,
it is possible to find solutions with non-extremal horizons which
are diffeomorphic to BTZ black holes, and that are endowed with not
trivial soft hair charges. These solutions were called ``black flowers''\footnote{A different type of black flower solution was found in ref. \cite{Barnich:2015dvt}
	in the context of new massive gravity \cite{Bergshoeff:2009hq,Bergshoeff:2009aq}.} \cite{Afshar:2016wfy,Afshar:2016kjj}.

\subsection{Gardner equation\label{sec:G-mixedmkdv}}

A different choice of boundary conditions that makes contact with
the Gardner (mixed KdV-mKdV) equation is $\zeta^{\pm}=\frac{3}{2}a\mathcal{J_{\pm}}^{2}+b\mathcal{J_{\pm}}^{3}-2\mathcal{J_{\pm}^{\prime\prime}}$,
which according to eq. \eqref{eq:G-zeta} corresponds to use the following
Hamiltonians
\begin{equation}
	H_{\left(1\right)}^{\pm}=\frac{\kappa}{4\pi}\oint d\phi\left(\frac{1}{2}a\mathcal{J}_{\pm}^{3}+\frac{1}{4}b\mathcal{J}_{\pm}^{4}+\mathcal{J}_{\pm}'^{2}\right).\label{eq:G-H(1)}
\end{equation}
With this particular choice of chemical potentials, Einstein equations
\eqref{eq:G-fieldeq} precisely reduce to two independent left and
right copies of the Gardner equation \eqref{eq:Gardner Equation}, that read
\begin{equation}
	\dot{\mathcal{J}_{\pm}}=\pm\left(3a\mathcal{J_{\pm}}\mathcal{J_{\pm}^{\prime}}+3b\mathcal{J_{\pm}}^{2}\mathcal{J_{\pm}^{\prime}}-2\mathcal{J_{\pm}^{\prime\prime\prime}}\right).\label{eq:G-mixedkdvmkdv}
\end{equation}
Equations \eqref{eq:G-etapunto}, describing the time evolution of
the gauge parameters $\eta^{\pm}$, now take the form 
\begin{equation}
	\dot{\eta}_{\pm}=\pm\left(3a\mathcal{J_{\pm}}\eta_{\pm}^{\prime}+3b\mathcal{J_{\pm}}^{2}\eta_{\pm}^{\prime}-2\eta_{\pm}^{\prime\prime\prime}\right),\label{eq:G-etapunto2}
\end{equation}
where the explicit dependence on $\mathcal{J_{\pm}}$ is manifest.
These equations are linear in $\eta_{\pm}$, and by virtue of the
integrability of the system, it is possible to find their general
solutions under the assumption that they depend locally on $\mathcal{J_{\pm}}$
and their spatial derivatives. Indeed, the gauge parameters $\eta_{\pm}$
obeying eq. \eqref{eq:G-etapunto2}, are expressed as a linear combination
of ``generalized Gelfand-Dickey polynomials'' $R_{\left(n\right)}^{\pm}$,
i.e.,
\begin{equation}
	\eta_{\pm}=\frac{4\pi}{\kappa}\sum_{n\geq0}\alpha_{\left(n\right)}^{\pm}R_{\left(n\right)}^{\pm},\label{eq:G-etasol}
\end{equation}
where $\alpha_{\left(n\right)}^{\pm}$ are arbitrary constants, and
the polynomials $R_{\left(n\right)}^{\pm}$ are defined through the
following recursion relation
\begin{equation}
	\partial_{\phi}R_{(n+1)}^{\pm}=\mathcal{D}_{\phi}R_{(n)}^{\pm},\label{eq:G-recurrenciarel}
\end{equation}
where $\mathcal{D}_{\phi}$ is a non--local operator given by

\begin{equation}
	\mathcal{D}_{\phi}:=a\left(\partial_{\phi}\mathcal{J_{\pm}}+2\mathcal{J_{\pm}}\partial_{\phi}\right)+2b\partial_{\phi}\left(\mathcal{J_{\pm}}\partial_{\phi}^{-1}\left(\mathcal{J_{\pm}}\partial_{\phi}\right)\right)-2\partial_{\phi}^{3}.\label{eq:G-Dcaligrafico}
\end{equation}
The polynomials $R_{\left(n\right)}^{\pm}$ can be expressed as the
``gradient'' of the Hamiltonians $H_{\left(n\right)}^{\pm}$ of
the integrable system, i.e., 
\begin{equation}
	R_{\left(n\right)}^{\pm}=\frac{\delta H_{\left(n\right)}^{\pm}}{\delta\mathcal{J_{\pm}}}.\label{eq:G-RHJ}
\end{equation}
The first generalized Gelfand-Dickey polynomials together with their
corresponding Hamiltonians are explicitly displayed in appendix \ref{Appendix 2: Gelfand-Dickey pol Gardner}.

If one replaces the general solution of eqs. \eqref{eq:G-etapunto2},
given by \eqref{eq:G-etasol}, into the expression for the variation
of the canonical generators \eqref{eq:G-deltaQ}, one can take immediately
the delta outside, and write the conserved charges as

\begin{equation}
	Q^{\pm}=\pm\sum_{n\geq0}\alpha_{\left(n\right)}^{\pm}H_{\left(n\right)}^{\pm},\label{eq:G-QHn}
\end{equation}
where the Hamiltonians $H_{\left(n\right)}^{\pm}$ are in involution
with respect to the Poisson brackets \eqref{eq:G-Poissonbracket},
\[
\left\{ H_{\left(n\right)}^{\pm},H_{\left(m\right)}^{\pm}\right\} =0.
\]

The presence of the operator $\mathcal{D}_{\phi}$ in the recurrence
relation for the generalized Gelfand-Dickey polynomials \eqref{eq:G-recurrenciarel},
is related to the fact that this integrable system is bi--Hamiltonian.
Consequently, it is possible to define a second Poisson bracket between
the dynamical fields
\begin{equation}
	\left\{ \mathcal{J_{\pm}}\left(\phi\right),\mathcal{J_{\pm}}\left(\bar{\phi}\right)\right\} _{2}:=\pm\frac{4\pi}{\kappa}\mathcal{D}_{\phi}\delta\left(\phi-\bar{\phi}\right),\label{eq:G-Poissonbracket2}
\end{equation}
so that the Gardner equations can be written as 
\[
\dot{\mathcal{J}_{\pm}}=\left\{ \mathcal{J_{\pm}},H_{\left(1\right)}^{\pm}\right\} =\left\{ \mathcal{J_{\pm}},H_{\left(0\right)}^{\pm}\right\} _{2},
\]
where the Hamiltonians $H_{\left(1\right)}^{\pm}$ are given by eq.
\eqref{eq:G-H(1)}, while the Hamiltonians $H_{\left(0\right)}^{\pm}$
take the form
\begin{equation}
	H_{(0)}^{\pm}=\frac{\kappa}{4\pi}\oint d\phi\left(\frac{1}{2}\mathcal{J}_{\pm}^{2}\right).\label{eq:G-H0}
\end{equation}

It is worth pointing out that only the first Poisson bracket structure,
given by eq. \eqref{eq:G-Poissonbracket}, is obtained from the gravitational
theory using the Dirac method for constrained systems. It is not clear
how to obtain the second Poisson structure \eqref{eq:G-Poissonbracket2}
directly from the canonical structure of General Relativity.

As it was discussed in the introduction, the (left/right) Gardner
equation in \eqref{eq:G-mixedkdvmkdv} has the special property that
when $a=0$ and $b\neq0$, the equation and its whole integrable structure,
precisely reduce to those of the mKdV integrable system. Conversely,
when $b=0$ and $a\neq0$, the integrable system reduces to KdV. In
this case, the operator $\mathcal{D}_{\phi}$ in \eqref{eq:G-Dcaligrafico}
becomes a local differential operator, and the second Poisson structure
associated to it through eq. \eqref{eq:G-Poissonbracket2} corresponds
to two copies of the Virasoro algebra with left and right central
charges given by $c^{\pm}=3\ell/\left(Ga^{2}\right)$. Note that these
central charges do not coincide with the ones of Brown and Henneaux.
This is not surprising because, as we say before, this Poisson structure
does not come directly from the canonical structure of the gravitational
theory.

The Gardner equation does not have additional symmetries besides the
ones generated by the infinite number of commuting Hamiltonians described
above. In particular, it is not invariant under Galilean boosts unless
$b=0$ (KdV case). Nevertheless, it is possible to show that there
exists a particular Galilean boost with parameter $\omega=3a^{2}/\left(4b\right)$,
together with a shift in $\mathcal{J_{\pm}}$ given by $\mathcal{J_{\pm}}\rightarrow\mathcal{J_{\pm}}-a/\left(2b\right)$,
such that the Gardner equation reduces to the mKdV equation. However,
the conserved charges are not mapped into each other, which is a direct
consequence of the fact that this transformation does not map the
higher members of the Gardner and mKdV hierarchies.

In the next section we will show how to extend the previous results
in order to incorporate an arbitrary member of the Gardner hierarchy.

\subsection{Extension to the Gardner hierarchy\label{sec:G-hierarchy}}

It is possible to find precise boundary conditions for General Relativity
on AdS$_{3}$, such that the dynamics of the boundary degrees of freedom
are described by the $k$-th member of the Gardner hierarchy. This
is achieved by choosing the chemical potentials $\zeta_{\pm}$ as
follows 
\begin{equation}
	\zeta_{\pm}=\frac{4\pi}{\kappa}R_{\left(k\right)}^{\pm},\label{eq:G-zeta2}
\end{equation}
where $R_{\left(k\right)}^{\pm}$ are the $k$-th generalized Gelfand-Dickey
polynomials. The functionals $H^{\pm}$ in eq. \eqref{eq:G-zeta},
are identified with the $k$-th Hamiltonian of the hierarchy defined
by eq. \eqref{eq:G-RHJ}. The particular case $k=1$ corresponds to
the one developed in the previous section.

With the boundary condition specified by eq. \eqref{eq:G-zeta2},
Einstein equations in \eqref{eq:G-fieldeq} take the form 
\begin{equation}
	\dot{\mathcal{J}_{\pm}}=\pm\frac{4\pi}{\kappa}\partial_{\phi}R_{\left(k\right)}^{\pm},\label{eq:G-JdR}
\end{equation}
which coincide with two (left/right) copies of the $k$-th element
of the Gardner hierarchy. By virtue of the bi-Hamiltonian character
of the system, eq. \eqref{eq:G-JdR} can also be written as
\begin{equation}
	\dot{\mathcal{J}_{\pm}}=\left\{ \mathcal{J_{\pm}},H_{\left(k\right)}^{\pm}\right\} =\left\{ \mathcal{J_{\pm}},H_{\left(k-1\right)}^{\pm}\right\} _{2}.\label{eq:G-jpuntobracket}
\end{equation}

The equations that describe the time evolution of the gauge parameters
$\eta_{\pm}$ take the form of eq. \eqref{eq:G-etapunto}, but with
$H^{\pm}\rightarrow H_{\left(k\right)}^{\pm}$, whose general solution
that depends locally on $\mathcal{J_{\pm}}$ and their spatial derivatives
coincides with eq. \eqref{eq:G-etasol}. Consequently, the global
charges integrate as \eqref{eq:G-QHn}, which as it was discussed
in the previous section, commute among them with both Poisson structures.

\subsubsection*{\emph{Lifshitz scaling.}}

The members of the Gardner hierarchy are not invariant under Lifshitz
scaling. However, in the particular cases when they belong to KdV
or mKdV hierarchies, the anisotropic scaling, with dynamical exponent
$z=2k+1$, is restored.

\emph{Case 1: $a=0$ (mKdV)}

When $a=0$, the $k$-th member of the hierarchy is invariant under

\begin{equation}
	t\rightarrow\lambda^{z}t\quad,\quad\phi\rightarrow\lambda\phi\quad,\quad\mathcal{J_{\pm}}\rightarrow\lambda^{-1}\mathcal{J_{\pm}}.\label{eq:LifshitzmKdV}
\end{equation}

\emph{Case 2: $b=0$ (KdV)}

When $b=0$, the $k$-th member of the hierarchy is invariant under

\begin{equation}
	t\rightarrow\lambda^{z}t\quad,\quad\phi\rightarrow\lambda\phi\quad,\quad\mathcal{J_{\pm}}\rightarrow\lambda^{-2}\mathcal{J_{\pm}}.\label{eq:LifshitzKdV}
\end{equation}

Note that the dynamical exponent is the same in both cases, but the
fields $\mathcal{J_{\pm}}$ scale in different ways.

\subsubsection*{\emph{Extending the hierarchy backwards.}}

The first nonlinear members of the Gardner hierarchy correspond to
the case $k=1$, and are given by eq. \eqref{eq:G-mixedkdvmkdv}.
However, if one uses the Hamiltonians $H_{\left(0\right)}^{\pm}$
defined in eq. \eqref{eq:G-H0} together with the first Poisson bracket
structure \eqref{eq:G-Poissonbracket}, one obtains a linear equation
describing left and right chiral movers 
\[
\dot{\mathcal{J_{\pm}}}=\pm\mathcal{J_{\pm}^{\prime}},
\]
which can be considered as the member with $k=0$ of the hierarchy.

From the point of view of the gravitational theory, it is useful to
extend the hierarchy an additional step backwards by using the recursion
relation \eqref{eq:G-recurrenciarel} in the opposite direction. Hence,
one obtains a new Hamiltonian for each copy of the form
\begin{equation}
	H_{\left(-1\right)}^{\pm}=\frac{\kappa}{4\pi}\oint d\phi\left(a^{-1}\mathcal{J_{\pm}}\right).\label{eq:G-Hm1}
\end{equation}
These Hamiltonians can only be defined when $a\neq0$, and their corresponding
generalized Gelfand-Dickey polynomials $R_{\left(-1\right)}^{\pm}=\kappa/\left(4\pi a\right)$
can be used as a seed that generates the whole hierarchy through the
recursion relation.

Using the Hamiltonians $H_{\left(-1\right)}^{\pm}$ in \eqref{eq:G-Hm1},
together with the first Poisson bracket structure \eqref{eq:G-Poissonbracket},
the soft hairy boundary conditions of refs. \cite{Afshar:2016wfy,Afshar:2016kjj},
reviewed in section \ref{sec:G-softhairy}, are recovered in a co-rotating
frame. In this case, the chemical potentials $\zeta_{\pm}$ are constants
and given by
\[
\zeta_{\pm}=\frac{4\pi}{\kappa}R_{\left(-1\right)}^{\pm}=a^{-1},
\]
while the equations of motion become $\dot{\mathcal{J_{\pm}}}=0$.
In this sense, one can consider the soft hairy boundary conditions
as being part of the hierarchy, and besides as the first member of
it.

\section{Metric formulation\label{sec:G-metric}}

In this section, we provide a metric description of the results previously
obtained in the context of the Chern--Simons formulation of General
Relativity on AdS$_{3}$. As was discussed in sec. \ref{subsec:Asymptotic-form-of1},
the boundary conditions that describe the Gardner hierarchy can be
interpreted as being defined either at infinity or in the near horizon
region. We will analyze these two possible interpretations in the
metric formalism following the lines of ref. \cite{Afshar:2016kjj}.

\subsection{Asymptotic behavior}

The spacetime metric can be directly reconstructed from the Chern--Simons
fields \eqref{eq:G-gaugetransform}, \eqref{eq:G-diagonalg}, provided
a particular gauge group element $b_{\pm}\left(r\right)$ is specified.
In order to describe the metric in the asymptotic region, it is useful
to choose 
\[
b_{\pm}\left(r\right)=\exp\left[\pm\frac{1}{2}\log\left(\frac{2r}{\ell}\right)\left(L_{1}-L_{-1}\right)\right].
\]
The expansion of the metric for $r\rightarrow\infty$ then reads

\begin{eqnarray}
	g_{tt} & = & -\zeta_{+}\zeta_{-}r^{2}+\frac{\ell^{2}}{4}\left(\zeta_{+}^{2}+\zeta_{-}^{2}\right)+\mathcal{O}\left(r^{-1}\right),\nonumber \\
	g_{tr} & = & \mathcal{O}\left(r^{-2}\right),\nonumber \\
	g_{t\phi} & = & \left(\zeta_{+}\mathcal{J}_{-}-\zeta_{-}\mathcal{J}_{+}\right)\frac{r^{2}}{2}+\frac{\ell^{2}}{4}\left(\zeta_{+}\mathcal{J}_{+}-\zeta_{-}\mathcal{J}_{-}\right)+\mathcal{O}\left(r^{-1}\right),\label{eq:AsymptoticMetric}\\
	g_{rr} & = & \frac{\ell^{2}}{r^{2}}+\mathcal{O}\left(r^{-5}\right),\nonumber \\
	g_{r\phi} & = & \mathcal{O}\left(r^{-2}\right),\nonumber \\
	g_{\phi\phi} & = & \mathcal{J_{+}}\mathcal{J_{-}}r^{2}+\frac{\ell^{2}}{4}\left(\mathcal{J}_{+}^{2}+\mathcal{J}_{-}^{2}\right)+\mathcal{O}\left(r^{-1}\right).\nonumber 
\end{eqnarray}

As it was explained in sec. \ref{sec:G-hierarchy}, in order to implement
the boundary conditions associated to the Gardner hierarchy we must
choose the chemical potentials $\zeta_{\pm}$ according to eq. \eqref{eq:G-zeta2},
i.e.,
\[
\zeta_{\pm}=\frac{4\pi}{\kappa}R_{\left(k\right)}^{\pm}.
\]
The differential equations associated to the (left/right) $k$-th
element of the hierarchy are precisely recovered if one imposes that
the metric in eq. \eqref{eq:AsymptoticMetric} obeys Einstein equations
with a negative cosmological constant in the asymptotic region of
spacetime.

The fall-off in \eqref{eq:AsymptoticMetric} is preserved under the
asymptotic symmetries generated by the following Killing vectors

\begin{eqnarray}
	\xi^{t} & = & \frac{\eta^{+}\mathcal{J}_{-}+\eta^{-}\mathcal{J}_{+}}{\zeta_{+}\mathcal{J}_{-}+\zeta_{-}\mathcal{J}_{+}}+\mathcal{O}\left(\frac{1}{r^{3}}\right),\nonumber \\
	\xi^{r} & = & \mathcal{O}\left(\frac{1}{r^{2}}\right),\\
	\xi^{\phi} & = & \frac{\eta^{+}\zeta_{-}-\eta^{-}\zeta_{+}}{\zeta_{+}\mathcal{J}_{-}+\zeta_{-}\mathcal{J}_{+}}+\mathcal{O}\left(\frac{1}{r^{3}}\right).\nonumber 
\end{eqnarray}
The conserved charges can be directly computed using the Regge-Teitelboim
approach \cite{Regge:1974zd}, and as expected coincide with the expression
in eq. \eqref{eq:G-deltaQ} obtained using the Chern--Simons formulation.

Note that the boundary metric

\[
d\bar{s}^{2}=r^{2}\left(-\zeta_{+}\zeta_{-}dt^{2}+\left(\zeta_{+}\mathcal{J}_{-}-\zeta_{-}\mathcal{J}_{+}\right)dtd\phi+\mathcal{J_{+}}\mathcal{J_{-}}d\phi^{2}\right),
\]
explicitly depends on the dynamical fields $\mathcal{J_{\pm}}$ and
consequently \emph{it is not fixed} at the boundary of spacetime,
i.e., it has a nontrivial functional variation.

\subsection{Near horizon behavior}

Following ref. \cite{Afshar:2016kjj}, the metric in the near horizon
region can be reconstructed using

\[
b_{\pm}\left(r\right)=\exp\left(\pm\frac{r}{2\ell}\left(L_{1}-L_{-1}\right)\right),
\]
and considering an expansion around $r=0$. The metric then reads

\begin{eqnarray}
	g_{tt} & = & \frac{\ell^{2}}{4}\left(\zeta_{+}-\zeta_{-}\right)^{2}-\zeta_{+}\zeta_{-}r^{2}+\mathcal{O}\left(r^{3}\right),\nonumber \\
	g_{tr} & = & \mathcal{O}\left(r^{2}\right),\nonumber \\
	g_{t\phi} & = & \frac{\ell^{2}}{4}\left(\mathcal{J}_{+}+\mathcal{J}_{-}\right)\left(\zeta_{+}-\zeta_{-}\right)+\left(\zeta_{+}\mathcal{J}_{-}-\zeta_{-}\mathcal{J}_{+}\right)\frac{r^{2}}{2}+\mathcal{O}\left(r^{3}\right),\label{eq:NearHorizonMetric}\\
	g_{rr} & = & 1+\mathcal{O}\left(r^{2}\right),\nonumber \\
	g_{r\phi} & = & \mathcal{O}\left(r^{2}\right),\nonumber \\
	g_{\phi\phi} & = & \frac{\ell^{2}}{4}\left(\mathcal{J}_{+}+\mathcal{J}_{-}\right)^{2}+\mathcal{J}_{+}\mathcal{J}_{-}r^{2}+\mathcal{O}\left(r^{3}\right).\nonumber 
\end{eqnarray}
Again, the chemical potentials $\zeta_{\pm}$ are expressed in terms
of the generalized Gelfand-Dickey polynomials according to eq. \eqref{eq:G-zeta2}\footnote{Note that with our choice of boundary conditions it is not possible
	to write the metric \eqref{eq:NearHorizonMetric} in a co-rotating
	frame ($\zeta_{+}=\zeta_{-}=const.$), because $\zeta_{\pm}$ have
	a very precise dependence on the fields $\mathcal{J_{\pm}}$, and
	generically cannot be set to be equal to constants.}.

The behavior of the metric near the horizon is preserved under the
action of the following Killing vectors

\begin{eqnarray}
	\xi^{t} & = & \frac{\eta^{+}\mathcal{J}_{-}+\eta^{-}\mathcal{J}_{+}}{\zeta_{+}\mathcal{J}_{-}+\zeta_{-}\mathcal{J}_{+}}+\mathcal{O}\left(r^{3}\right),\nonumber \\
	\xi^{r} & = & \mathcal{O}\left(r^{3}\right),\\
	\xi^{\phi} & = & \frac{\eta^{+}\zeta_{-}-\eta^{-}\zeta_{+}}{\zeta_{+}\mathcal{J}_{-}+\zeta_{-}\mathcal{J}_{+}}+\mathcal{O}\left(r^{3}\right).\nonumber 
\end{eqnarray}
The conserved charges can be obtained using the Regge-Teitelboim approach,
and evaluating them at $r=0$. The results coincide with eq. \eqref{eq:G-deltaQ},
as expected.

\subsection{General solution}

In ref. \cite{Afshar:2016wfy,Afshar:2016kjj}, it was shown that it
is possible to construct the general solution of Einstein equations
obeying the fall-off described in \eqref{eq:AsymptoticMetric}. It
is given by
\begin{equation}
	\begin{array}{ccc}
		ds^{2} & = & dr^{2}+\frac{\ell^{2}}{4}\cosh^{2}\left(r/\ell\right)\left[\left(\zeta_{+}-\zeta_{-}\right)dt+\left(\mathcal{J}_{+}+\mathcal{J}_{-}\right)d\phi\right]^{2}\\
		&  & -\frac{\ell^{2}}{4}\sinh^{2}\left(r/\ell\right)\left[\left(\zeta_{+}+\zeta_{-}\right)dt+\left(\mathcal{J}_{+}-\mathcal{J}_{-}\right)d\phi\right]^{2},
	\end{array}\label{eq:metriccompleta-1}
\end{equation}
and satisfies Einstein equations provided that $\mathcal{J}_{\pm}$
obey the differential equations associated to the $k$-th member of
the hierarchy when $\zeta_{\pm}$ is fixed according to eq. \eqref{eq:G-zeta2}.
In the near horizon region, this solution also obeys the fall-off
in \eqref{eq:NearHorizonMetric}. Note that the metric \eqref{eq:metriccompleta-1}
is diffeomorphic to a BTZ geometry, but as we will show in the next
section, it carries nontrivial charges associated to improper (large)
gauge transformations \cite{Benguria:1977in}, and consequently describes
a different physical state.

It is worth emphasizing that there is a one-to-one map between three--dimensional
geometries described by eq. \eqref{eq:metriccompleta-1}, and solutions
of the members of the Gardner hierarchy. In this sense, we can say
that this integrable system was ``fully geometrized'' in terms of
certain three--dimensional spacetimes which are locally of constant
curvature.

\section{Black holes\label{sec:blackholes}}

\subsection{Regularity conditions and thermodynamics}

Euclidean black holes solutions are obtained by requiring regularity
of the Euclidean geometries associated to the family of metrics in
\eqref{eq:metriccompleta-1}. This fixes the inverse of left and right
temperatures $\beta_{\pm}=T_{\pm}^{-1}$ in terms of the fields $\zeta_{\pm}$
according to
\begin{equation}
	\beta_{\pm}=\frac{2\pi}{\zeta_{\pm}}.\label{eq:betamasmenos}
\end{equation}
These conditions can also be obtained by requiring that the holonomy
around the thermal cycle for the gauge connections \eqref{eq:G-diagonalg}
be trivial.

A direct consequence of eq. \eqref{eq:betamasmenos} is that the chemical
potentials $\zeta_{\pm}$ are now constants, and hence from the field
equations \eqref{eq:G-fieldeq}, the regular Euclidean solutions are
characterized by $\dot{\mathcal{J_{\pm}}}=0$, i.e., by static solutions
of the members of the Gardner hierarchy. In sum, in order to obtain
an explicit black hole solution, the following equations must be solved

\begin{equation}
	\partial_{\phi}R_{\left(k\right)}^{\pm}=0,\label{eq:DR=00003D0}
\end{equation}
restricted to the conditions

\begin{equation}
	T_{\pm}=\frac{2}{\kappa}R_{\left(k\right)}^{\pm}.\label{eq:Temperature}
\end{equation}

The Bekenstein--Hawking entropy can be directly obtained from the
near horizon expansion \eqref{eq:NearHorizonMetric}, and gives
\begin{equation}
	S=\frac{A}{4G}=\frac{\kappa}{2}\oint d\phi\left(\mathcal{J_{+}}+\mathcal{J_{-}}\right).\label{eq:HawkingBek-Entropy}
\end{equation}
As expected, the first law is automatically fulfilled. Indeed, using
\eqref{eq:Temperature} one obtains

\[
\beta_{+}\delta H_{\left(k\right)}^{+}+\beta_{-}\delta H_{\left(k\right)}^{-}=\oint d\phi\left(\beta_{+}R_{\left(k\right)}^{+}\delta\mathcal{J_{+}}+\beta_{-}R_{\left(k\right)}^{-}\delta\mathcal{J_{-}}\right)=\delta\left[\frac{\kappa}{2}\oint d\phi\left(\mathcal{J_{+}}+\mathcal{J_{-}}\right)\right]=\delta S.
\]
Here, $\beta_{\pm}$ turn out to be the conjugates to the left and
right energies $H_{\left(k\right)}^{\pm}$. The inverse temperature,
conjugate to the energy $E$ in eq. \eqref{eq:G-Energia}, is expressed
in terms of the left and right temperatures according to $T^{-1}=\frac{1}{2}\left(T_{+}^{-1}+T_{-}^{-1}\right)$.

The previous analysis was performed in a rather abstract form without
using an explicit solution to eq. \eqref{eq:DR=00003D0}, which in
general are very hard to find. A simple solution corresponds to $\mathcal{J_{\pm}}=const.$,
which describes a BTZ configuration. In this case the Hamiltonians
take the form

\[
H_{\left(k\right)}^{\pm}=\sum_{n=k+2}^{2k+2}\alpha_{n}^{\pm}\mathcal{J}_{\pm}^{n},
\]
where $\alpha_{n}^{\pm}$ are constant coefficients which are not
specified in general, but whose values can be determined once the
corresponding Hamiltonians are explicitly computed through the recursion
relation for the generalized Gelfand-Dickey polynomials.

Some simplifications occur when we turn off either $a$ or $b$ (mKdV
and KdV cases), that we discuss next.

\subsection{mKdV case ($a=0$)}

When $a=0$, the metric associated to the black hole solution with
$\mathcal{J_{\pm}}=const.$ can be written as
\begin{equation}
	\begin{array}{ccc}
		ds^{2} & = & dr^{2}+\frac{\ell^{2}}{4}\cosh^{2}\left(r/\ell\right)\left[4\pi^{2}\left(\frac{\pi\kappa}{2\sigma_{\left(k\right)}\left(k+1\right)}\right)^{2k+1}\left(\mathcal{J}_{+}^{2k+1}-\mathcal{J}_{-}^{2k+1}\right)dt+\left(\mathcal{J}_{+}+\mathcal{J}_{-}\right)d\phi\right]^{2}\\
		&  & -\frac{\ell^{2}}{4}\sinh^{2}\left(r/\ell\right)\left[4\pi^{2}\left(\frac{\pi\kappa}{2\sigma_{\left(k\right)}\left(k+1\right)}\right)^{2k+1}\left(\mathcal{J}_{+}^{2k+1}+\mathcal{J}_{-}^{2k+1}\right)dt+\left(\mathcal{J}_{+}-\mathcal{J}_{-}\right)d\phi\right]^{2},
	\end{array}\label{eq:metriccompleta-1-1-1}
\end{equation}
where the constant $\sigma_{\left(k\right)}$, given by
\[
\sigma_{\left(k\right)}:=\left(\frac{\pi\kappa}{2k+2}\right)^{\frac{k+1}{k+\frac{1}{2}}}\left(\frac{\sqrt{\pi}}{\kappa2^{k-2}b^{k}}\frac{\Gamma\left(k+2\right)}{\Gamma\left(k+\frac{1}{2}\right)}\right)^{\frac{1}{2k+1}},
\]
will play the role of the anisotropic Stefan--Boltzmann constant
of the system. The metric can be written in Schwarzschild--like coordinates
using the following coordinate transformation
\begin{equation}
	r=\frac{\ell}{2}\log\left(\frac{4\sqrt{\left(\bar{r}^{2}-\frac{\ell^{2}}{4}\left(\mathcal{J}_{+}-\mathcal{J}_{-}\right)^{2}\right)\left(\bar{r}^{2}-\frac{\ell^{2}}{4}\left(\mathcal{J}_{+}+\mathcal{J}_{-}\right)^{2}\right)}-\ell^{2}\left(\mathcal{J}_{+}^{2}+\mathcal{J}_{-}^{2}\right)+4\bar{r}^{2}}{2\ell^{2}\mathcal{J}_{+}\mathcal{J}_{-}}\right).\label{eq:changeinr}
\end{equation}
It coincides with the metric of a BTZ black hole in a rotating frame,
with outer and inner horizons located at $\bar{r}_{\pm}=\frac{\ell}{2}\left(\mathcal{J}_{+}\pm\mathcal{J}_{-}\right)$.

With the choice $a=0$, the expression for the left and right energies
written in terms of the constants $\mathcal{J}_{\pm}$ becomes simpler
than the one in the general case. Indeed, it can be written in a closed
form as
\begin{equation}
	H_{\left(k\right)}^{\pm}=\sigma_{\left(k\right)}^{-z}\left(\frac{\pi\kappa}{\left(z+1\right)}\mathcal{J}_{\pm}\right)^{z+1},\label{eq:HdeJmKdV}
\end{equation}
where $z=2k+1$ is the dynamical exponent of the Lifshitz scale symmetry
\eqref{eq:LifshitzmKdV} of the $k$-th element of the mKdV hierarchy.

Using eqs. \eqref{eq:Temperature} and \eqref{eq:HdeJmKdV}, the left
and right energies $H_{\left(k\right)}^{\pm}$ can be expressed in
terms of the left and right temperatures $T_{\pm}$, acquiring the
form dictated by the Stefan--Boltzmann law for a two--dimensional
system with anisotropic Lifshitz scaling \cite{Gonzalez:2011nz}

\[
H_{\left(k\right)}^{\pm}=\sigma_{\left(k\right)}T_{\pm}^{1+\frac{1}{z}}.
\]

The Bekenstein--Hawking entropy, given by eq. \eqref{eq:HawkingBek-Entropy},
can be expressed in terms of the left and right energies $H_{\left(k\right)}^{\pm}$
as follows
\begin{equation}
	S=\left(1+z\right)\sigma_{\left(k\right)}^{\frac{z}{1+z}}\left(\left(H_{\left(k\right)}^{+}\right)^{\frac{1}{z+1}}+\left(H_{\left(k\right)}^{-}\right)^{\frac{1}{z+1}}\right).\label{eq:EntropyEnergy}
\end{equation}
Note that the dependence of the entropy in terms of the left/right
energies is consistent with the Lifshitz scaling of the $k$--th
element of the mKdV hierarchy.

\subsubsection*{\emph{Power partitions and microstate counting}}

The dependence of the entropy in terms of the left/right energies
in eq. \eqref{eq:EntropyEnergy} might be understood from a microscopic
point of view if, following \cite{Melnikov:2018fhb}, we assume that
there exists a two--dimensional field theory with Lifshitz scaling,
defined on a circle, whose dispersion relation for very high energies
takes the form

\begin{equation}
	E_{n}^{\pm}=\varepsilon_{\left(z\right)}^{\pm}n^{z},\label{eq:Dispersion}
\end{equation}
where $n$ is a non--negative integer, and $\varepsilon_{\left(z\right)}^{\pm}$
denote the characteristic energy of the left/right modes. The problem
of computing the entropy in the microcanonical ensemble is then equivalent
to compute the power partitions of given integers $N_{\pm}=E^{\pm}/\varepsilon_{\left(z\right)}^{\pm}$.
Here $E^{\pm}$ are the left/right energies given by 
\[
E^{\pm}=\varepsilon_{\left(z\right)}^{\pm}\sum_{i}n_{i}^{z}.
\]
This problem was solved long ago by Hardy and Ramanujan in \cite{hardy1918},
where at the end of their paper they conjecture that the asymptotic
growth of power partitions is given by

\[
p_{z}\left(N_{\pm}\right)\approx\exp\left[\left(1+z\right)\left(\frac{\Gamma\left(1+\frac{1}{z}\right)\zeta\left(1+\frac{1}{z}\right)}{z}\right)^{\frac{z}{1+z}}N_{\pm}^{\frac{1}{1+z}}\right],
\]
result that was proven later by Wright in 1934 \cite{wright1934asymptotic}.

The (left/right) entropies then reads

\begin{equation}
	S^{\pm}=\log\left[p_{z}\left(N_{\pm}\right)\right]=\left(1+z\right)\left(\frac{\Gamma\left(1+\frac{1}{z}\right)\zeta\left(1+\frac{1}{z}\right)}{z}\right)^{\frac{z}{1+z}}\left(\frac{E^{\pm}}{\varepsilon_{\left(z\right)}^{\pm}}\right)^{\frac{1}{1+z}}.\label{eq:EntropyWright}
\end{equation}
This expression precisely coincides with the entropy of the black
hole in eq. \eqref{eq:EntropyEnergy}, provided $E^{\pm}=H_{\left(k\right)}^{\pm}$,
and
\[
\varepsilon_{\left(z\right)}^{\pm}=\left(\frac{\Gamma\left(1+\frac{1}{z}\right)\zeta\left(1+\frac{1}{z}\right)}{\sigma_{\left(k\right)}z}\right)^{z}.
\]

Note that AdS spacetime is not contained within the spectrum of our
boundary conditions, and consequently one can naively think that the
anisotropic extension of Cardy formula of refs. \cite{Gonzalez:2011nz,Perez:2016vqo}
cannot be used to reproduce the entropy of the black hole \eqref{eq:EntropyEnergy}.
However, there is a known case \cite{Gonzalez:2011nz}, where the
anisotropic extension of Cardy formula can still be used in spite
of the fact that ground state, given by a gravitational soliton, does
not fit within the boundary conditions that accommodate the Lifshitz
black hole. This approach is based on the use of an anisotropic extension
of modular invariance that relates the Euclidean black hole and its
corresponding soliton, which turn out to be diffeomorphic. It would
be interesting to explore in the future whether this approach could
be applied to the BTZ black holes in the context of our boundary conditions.

\subsection{KdV case ($b=0$)}

When $b=0$, the metric associated to the black hole solution with
$\mathcal{J_{\pm}}=const.$ takes the form

\begin{equation}
	\begin{array}{ccc}
		ds^{2} & = & dr^{2}+\frac{\ell^{2}}{4}\cosh^{2}\left(r/\ell\right)\left[4\pi^{2}\left(\frac{\pi\kappa}{\bar{\sigma}_{\left(k\right)}\left(k+2\right)}\right)^{k+1}\left(\mathcal{J}_{+}^{k+1}-\mathcal{J}_{-}^{k+1}\right)dt+\left(\mathcal{J}_{+}+\mathcal{J}_{-}\right)d\phi\right]^{2}\\
		&  & -\frac{\ell^{2}}{4}\sinh^{2}\left(r/\ell\right)\left[4\pi^{2}\left(\frac{\pi\kappa}{\bar{\sigma}_{\left(k\right)}\left(k+2\right)}\right)^{k+1}\left(\mathcal{J}_{+}^{k+1}+\mathcal{J}_{-}^{k+1}\right)dt+\left(\mathcal{J}_{+}-\mathcal{J}_{-}\right)d\phi\right]^{2},
	\end{array}\label{eq:metriccompleta-1-1}
\end{equation}
where

\[
\bar{\sigma}_{\left(k\right)}=\left(\frac{\pi\kappa}{k+2}\right)^{\frac{k+2}{k+1}}\left(\frac{\sqrt{\pi}}{\kappa\left(2a\right)^{k}}\frac{\Gamma\left(k+3\right)}{\Gamma\left(k+\frac{3}{2}\right)}\right)^{\frac{1}{k+1}}.
\]
Using the change of coordinates \eqref{eq:changeinr}, the metric
\eqref{eq:metriccompleta-1-1} can be written in Schwarzschild--like
form, and coincides with the one of a BTZ black hole with outer and
inner horizons located at $\bar{r}_{\pm}=\frac{\ell}{2}\left(\mathcal{J}_{+}\pm\mathcal{J}_{-}\right)$.

The left and right energies $H_{\left(k\right)}^{\pm}$ can then be
expressed in terms of the constants $\mathcal{J}_{\pm}$ according
to
\begin{equation}
	H_{\left(k\right)}^{\pm}=\bar{\sigma}_{\left(k\right)}^{-\frac{z+1}{2}}\left(\frac{2\pi\kappa}{z+3}\mathcal{J}_{\pm}\right)^{\frac{z+3}{2}},\label{eq:HdeJKdV}
\end{equation}
where $z=2k+1$ is the dynamical exponent associated to the Lifshitz
symmetry \eqref{eq:LifshitzKdV} of the $k$-th member of the KdV
hierarchy.

The expression for the left/right energies $H_{\left(k\right)}^{\pm}$
in terms of the left/right temperatures $T_{\pm}$, is then given
by

\[
H_{\left(k\right)}^{\pm}=\bar{\sigma}_{\left(k\right)}T_{\pm}^{\frac{z+3}{z+1}}.
\]
In spite of the fact that the $k$-th equation of the KdV hierarchy
is invariant under Lifshitz scaling with dynamical exponent $z$,
the power in the temperature is not the one expected for a two--dimensional
theory with this symmetry. Furthermore, this is inherited to the expression
for the entropy written in terms of the left/right energies $H_{\left(k\right)}^{\pm}$
\[
S=\left(\frac{z+3}{2}\right)\bar{\sigma}_{\left(k\right)}^{\frac{z+1}{z+3}}\left(\left(H_{\left(k\right)}^{+}\right)^{\frac{2}{z+3}}+\left(H_{\left(k\right)}^{-}\right)^{\frac{2}{z+3}}\right),
\]
which is not of the expected form \eqref{eq:EntropyWright}.

Remarkably, if instead of the Hamiltonians $H_{\left(k\right)}^{\pm}$,
one uses the Hamiltonians $H_{\left(2k\right)}^{\pm}$, this naive
incompatibility with the Lifshitz symmetry disappears. Indeed, the
relation between $H_{\left(2k\right)}^{\pm}$ and the left/right temperatures
takes the form

\[
H_{\left(2k\right)}^{\pm}=\bar{\sigma}_{\left(2k\right)}T_{\pm}^{1+\frac{1}{z}},
\]
while the expression for the entropy in terms of the extensive quantities
$H_{\left(2k\right)}^{\pm}$ reads
\[
S=\left(1+z\right)\bar{\sigma}_{\left(2k\right)}^{\frac{z}{1+z}}\left(\left(H_{\left(2k\right)}^{+}\right)^{\frac{1}{z+1}}+\left(H_{\left(2k\right)}^{-}\right)^{\frac{1}{z+1}}\right).
\]
The entropy then takes the expected Hardy--Ramanujan form \eqref{eq:EntropyWright},
with the characteristic energy of the dispersion relation given by
\[
\varepsilon_{\left(z\right)}^{\pm}=\left(\frac{\Gamma\left(1+\frac{1}{z}\right)\zeta\left(1+\frac{1}{z}\right)}{\bar{\sigma}_{\left(2k\right)}z}\right)^{z}.
\]

\subsubsection*{\emph{Black hole with nonconstants $\mathcal{J_{\pm}}$}}

In the particular case when $b=0$ and $k=1$, it is possible to find
explicit nonconstants solutions to eq. \eqref{eq:DR=00003D0} that
characterize a regular Euclidean black hole. These are static solutions
of the left/right KdV equations, which take the form of periodic cnoidal
waves. The solutions are then given by
\begin{equation}
	\mathcal{J_{\pm}}=-\frac{8K^{2}\left(m_{\pm}\right)}{3a\pi^{2}}\left(1-2m_{\pm}+3m_{\pm}\text{cn}^{2}\left(\frac{K\left(m_{\pm}\right)}{\pi}\phi|m^{\pm}\right)\right),
\end{equation}
where $m_{\pm}$ are constants in the range $0\leq m_{\pm}<1$, $\text{cn}$
denotes the Jacobi elliptic cosine function, and $K\left(m\right)$
is the complete elliptic integral of the first kind defined as

\begin{equation}
	K(m)=\int_{0}^{\frac{\pi}{2}}\frac{d\theta}{\sqrt{1-m\sin^{2}\theta}}.
\end{equation}
The regularity conditions \eqref{eq:Temperature} fix the left and
right temperatures $T_{\pm}$ in terms of the constant $m_{\pm}$
according to
\[
T_{\pm}=\frac{16\left(m_{\pm}^{2}-m_{\pm}+1\right)}{3\pi^{5}a\left(m_{\pm}-1\right)^{2}}K\left(\frac{m_{\pm}}{m_{\pm}-1}\right)^{4}.
\]

\newpage{}

\chapter{Boundary conditions for higher spin gravity on AdS$_{3}$ and the modified Gelfand-Dickey hierarchy  \label{chapter:Gelfand-Dickey hierarchy Part}}

Here we discuss the generalization of the results described in the previous chapter to spin-$N$ gravity. In this case, the corresponding integrable system is the $N$-th modified Gelfand-Dickey hierarchy, which for $N=2$ reduces to the modified KdV hierarchy, while for $N=3$ describe to the modified Boussinesq hierarchy. Indeed, in ref. \cite{Compere:2013gja} a connection between spin-3 gravity
and a ``Boussinesq equation in the light-cone'' was pointed out.
In refs. \cite{Gutperle:2014aja,Beccaria:2015iwa}, some particular
boundary conditions for higher spin gravity with gauge group $SL(N,\mathbb{R})\times SL(N,\mathbb{R})$
were associated to a generalized KdV hierarchy, and some particular
cases, including the Boussinesq equation, were explicitly worked out.
On the other hand, in ref. \cite{Perez:2016vqo} a very precise link
with the Boussinesq hierarchy was described for spin-3 gravity, as
well as for its extension including fields with arbitrary higher spins.
The analysis was based on boundary conditions defined in the highest
weight gauge with a particular choice of chemical potentials, generalizing
the results obtained for the KdV hierarchy in pure gravity.

The Boussinesq equation \eqref{eq:Lax - Boussinesq equation} was first introduced by Joseph Boussinesq
in 1872 in the context of the study of the propagation of one-dimensional
long waves in shallow water, moving in both directions\textcolor{red}{{}
}\cite{Boussinesq1872}. Long after, in 1974, it was realized that
the equation was integrable, and that belongs to a hierarchy of differential
equations \cite{Zakharov:1974}. Most of the important properties
of the Boussinesq equation, including its infinite set of commuting
conserved charges, can be easily derived when a potential equation,
called ``modified Boussinesq'' (mBoussinesq), is introduced (see
e.g \cite{Kaup:1975}). Both equations are then related by an appropriate
generalization of the Miura transformation \eqref{eq: Lax - Miura Transformation} \cite{Hirota:1977,Fordy:1981,Drinfeld:1984qv}.

In this chapter we show that the asymptotic dynamics of spin-3 gravity on AdS$_{3}$ endowed with a special class of boundary
conditions, is precisely described by the members of the mBoussinesq
hierarchy. In this framework, the gauge fields are defined in the
``diagonal gauge,'' where the excitations go along the generators
of the Cartan subalgebra of $sl(3,\mathbb{R})\oplus sl(3,\mathbb{R})$
\cite{Grumiller:2016kcp}. The link with the integrable system is
then obtained by choosing the chemical potentials as precise functionals
of the dynamical fields, in a way consistent with the action principle.
Hence, the entire integrable structure of the mBoussinesq hierarchy,
i.e., the phase space, the fundamental Poisson brackets given by two
independent $\hat{u}\left(1\right)$ current algebras, and the infinite
set of Hamiltonians in involution, are obtained from the asymptotic
structure of the higher spin theory in the bulk. Furthermore, the
relation with the Boussinesq hierarchy previously found in ref. \cite{Perez:2016vqo}
is inherited from our analysis once the asymptotic conditions are
re-expressed in the highest weight gauge along the lines of ref. \cite{Grumiller:2016kcp}.
Thus, the Miura map is recovered from a purely geometric construction
in the higher spin theory. Black hole solutions that fit within our
boundary conditions, the Hamiltonian reduction at the boundary, and
the generalization to higher spin gravity with gauge group $SL(N,\mathbb{R})\times SL(N,\mathbb{R})$
are also discussed.

\section{Review of the modified Boussinesq hierarchy\label{sec:Review-of-the}}

The mBoussinesq equation is the first member of the mBoussinesq hierarchy
and is given by the following set of differential equations

\begin{eqnarray}
	\dot{\mathcal{J}} & = & \lambda_{1}\mathcal{J}^{\prime}-\lambda_{2}\left(2\left(\mathcal{J}\mathcal{U}\right)^{\prime}+\mathcal{U}^{\prime\prime}\right),\nonumber \\
	\dot{{\cal U}} & = & \lambda_{1}\mathcal{U}^{\prime}+\lambda_{2}\left(\mathcal{U}^{2\prime}-\mathcal{J}^{2\prime}+\mathcal{J}^{\prime\prime}\right).\label{eq:mBsq}
\end{eqnarray}
Here dots and primes denote derivatives with respect to the time $t$
and the angle $\phi$ respectively, and $\lambda_{1}$, $\lambda_{2}$
are arbitrary constants associated to the two different flows of the
hierarchy \cite{Fordy:1981}. The case with $\lambda_{1}=0$ and $\lambda_{2}=1$
is known as the mBoussinesq equation, while the case with $\lambda_{1}=1$
and $\lambda_{2}=0$ describes two independent chiral fields.

As was described in chapter \ref{chapter:Integrable systems}, the dynamics
of the above equations may be described using the Hamiltonian formalism.
If the Poisson brackets of two arbitrary functional $F$ and $G$
is given by

\begin{equation}
	\left\{ F,G\right\} =\frac{4\pi}{\hat{\kappa}}\int d\phi\left(\frac{\delta F}{\delta\mathcal{J}}\partial_{\phi}\frac{\delta G}{\delta\mathcal{J}}+\frac{\delta F}{\delta\mathcal{U}}\partial_{\phi}\frac{\delta G}{\delta\mathcal{U}}\right),\label{eq:Poissonbrack}
\end{equation}
together with the Hamiltonian

\begin{equation}
	H_{\left(1\right)}=\frac{\hat{\kappa}}{4\pi}\int d\phi\left\{ \frac{\lambda_{1}}{2}\left(\mathcal{J}^{2}+\mathcal{U}^{2}\right)+\lambda_{2}\left(\frac{1}{3}\mathcal{U}^{3}-\mathcal{J}^{2}\mathcal{U}-\mathcal{J}\mathcal{U}^{\prime}\right)\right\} ,\label{eq:Ham1}
\end{equation}
then eq. \eqref{eq:mBsq} can be rewritten as
\[
\dot{\mathcal{J}}=\left\{ \mathcal{J},H_{\left(1\right)}\right\} \,,\qquad\qquad\dot{{\cal U}}=\left\{ {\cal U},H_{\left(1\right)}\right\} .
\]
Note that the arbitrary constant $\hat{\kappa}$ does not appear in
the differential equations \eqref{eq:mBsq}, however it is useful
to introduce it in \eqref{eq:Poissonbrack} and \eqref{eq:Ham1} for
later convenience.

Alternatively, if we define the operator 
\begin{equation}
	\mathcal{D}:=\frac{4\pi}{\hat{\kappa}}\left(\begin{array}{cc}
		\partial_{\phi} & 0\\
		0 & \partial_{\phi}
	\end{array}\right),\label{eq:Symp1}
\end{equation}
the equations in \eqref{eq:mBsq} can be re-written in vector form
as follows

\[
\left(\begin{array}{c}
	\dot{\mathcal{J}}\\
	\dot{{\cal U}}
\end{array}\right)=\mathcal{D}\left(\begin{array}{c}
	\frac{\delta H_{\left(1\right)}}{\delta\mathcal{J}}\\
	\frac{\delta H_{\left(1\right)}}{\delta\mathcal{U}}
\end{array}\right).
\]
The operator $\mathcal{D}$ in \eqref{eq:Symp1} defines the symplectic
structure in eq. \eqref{eq:Poissonbrack}.

It is worth to emphasize that one of the key points in the relation
of this integrable system with higher spin gravity comes from the
fact that, according to eq. \eqref{eq:Poissonbrack}, the fundamental
Poisson brackets are described by two independent $\hat{u}\left(1\right)$
current algebras

\begin{eqnarray}
	\left\{ \mathcal{J}\left(\phi\right),\mathcal{J}\left(\phi^{\prime}\right)\right\}  & = & \frac{4\pi}{\hat{\kappa}}\partial_{\phi}\delta\left(\phi-\phi^{\prime}\right),\nonumber \\
	\left\{ \mathcal{U}\left(\phi\right),\mathcal{U}\left(\phi^{\prime}\right)\right\}  & = & \frac{4\pi}{\hat{\kappa}}\partial_{\phi}\delta\left(\phi-\phi^{\prime}\right).\label{eq:FUndbrack}
\end{eqnarray}
As we will show below, once appropriate boundary conditions are imposed,
this Poisson bracket algebra is obtained from the Dirac brackets in
the higher spin theory.

The integrability of \eqref{eq:mBsq} and the existence of a hierarchy
of equations, rely on the fact that this system is actually bi-Hamiltonian.
Indeed, there exists an alternative symplectic structure characterized
by the non-local operator given from the eq. \eqref{eq:modified Second Hamiltonian Op}

\begin{equation}
	\mathcal{D}_{\left(2\right)}=\mathcal{D}M^{\dagger}\mathcal{O}M\mathcal{D}.\label{eq:secondPstr}
\end{equation}
Here,

\begin{equation}
	M=\left(\begin{array}{cc}
		{\cal J}+\partial_{\phi}\qquad & \mathcal{U}\\
		-2{\cal J}\mathcal{U}-\frac{1}{2}\mathcal{U}\partial_{\phi}-\frac{3}{2}{\cal U}^{\prime}\qquad & \mathcal{U}^{2}-{\cal J}^{2}-\frac{1}{2}{\cal J}^{\prime}-\frac{3}{2}{\cal J}\partial_{\phi}-\frac{1}{2}\partial_{\phi}^{2}
	\end{array}\right),\label{eq:M}
\end{equation}
and $M^{\dagger}$ is the formal adjoint of $M$ (see e.g. \cite{Mathieu:1991}).
The operator $\mathcal{O}$ corresponds to the inverse of the first
Poisson structure associated to the Boussinesq hierarchy and is given by
\[
\mathcal{O}=\frac{2\hat{\kappa}}{\pi}\left(\begin{array}{cc}
	0 & \partial_{\phi}^{-1}\\
	\partial_{\phi}^{-1} & 0
\end{array}\right).
\]
Consequently, the Poisson bracket of two arbitrary functionals $F$
and $G$ associated with the operator $\mathcal{D}_{\left(2\right)}$
is

\begin{equation}
	\left\{ F,G\right\} _{2}=\int d\phi\left(\begin{array}{cc}
		\frac{\delta F}{\delta\mathcal{J}} & \frac{\delta F}{\delta\mathcal{U}}\end{array}\right)\mathcal{D}_{(2)}\left(\begin{array}{c}
		\frac{\delta G}{\delta\mathcal{J}}\\
		\frac{\delta G}{\delta\mathcal{U}}
	\end{array}\right).\label{eq:Poisson2}
\end{equation}
The explicit components of $\mathcal{D}_{\left(2\right)}$ are exhibited
in appendix \ref{Appendix 3: Second Hamiltonian structure mBsq}.

The modified Boussinesq equations \eqref{eq:mBsq} can then be recovered using the Poisson
bracket \eqref{eq:Poisson2}, together with the Hamiltonian\footnote{The coefficients $\lambda_{1}$ and $\lambda_{2}$ in eq. \eqref{eq:mBsq}
	are determined by the integration constants obtained by the action
	of $\mathcal{D}_{\left(2\right)}$. In the case of higher members
	of the hierarchy, the subsequent integration constants may be consistently
	set to zero as they contribute nothing new (see e.g \cite{Fordy:1981}).}

\[
H_{\left(0\right)}=\frac{\hat{\kappa}}{4\pi}\int d\phi\left(\lambda_{1}\mathcal{J}+\lambda_{2}\mathcal{U}\right).
\]
This system possess an infinite number of conserved charges in involution
that can be constructed from the recursion relation \eqref{eq: Modified GD u dot =00003D M v dot}
\begin{equation}
	R_{\left(n+1\right)}=\mathcal{D}^{-1}\mathcal{D}_{\left(2\right)}R_{\left(n\right)},\label{eq:rec}
\end{equation}
where
\begin{equation}
	R_{\left(n\right)}=\left(\begin{array}{c}
		\frac{\delta H_{\left(n\right)}}{\delta\mathcal{J}}\\
		\frac{\delta H_{\left(n\right)}}{\delta\mathcal{U}}
	\end{array}\right),\label{eq:GelfandDickey}
\end{equation}
are the Gelfand-Dickey polynomials associated to the hierarchy. The
conserved quantities $H_{\left(n\right)}$, with $n$ being a nonnegative
integer, are generically decomposed into two flows proportional to
the constants $\lambda_{1}$ and $\lambda_{2}$ respectively
\begin{equation}
	H_{\left(n\right)}=\sum_{I=1}^{2}\lambda_{I}H_{\left(n\right)}^{I}.\label{eq: decomp}
\end{equation}
Then one can prove that the $H_{\left(k\right)}^{I}$ are in involution
with both Poisson brackets, i.e.,
\[
\left\{ H_{\left(n\right)}^{I},H_{\left(m\right)}^{J}\right\} =\left\{ H_{\left(n\right)}^{I},H_{\left(m\right)}^{J}\right\} _{2}=0.
\]
Furthermore, if we one uses the conserved quantities $H_{\left(k\right)}^{I}$
as new Hamiltonians, it is then possible to define a hierarchy of
integrable equations labelled by the nonnegative integer $k$ of the
form

\begin{eqnarray}
	\dot{\mathcal{J}} & = & \left\{ \mathcal{J},H_{\left(k\right)}\right\} =\left\{ \mathcal{J},H_{\left(k-1\right)}\right\} _{2},\nonumber \\
	\dot{{\cal U}} & = & \left\{ {\cal U},H_{\left(k\right)}\right\} =\left\{ {\cal U},H_{\left(k-1\right)}\right\} _{2}.\label{eq:hierarchy}
\end{eqnarray}

The equations associated to each flow, labelled by the index $I=1,2$,
have different scaling properties. Under a Lifshitz scaling transformation
with dynamical exponent $z$

\[
t\rightarrow\varepsilon^{z}t\,,\quad\quad\phi\rightarrow\varepsilon\phi\,,\quad\quad\mathcal{J}\rightarrow\varepsilon^{-1}\mathcal{J}\,,\quad\quad\mathcal{U}\rightarrow\varepsilon^{-1}\mathcal{U},
\]
the flow with $I=1$ is invariant for $z=3k-2$, while the flow with
$I=2$ is invariant for $z=3k-1$.

As explained above, the mBoussinesq equation is a ``potential equation''
for the Boussinesq one. Indeed, if $\mathcal{U}$ and $\mathcal{J}$
obey the mBoussinesq equation, then the fields $\mathcal{L}$ and
$\mathcal{W}$ defined by the Miura transformation as

\begin{align}
	\mathcal{L} & =\frac{1}{2}{\cal J}^{2}+\frac{1}{2}\mathcal{U}^{2}+{\cal J}^{\prime},\nonumber \\
	\mathcal{W} & =\frac{1}{3}\mathcal{U}^{3}-{\cal J}^{2}\mathcal{U}-\frac{1}{2}\mathcal{U}{\cal J}^{\prime}-\frac{3}{2}{\cal J}{\cal U}^{\prime}-\frac{1}{2}{\cal U}^{\prime\prime},\label{eq:Miura}
\end{align}
obey the Boussinesq equation given by

\begin{eqnarray}
	\dot{\mathcal{L}} & = & 2\mathcal{W}^{\prime},\nonumber \\
	\dot{\mathcal{W}} & = & 2\mathcal{L}^{2\prime}-\frac{1}{2}\mathcal{L}^{\prime\prime\prime}.\label{eq:Bsqeq}
\end{eqnarray}

Combining both equations and eliminating the field $\mathcal{W}$, one finds that the field $\mathcal{L}$ must satisfy the ``Good'' Boussinesq
equation.\footnote{Eq. \eqref{eq:Lax - Boussinesq equation} is recovered by applying the following rescaling $\mathcal{L} \rightarrow -\frac{1}{2}\mathcal{L}$, $t \rightarrow \frac{1}{\sqrt{3}}t $,  $\phi  \rightarrow \phi  $.}

The entire Boussinesq hierarchy, including the infinite set of charges
in involution, is obtained from the mBoussinesq one by using the Miura
transformation (see appendix \ref{Appendix 5: Bsq and mBsq}\textcolor{red}{{}
}for more details on the Boussinesq hierarchy). It is worth noting
that the Miura transformation \eqref{eq:Miura} coincides the twisted
Sugawara construction of the stress tensor and a spin-3 current in
terms of two independent $U\left(1\right)$ currents in a two-dimensional
CFT. Hence, using the Poisson brackets \eqref{eq:FUndbrack} and the
Miura map \eqref{eq:Miura}, one can show that the (first) Poisson
brackets for $\mathcal{L}$ and $\mathcal{W}$ are precisely given
by the classical $W_{3}$ algebra.

\section{Modified Boussinesq hierarchy from spin-3 gravity on AdS$_{3}$\label{sec:2 mBq hierarchy from gravity on AdS}}

\subsection{Chern-Simons formulation of spin-3 gravity on AdS$_{3}$}

Higher spin gravity in 3D has the very special property that, in contrast
with their higher dimensional counterparts \cite{Fradkin:1987ks,Vasiliev:1990en,Vasiliev:2003ev},
its spectrum can be consistently truncated to a finite number of higher
spin fields \cite{Blencowe:1988gj,Bergshoeff:1989ns,Henneaux:2010xg,Campoleoni:2010zq}.
One of the simplest cases corresponds to a spin-two field non-minimally
coupled to a spin-three field, that may be described by a Chern-Simons
action for the gauge group $SL\left(3,\mathbb{R}\right)\times SL\left(3,\mathbb{R}\right)$,

\begin{equation}
	I=I_{CS}\left[A^{+}\right]-I_{CS}\left[A^{-}\right],\label{eq:ICS}
\end{equation}
where

\begin{equation}
	I_{CS}[A]=\frac{\kappa_{3}}{4\pi}\int_{\mathcal{M}}\text{tr}\left(AdA+\frac{2}{3}A^{3}\right).\label{eq:Chern-Simons Action}
\end{equation}
Here, the level is given by \emph{$\kappa_{3}=\kappa/4=l/16G$}, where
$l$ and $G$ correspond to the AdS radius and the three-dimensional
Newton constant respectively, and the trace is in the fundamental
representation of the $sl\left(3,\mathbb{R}\right)$ algebra in the
principal embedding (see appendix \ref{Appendix 6: Fundamental rep}).
The field equations are then given by the vanishing of the field strength
\begin{equation}
	F^{\pm}=dA^{\pm}+A^{\pm2}=0.\label{eq:Field Strength}
\end{equation}
The metric and the spin-three field are reconstructed in terms of
a generalized dreibein $e:=\frac{l}{2}\left(A^{+}-A^{-}\right)$ according
to

\[
g_{\mu\nu}=\frac{1}{2}\textrm{tr}\left(e_{\mu}e_{\nu}\right)\,,\qquad\qquad\varphi_{\mu\nu\rho}=\frac{1}{3!}\textrm{tr}\left(e_{\left(\mu\right.}e_{\nu}e_{\left.\rho\right)}\right).
\]

\subsection{Asymptotic behavior of the fields. Diagonal gauge\label{subsec:2.1 Asymptotic behavior of the fields. Diagonal gauge}}

Following refs. \cite{Coussaert:1995zp,Henneaux:2010xg,Campoleoni:2010zq},
it is convenient to perform the analysis of the asymptotic symmetries
of spin-three gravity in terms of an auxiliary connection depending
only on $t$ and $\phi$. For simplicity, and without loss of generality,
hereafter we will consider only the ``plus copy,'' and hence the
superscript ``$+$'' will be omitted. The gauge field $A$ is then
written as

\begin{equation}
	A=b^{-1}\left(d+a\right)b,\label{eq:A Connections}
\end{equation}
where $a=a_{t}dt+a_{\phi}d\phi$ is the auxiliary connection, and
$b=b\left(r\right)$ is a gauge group element which captures the whole
the radial dependence of the gauge connection. The asymptotic analysis
will be insensitive to the precise form of $b\left(r\right)$.

We will consider asymptotic conditions in the ``diagonal gauge,''
i.e., where all the permissible excitations in the auxiliary connection
go along the generators of the Cartan subalgebra of $sl\left(3,\mathbb{R}\right)$
\cite{Grumiller:2016kcp}. Then, it takes the form

\begin{equation}
	a=\left({\cal J}d\phi+\zeta dt\right)L_{0}+\frac{\sqrt{3}}{2}\left(\mathcal{U}d\phi+\zeta_{\mathcal{U}}dt\right)W_{0}.\label{eq:Auxiliary Connections}
\end{equation}
The fields \emph{${\cal J}$} and\emph{ $\mathcal{U}$} belong to
the spatial components of the auxiliary connection, and hence they
are identified as the dynamical fields. On the other hand, \emph{$\zeta$
}and\emph{ $\zeta_{\mathcal{U}}$} are defined along the temporal
components, and therefore they correspond to the boundary values of
the Lagrange multipliers. In ref. \cite{Grumiller:2016kcp} the same
asymptotic form for the auxiliary connection was used, with the replacement
$\mathcal{J}_{\left(3\right)}\rightarrow\frac{\sqrt{3}}{2}\mathcal{U}$
and $\zeta_{\left(3\right)}\rightarrow\frac{\sqrt{3}}{2}\zeta_{\mathcal{U}}$.
However, the boundary conditions will be different. In \cite{Grumiller:2016kcp}
it was assumed that $\zeta$ and $\zeta_{\left(3\right)}$ are kept
fixed at the boundary, while here, as we will show in the next subsection,
they will acquire a precise functional dependence on the dynamical
fields ${\cal J}$, ${\cal U}$ and their spatial derivatives.

\subsection{Boundary conditions for spin-3 gravity and the modified Boussinesq
	hierarchy\label{subsec:2.2 Consistency-with-the}}

A fundamental requirement in the study of the asymptotic structure
of spacetime is that the boundary conditions must be compatible with
the action principle. In the canonical formalism one has to add an
appropriate boundary term $B_{\infty}$ to the canonical action in
order to guarantee that the action principle attains an extremum \cite{Regge:1974zd}
\begin{equation}
	I_{can}\left[A\right]=-\frac{\kappa}{16\pi}\int dtd^{2}x\epsilon^{ij}\left\langle A_{i}\dot{A_{j}}-A_{t}F_{ij}\right\rangle +B_{\infty}.\label{eq:Action in hamiltonian form}
\end{equation}
Following \cite{Grumiller:2016kcp}, for the action \eqref{eq:Action in hamiltonian form}
and the asymptotic conditions \eqref{eq:A Connections}, \eqref{eq:Auxiliary Connections},
the variation of the boundary term is given by
\begin{equation}
	\delta B_{\infty}=-\frac{\kappa}{4\pi}\int dtd\phi\left(\zeta\delta\mathcal{J}+\zeta_{\mathcal{U}}\delta\mathcal{U}\right).\label{eq:Variation of boundary terms}
\end{equation}
In the absence of ingoing or outgoing radiation, as is the case in
three-dimensional higher spin gravity, the boundary term $B_{\infty}$
has to be integrable in a functional sense, i.e., one must be able
to ``take the delta outside'' in \eqref{eq:Variation of boundary terms}.
The precise way in which $\zeta$ and $\zeta_{\mathcal{U}}$ are fixed
at the boundary is what defines the boundary conditions. Thus, following
\cite{Perez:2016vqo}, in order to make contact with the Boussinesq
hierarchy in a way consistent with the action principle, we choose
the Lagrange multipliers as
\begin{equation}
	\zeta=\frac{4\pi}{\kappa}\frac{\delta H_{\left(k\right)}}{\delta\mathcal{J}}\,,\qquad\qquad\zeta_{\mathcal{U}}=\frac{4\pi}{\kappa}\frac{\delta H_{\left(k\right)}}{\delta\mathcal{U}},\label{eq:chemical potentials}
\end{equation}
where $H_{\left(k\right)}$ is the Hamiltonian associated to the $k$-th
element of the mBoussinesq hierarchy. With this choice, the boundary
term can be readily integrated, and yields

\begin{equation}
	B_{\infty}=-\int dtH_{\left(k\right)}.\label{eq:Boundary terms in terms of H}
\end{equation}
Thus, the Hamiltonian of the gravitational theory in the reduced phase
space precisely matches the one of the integrable system.

On the other hand, the field equations in the higher spin theory given
by the vanishing of the field strength become

\begin{equation}
	\dot{\mathcal{J}}=\zeta^{\prime}\,,\qquad\qquad\dot{\mathcal{U}}=\zeta_{\mathcal{U}}^{\prime}\,,\label{eq:Field equations}
\end{equation}
which, by virtue of \eqref{eq:chemical potentials}, precisely coincide
with the differential equations associated to the $k$-th element
of the mBoussinesq hierarchy in eq. \eqref{eq:hierarchy}, provided
the constant $\hat{\kappa}$ in eqs. \eqref{eq:Poissonbrack} and
\eqref{eq:Ham1} is assumed to depend on the cosmological and Newton
constants according to $\hat{\kappa}=\kappa$. Then,

\begin{equation}
	\left(\begin{array}{c}
		\dot{\mathcal{J}}\\
		\dot{{\cal U}}
	\end{array}\right)=\mathcal{D}\left(\begin{array}{c}
		\frac{\delta H_{\left(k\right)}}{\delta\mathcal{J}}\\
		\frac{\delta H_{\left(k\right)}}{\delta\mathcal{U}}
	\end{array}\right)=\left(\begin{array}{c}
		\left\{ \mathcal{J},H_{\left(k\right)}\right\} \\
		\left\{ {\cal U},H_{\left(k\right)}\right\} 
	\end{array}\right).\label{eq:EOM}
\end{equation}

\subsection{Asymptotic symmetries and conserved charges\label{subsec:2.3 Asymptotic-symmetries-and}}

The asymptotic symmetries are determined by set of gauge transformations
that preserve the asymptotic form of the gauge connection, with non-vanishing
associated charges. The form of the the auxiliary connection in eq.
\eqref{eq:Auxiliary Connections} is preserved by gauge transformations
$\delta a=d\lambda+\left[a,\lambda\right]$, with parameter

\[
\lambda=\eta L_{0}+\frac{\sqrt{3}}{2}\eta_{{\cal U}}W_{0}.
\]
There could be some additional terms in the non-diagonal components
of the gauge parameter $\lambda$, but they are pure gauge in the
sense that there are no generators associated to them, so they can
be consistently set to zero.

The preservation of the angular components of the auxiliary connection
gives the transformation law of the dynamical fields
\begin{equation}
	\delta\mathcal{J}=\eta^{\prime}\,,\qquad\qquad\delta{\cal U}=\eta_{{\cal U}}^{\prime},\label{eq:tarnsflawfields}
\end{equation}
while that the preservation of the temporal components provides the
transformation law of the Lagrange multipliers
\begin{equation}
	\delta\zeta=\dot{\eta}\,,\qquad\qquad\delta\mathcal{\zeta}_{{\cal U}}=\dot{\eta}_{{\cal U}}.\label{eq:transflachem}
\end{equation}
The variation of the conserved charges can be computed using the Regge-Teitelboim
method \cite{Regge:1974zd}, and they are given by the following surface
integral

\begin{equation}
	\delta Q\left[\eta,\eta_{{\cal U}}\right]=\frac{\kappa}{4\pi}\int d\phi\left(\eta\delta\mathcal{J}+\eta_{{\cal U}}\delta{\cal U}\right).\label{eq:Variation of the charges}
\end{equation}
The Dirac brackets of the dynamical fields $\mathcal{J}$ and $\mathcal{U}$
induced by the asymptotic conditions may be obtained from the relation
$\delta_{Y}Q\left[X\right]=\left\{ Q\left[X\right],Q\left[Y\right]\right\} $,
and is given by two independent $\hat{u}\left(1\right)$ current algebras

\begin{eqnarray}
	\left\{ \mathcal{J}\left(\phi\right),\mathcal{J}\left(\phi^{\prime}\right)\right\} ^{\star} & = & \frac{4\pi}{\kappa}\partial_{\phi}\delta\left(\phi-\phi^{\prime}\right),\nonumber \\
	\left\{ \mathcal{U}\left(\phi\right),\mathcal{U}\left(\phi^{\prime}\right)\right\} ^{\star} & = & \frac{4\pi}{\kappa}\partial_{\phi}\delta\left(\phi-\phi^{\prime}\right),\label{eq:Dirac}
\end{eqnarray}
expression that coincides with the first Poisson bracket of the mBoussinesq
hierarchy given by eq. \eqref{eq:FUndbrack}. Furthermore, the infinite
set of commuting charges of the hierarchy is obtained from the surface
integral \eqref{eq:Variation of the charges} as follows: if we take
into account that due to eq. \eqref{eq:chemical potentials} the Lagrange
multipliers are field dependent, then the consistency with their transformation
law \eqref{eq:transflachem} implies the following differential equation
that must be obeyed by $\eta$ and $\eta_{{\cal U}}$

\[
\left(\begin{array}{c}
	\dot{\eta}\left(t,\theta\right)\\
	\dot{\eta}_{{\cal U}}\left(t,\theta\right)
\end{array}\right)=\int d\phi\left(\begin{array}{cc}
	\frac{\delta^{2}H_{\left(k\right)}}{\delta\mathcal{J}\left(t,\theta\right)\delta\mathcal{J}\left(t,\phi\right)} & \frac{\delta^{2}H_{\left(k\right)}}{\delta{\cal U}\left(t,\theta\right)\delta\mathcal{J}\left(t,\phi\right)}\\
	\frac{\delta^{2}H_{\left(k\right)}}{\delta\mathcal{J}\left(t,\theta\right)\delta{\cal U}\left(t,\phi\right)} & \frac{\delta^{2}H_{\left(k\right)}}{\delta{\cal U}\left(t,\theta\right)\delta{\cal U}\left(t,\phi\right)}
\end{array}\right)\mathcal{D}\left(\begin{array}{c}
	\eta\\
	\eta_{{\cal U}}
\end{array}\right).
\]
By virtue of the integrability of the system, the most general solution
of this equation, under the assumption that $\eta$ and $\eta_{{\cal U}}$
depend locally on $\mathcal{J}$, $\mathcal{U}$ and their spatial
derivatives, is given by

\begin{equation}
	\left(\begin{array}{c}
		\eta\\
		\eta_{{\cal U}}
	\end{array}\right)=\frac{4\pi}{\kappa}\sum_{n=0}^{\infty}\alpha_{\left(n\right)}\left(\begin{array}{c}
		\frac{\delta H_{\left(n\right)}}{\delta\mathcal{J}}\\
		\frac{\delta H_{\left(n\right)}}{\delta\mathcal{U}}
	\end{array}\right),\label{eq:etas}
\end{equation}
where the $\alpha_{\left(n\right)}$ are arbitrary constants. Therefore,
replacing the solution \eqref{eq:etas} in \eqref{eq:Variation of the charges}
one can integrate the charge in the functional sense (taking the delta
outside), obtaining

\[
Q=\sum_{n=0}^{\infty}\alpha_{\left(n\right)}H_{\left(n\right)}.
\]
Thus, the conserved charges in the higher spin theory are precisely
given by a linear combination of the Hamiltonians of the mBoussinesq
hierarchy. Indeed, using the transformation law \eqref{eq:tarnsflawfields},
one can show the Hamiltonians $H_{\left(n\right)}^{I}$ are in involution
with respect to the Dirac bracket \eqref{eq:Dirac}

\[
\left\{ H_{\left(m\right)}^{I},H_{\left(n\right)}^{J}\right\} ^{\star}=0,
\]
as expected.

In sum, all the relevant properties of the integrable mBoussinesq
hierarchy described in section \ref{sec:Review-of-the} are derived
from spin-3 gravity endowed with the boundary conditions defined in
eqs. \eqref{eq:A Connections}, \eqref{eq:Auxiliary Connections}
and \eqref{eq:chemical potentials}. Thus, the reduced phase space
of spin-3 gravity and its boundary dynamics are equivalent to the
ones of the mBoussinesq hierarchy. In particular, this provides an
explicit one-to-one map between solutions of the integrable system
at the boundary and solutions of the higher spin gravity theory in
the bulk.

\subsection{Highest weight gauge, Miura map and the Boussinesq hierarchy\label{subsec:Relation-with-the}}

Asymptotic conditions for spin-3 gravity on AdS$_{3}$ in the highest
weight gauge were first given in refs. \cite{Campoleoni:2010zq,Henneaux:2010xg},
where it was shown that the asymptotic symmetries are spanned by two
copies of the classical $W_{3}$ algebra with the Brown-Henneaux central
charge. In what follows we consider the generalization introduced
in refs. \cite{Henneaux:2013dra,Bunster:2014mua}, where the most
general form of the Lagrange multipliers $a_{t}$, compatible with
the $W_{3}$ symmetry, is allowed. This generalization has the important
property that it accommodates black holes carrying non-trivial higher
spin charges.

In this subsection we show that, with a particular gauge transformation,
the auxiliary connection in the diagonal gauge \eqref{eq:Auxiliary Connections}
can be mapped to an auxiliary connection in the highest weight gauge,
such that the Miura transformation in eq. \eqref{eq:Miura}, that
relates the mBoussinesq with the Boussinesq hierarchies, is recovered
from a purely geometric construction in the higher spin theory. The
analysis is very close to the one in ref. \cite{Grumiller:2016kcp},
with the main difference that now the Lagrange multipliers $\zeta$
and $\zeta_{\mathcal{U}}$ depend on the dynamical fields according
to eq. \eqref{eq:chemical potentials}.

The angular components of the auxiliary connection in the highest
weight gauge $\hat{a}$ is assumed to be of the form

\begin{equation}
	\hat{a}_{\varphi}=L_{1}-\frac{1}{2}\mathcal{L}L_{-1}-\frac{1}{4\sqrt{3}}\mathcal{W}W_{-2}.\label{eq:Auxiliary connection HWg}
\end{equation}
Following \cite{Henneaux:2013dra,Bunster:2014mua}, the most general
form of $\hat{a}_{t}$ which is compatible with the field equations
is

\begin{align}
	\hat{a}_{t} & =\mu L_{1}-\frac{\sqrt{3}}{2}\nu W_{2}-\mu^{\prime}L_{0}+\frac{\sqrt{3}}{2}\nu^{\prime}W_{1}+\frac{1}{2}\left(\mu^{\prime\prime}-\mu\mathcal{L}-2\mathcal{W}\nu\right)L_{-1}\nonumber \\
	& \quad-\frac{\sqrt{3}}{48}\left(4\mathcal{W}\mu-7\mathcal{L}^{\prime}\nu^{\prime}-2\nu\mathcal{L}^{\prime\prime}-8\mathcal{L}\nu^{\prime\prime}+6\mathcal{L}^{2}\nu+\nu^{\prime\prime\prime\prime}\right)W_{-2}\nonumber \\
	& \quad-\frac{\sqrt{3}}{4}\left(\nu^{\prime\prime}-2\mathcal{L}\nu\right)W_{0}+\frac{\sqrt{3}}{12}\left(\nu^{\prime\prime\prime}-2\nu\mathcal{L}^{\prime}-5\mathcal{L}\nu^{\prime}\right)W_{-1}\ .\label{eq:AtHWG}
\end{align}
It is possible to find a gauge group element $g=g^{(1)}g^{(2)}$,
such that the auxiliary connection in the diagonal gauge $a$ is mapped
to the auxiliary connection in the highest weight gauge $\hat{a}$,
by a gauge transformation of the form $\hat{a}=g^{-1}\left(d+a\right)g$,
with
\[
\begin{array}{l}
	g^{(1)}=\exp\left[xL_{1}+yW_{1}+zW_{2}\right],\\
	g^{(2)}=\exp\left[-\frac{1}{2}\mathcal{J}L_{-1}-\frac{\sqrt{3}}{6}\mathcal{U}W_{-1}+\frac{\sqrt{3}}{12}\left(\mathcal{J}\mathcal{U}+\frac{1}{2}\mathcal{U}^{\prime}\right)\mathrm{W}_{-2}\right].
\end{array}
\]
Here, the functions $x$, $y$, $z$ are restricted to obey the following
differential equations

\[
\begin{array}{l}
	x^{\prime}=1+x\,\mathcal{J}+\sqrt{3}y\,\mathcal{U},\\
	y^{\prime}=y\,\mathcal{J}+\sqrt{3}x\,\mathcal{U},\\
	z^{\prime}=-\frac{1}{2}y+2z\,\mathcal{J}.
\end{array}
\]
The fields $\mathcal{L}$ and $\mathcal{W}$ are then related to the
fields $\mathcal{J}$ and $\mathcal{U}$ precisely by the Miura transformation
\eqref{eq:Miura}
\begin{align}
	\mathcal{L} & =\frac{1}{2}{\cal J}^{2}+\frac{1}{2}\mathcal{U}^{2}+{\cal J}^{\prime},\nonumber \\
	\mathcal{W} & =\frac{1}{3}\mathcal{U}^{3}-{\cal J}^{2}\mathcal{U}-\frac{1}{2}\mathcal{U}{\cal J}^{\prime}-\frac{3}{2}{\cal J}{\cal U}^{\prime}-\frac{1}{2}{\cal U}^{\prime\prime}.\label{eq:Miura2}
\end{align}
The Lagrange multipliers in the highest weight gauge, given by $\mu$
and $\nu$, are related to the variables in the diagonal gauge through
the following equations

\begin{align}
	\zeta & =\mathcal{J}\mu-\mu^{\prime}-2\left(\mathcal{J}\mathcal{U}+\frac{1}{2}\mathcal{U}^{\prime}\right)\nu+\frac{1}{2}\mathcal{U}\nu^{\prime},\nonumber \\
	\zeta_{\mathcal{U}} & =\mathcal{U}\mu-\left(\mathcal{J}^{2}-\mathcal{U}^{2}-\mathcal{J}^{\prime}\right)\nu+\frac{3}{2}\mathcal{J}\nu^{\prime}-\frac{1}{2}\nu^{\prime\prime}.\label{eq:zetas}
\end{align}
The complete Boussinesq hierarchy is then obtained from the mBoussinesq
one by virtue of the relations \eqref{eq:Miura2} and \eqref{eq:zetas}.
Indeed, from \eqref{eq:zetas}, one can prove that the chemical potentials
in the highest weight gauge take the form

\[
\mu=\frac{4\pi}{\kappa}\frac{\delta H_{\left(k\right)}^{\text{Bsq}}}{\delta\mathcal{L}},\quad\quad\nu=\frac{4\pi}{\kappa}\frac{\delta H_{\left(k\right)}^{\text{Bsq}}}{\delta\mathcal{W}},
\]
where $H_{\left(k\right)}^{\text{Bsq}}$ corresponds the $k$-th Hamiltonian
of the Boussinesq hierarchy (see appendix \ref{Appendix 5: Bsq and mBsq}
for more details on the Boussinesq hierarchy). For example, for the
first member given by $k=1$, one has

\[
H_{\left(1\right)}^{\text{Bsq}}=\frac{\kappa}{4\pi}\int d\phi\left(\lambda_{1}\mathcal{L}+\lambda_{2}\mathcal{W}\right),
\]
and hence the chemical potentials in the highest weight gauge become

\[
\mu=\lambda_{1},\quad\quad\nu=\lambda_{2}.
\]
In the particular case with $\lambda_{1}=1$ and $\lambda_{2}=0$
(first flow), the equations of motion are given by two chiral movers,
and the asymptotic conditions in eqs. \eqref{eq:Auxiliary connection HWg},
\eqref{eq:AtHWG} reduce to the ones in refs. \cite{Campoleoni:2010zq,Henneaux:2010xg}
but written in terms of the composite fields $\mathcal{L}$ and $\mathcal{W}$
according to \eqref{eq:Miura}. On the other hand, for the second
flow with $\lambda_{1}=0$ and $\lambda_{2}=1$, the field equation
in the bulk become equivalent to the Boussinesq equation in \eqref{eq:Bsqeq},
in agreement with the result found in ref. \cite{Perez:2016vqo}.

\section{Higher spin Black holes\label{sec:3 Black holes}}

The line element is no longer gauge invariant in higher spin gravity,
since it generically changes under the action of a higher spin gauge
transformation. Therefore, the spacetime geometry and the causal structure
cannot be directly used to define black holes. In refs. \cite{Gutperle:2011kf,Ammon:2011nk},
a new notion of higher spin black hole was introduced in the Euclidean
formulation of the theory, by requiring trivial holonomies for the
gauge connection around a thermal cycle $\mathcal{C}$, i.e.,

\begin{equation}
	\mathcal{H}_{\mathcal{C}}=\mathcal{P}e^{\int_{\mathcal{C}}a^{\pm}}=1.\label{eq:holocond}
\end{equation}
Here we have restored the $\pm$ superscript to denote the plus/minus
copy of the gauge field. If we assume that the Euclidean time is in
the range $0\leq t_{E}<1$ then, for time-independent configurations
in \eqref{eq:Auxiliary Connections}, the regularity condition \eqref{eq:holocond}
imposes the following restrictions on $\zeta^{\pm}$ and $\zeta_{\mathcal{U}}^{\pm}$
\begin{equation}
	\zeta^{\pm}=\pi\left(2n+m\right)\,,\qquad\qquad\zeta_{\mathcal{U}}^{\pm}=\sqrt{3}\pi m,\label{eq:chemreg}
\end{equation}
with $m$ and $n$ being integers. Static configurations that obey
\eqref{eq:chemreg} are regular Euclidean solutions and consequently
we call them ``black holes.''

According to ref. \cite{Grumiller:2016kcp}, the entropy takes the
form
\begin{equation}
	S=\frac{\kappa}{4}\int d\phi\left(\left(2n+m\right)\left(\mathcal{J}^{+}+\mathcal{J}^{-}\right)+\sqrt{3}m\left(\mathcal{U}^{+}+\mathcal{U}^{-}\right)\right),\label{eq:Entropy}
\end{equation}
which, by virtue of \eqref{eq:chemical potentials}, \eqref{eq:chemreg}
and \eqref{eq: decomp}, obeys the following first law 
\[
\delta S=\sum_{I=1}^{2}\left(\lambda_{I}^{+}\delta H_{\left(k\right)}^{I+}+\lambda_{I}^{-}\delta H_{\left(k\right)}^{I-}\right),
\]
where $H_{\left(k\right)}^{I\pm}$ are the left/right $k$-th Hamiltonian
of the mBoussinesq hierarchy associated to the flows labelled by $I=1,2$.
Note that the constants $\lambda_{I}^{\pm}$ correspond to the chemical
potentials conjugate to the extensive quantities $H_{\left(k\right)}^{I\pm}$.

When the integers $n$ and $m$ acquire the values $n=1$, $m=0$,
we obtain a branch which is connected with the pure gravitational
sector and the BTZ black hole. In that case, the entropy acquires
the simple expression 
\begin{equation}
	S=\frac{\kappa}{2}\int d\phi\left(\mathcal{J}^{+}+\mathcal{J}^{-}\right).\label{eq:Entropy-1}
\end{equation}

As was pointed out in \cite{Grumiller:2016kcp}, for constants $\mathcal{J}^{\pm}$
and $\mathcal{U}^{\pm}$, the entropy for this branch acquires the
expected form found in \cite{Henneaux:2013dra,Bunster:2014mua} for
a higher spin black hole, once it is written in terms of the charges
of the $W$-algebra.

In sum, black holes that fit within our boundary conditions in eqs.
\eqref{eq:Auxiliary Connections} and \eqref{eq:chemical potentials}
are identified with static solutions of the $k$-th element of the
mBoussinesq hierarchy. It is worth noting that it is of fundamental
importance to consider both flows to admit generic black hole configurations
without restricting the possible space of solutions. To illustrate
some of their properties, we will study the particular cases with
$k=0,1,2$ in the branch connected with the BTZ black hole. For simplicity
we consider only the plus copy.

\emph{Case with $k=0$}. The general solution of the field equations
\eqref{eq:Field equations} that obeys \eqref{eq:chemreg} is given
by two arbitrary functions of $\phi$, i.e., $\mathcal{J}=\mathcal{J}\left(\phi\right)$
and $\mathcal{U}=\mathcal{U}\left(\phi\right)$. This case corresponds
to the ``higher spin black flower'' described in ref. \cite{Grumiller:2016kcp},
carrying an infinite set of $\hat{u}\left(1\right)$ soft hairy charges.
The particular configuration with constant $\mathcal{J}$ and vanishing
$\mathcal{U}$ in the branch with $n=1$, $m=0$, corresponds to the
BTZ black hole embedded within this set of boundary conditions.

\emph{Case with $k=1$}. The choice of boundary conditions \eqref{eq:chemical potentials},
together with the regularity conditions, imply the following differential
equations

\begin{eqnarray}
	\lambda_{1}\mathcal{J}-\lambda_{2}\left(\mathcal{U}^{\prime}+2\mathcal{J}\mathcal{U}\right) & = & 2\pi,\nonumber \\
	\lambda_{1}\mathcal{U}+\lambda_{2}\left(\mathcal{J}^{\prime}-\mathcal{J}^{2}+\mathcal{U}^{2}\right) & = & 0.\label{eq:systeq}
\end{eqnarray}
These equations relate the chemical potentials $\lambda_{1}$ and
$\lambda_{2}$ with the fields $\mathcal{J}$ and $\mathcal{U}$.
In particular, for constants $\mathcal{J}$ and $\mathcal{U}$ we
obtain
\begin{equation}
	\lambda_{1}=\frac{2\pi\left(\mathcal{J}^{2}-\mathcal{U}^{2}\right)}{\mathcal{J}\left(\mathcal{J}^{2}-3\mathcal{U}^{2}\right)}\,,\qquad\qquad\lambda_{2}=\frac{2\pi\mathcal{U}}{\mathcal{J}\left(\mathcal{J}^{2}-3\mathcal{U}^{2}\right)}.\label{eq:lambdak1}
\end{equation}
The regularity conditions for the BTZ solution in the pure gravity
sector is recovered when $\mathcal{U}=0$, as expected. On the other
hand, the configurations with $\mathcal{J}\left(\mathcal{J}^{2}-3\mathcal{U}^{2}\right)=0$
possess non-trivial holonomies along the thermal cycle. Hence, one
might identify them with extremal configurations, along the lines
of ref. \cite{Henneaux:2015ywa} (see also \cite{Banados:2015tft}).
Indeed, when $\mathcal{U}^{\pm}=0$, solutions with vanishing $\mathcal{J}^{+}$
or $\mathcal{J}^{-}$ correspond to extremal BTZ black holes.

\emph{Case with $k=2$}. For constants $\mathcal{J}$ and $\mathcal{U}$,
the regularity conditions \eqref{eq:chemreg} fix the chemical potentials
$\lambda_{I}$ according to

\begin{eqnarray}
	\lambda_{1} & = & \frac{3\pi\mathcal{U}\left(15\mathcal{J}^{4}-10\mathcal{J}^{2}\mathcal{U}^{2}+7\mathcal{U}^{4}\right)}{2\mathcal{J}\left(9\mathcal{J}^{8}-72\mathcal{J}^{6}\mathcal{U}^{2}+210\mathcal{J}^{4}\mathcal{U}^{4}-224\mathcal{J}^{2}\mathcal{U}^{6}-3\mathcal{U}^{8}\right)},\nonumber \\
	\lambda_{2} & = & \frac{3\pi\left(3\mathcal{J}^{4}+6\mathcal{J}^{2}\mathcal{U}^{2}-5\mathcal{U}^{4}\right)}{2\mathcal{J}\left(9\mathcal{J}^{8}-72\mathcal{J}^{6}\mathcal{U}^{2}+210\mathcal{J}^{4}\mathcal{U}^{4}-224\mathcal{J}^{2}\mathcal{U}^{6}-3\mathcal{U}^{8}\right)}.\label{eq:lambdak2}
\end{eqnarray}
When $\mathcal{U}=0$, the auxiliary connection \eqref{eq:Auxiliary Connections}
reduces to the one that describes the BTZ geometry \cite{Afshar:2016wfy}.
However, in contrast to the case with $k=1$, the chemical potential
$\lambda_{1}$ associated to the first flow now vanishes, and hence,
the information of the BTZ black hole is completely encoded in the
second flow.

Note that for the cases with $k=1$ and $k=2$ described above, it
is of fundamental importance to take into account both flows to have
black holes characterized by two independent constants $\mathcal{J}^{\pm}$
and $\mathcal{U}^{\pm}$ for each copy. If one of the chemical potentials
$\lambda_{1}^{\pm}$ or $\lambda_{2}^{\pm}$ is set to zero, then
eqs. \eqref{eq:lambdak1} and \eqref{eq:lambdak2} would imply non-trivial
restrictions in the values of $\mathcal{J}^{\pm}$ and $\mathcal{U}^{\pm}$,
truncating in this way the spectrum of allowed black hole configurations.

\section{Hamiltonian reduction and boundary dynamics\label{sec:3 Reduction-of-the-Chern-Simons}}

In this section we will discuss the Hamiltonian reduction and the
boundary dynamics of the Chern-Simons action that describes spin-3
gravity endowed with the boundary conditions associated to the mBoussinesq
hierarchy. In the case of pure gravity with Brown-Henneaux asymptotic
conditions, the boundary dynamics is described by two left and right
chiral bosons which, by virtue of a Bäcklund transformation, turn
out to be equivalent to a Liouville theory \cite{Coussaert:1995zp,Henneaux:1999ib}.
The analysis was done by performing a Hamiltonian reduction of the
Wess-Zumino-Witten (WZW) theory at the boundary \cite{Forgacs:1989ac,Alekseev:1988ce,Witten:1989hf,Elitzur:1989nr}.

Here we follow an approach similar to the one proposed in ref. \cite{Gonzalez:2018jgp}
for KdV-type boundary conditions in pure gravity, where instead of
passing through the WZW theory, the boundary conditions are implemented
directly in the Hamiltonian action.

Let us consider the Hamiltonian action with the appropriate boundary
term in eq. \eqref{eq:Action in hamiltonian form}. The constraints
$F_{ij}=0$ are locally solved by expressing the spatial components
of the gauge connection in terms of a group element $G$ as $A_{i}=G^{-1}\partial_{i}G$,
provided that there are no holes in the spatial section. After replacing
the solution of the constraints into the action \eqref{eq:Action in hamiltonian form},
the following decomposition is obtained

\begin{equation}
	I_{can}\left[A\right]=I_{1}+I_{2}+B_{\infty},\label{eq: Total actions WZW}
\end{equation}
where

\begin{align}
	I_{1} & =\frac{\kappa}{16\pi}\int dtdrd\phi\epsilon^{ij}\left\langle \partial_{t}\left(G^{-1}\right)\partial_{i}GG^{-1}\partial_{j}G\right\rangle ,\label{eq: I1 WZW action}\\
	I_{2} & =-\frac{\kappa}{16\pi}\int d\phi dt\left\langle \partial_{t}G\partial_{\phi}\left(G^{-1}\right)\right\rangle .\label{eq:I2 WZW action}
\end{align}
The Wess-Zumino term $I_{1}$ reduces to a boundary term that vanishes
provided the group element is decomposed as $G=g\left(t,\phi\right)b\left(r\right)$
near the boundary (see appendix \ref{Appendix 7: Wess-Zumino-term}\textcolor{red}{{}
}for a detailed proof). Here $b\left(r\right)$ is the group element
that depends on the radial coordinate in eq. \eqref{eq:A Connections},
while $g\left(t,\phi\right)$ is such that the auxiliary connection
$a$ in \eqref{eq:Auxiliary Connections} is written as $a_{i}=g^{-1}\partial_{i}g$.
With this decomposition the term $I_{2}$ becomes

\begin{equation}
	I_{2}=-\frac{\kappa}{16\pi}\int d\phi dt\left\langle \dot{g}\partial_{\phi}g^{-1}\right\rangle .\label{eq:I1 WZW 2}
\end{equation}
Since the auxiliary connection in \eqref{eq:Auxiliary Connections}
is a diagonal matrix, we can write the group element $g$ as follows

\begin{equation}
	g=\exp\left[\sqrt{\frac{8\pi}{\kappa}}\varphi L_{0}+\sqrt{\frac{6\pi}{\kappa}}\psi W_{0}\right],\label{eq:Gauss decomposition}
\end{equation}
where $\varphi$ and $\psi$ are functions that depend only on $t$
and $\phi$. In what follows we will assume that these fields are
periodic in the angle, and consequently possible contributions coming
from non-trivial holonomies around $\phi$ are not considered in this
analysis.

Consistency with the asymptotic form of the auxiliary connection \eqref{eq:Auxiliary Connections}
then implies

\begin{equation}
	\mathcal{J}=\sqrt{\frac{8\pi}{\kappa}}\varphi^{\prime}\,,\qquad\qquad\mathcal{U}=\sqrt{\frac{8\pi}{\kappa}}\psi^{\prime}.\label{eq: Dynamical fields of Gauss decomposition-2}
\end{equation}
Replacing \eqref{eq:Gauss decomposition} in \eqref{eq:I1 WZW 2}
we find

\begin{eqnarray}
	I_{2} & = & \int d\phi dt\left(\varphi^{\prime}\dot{\varphi}+\psi^{\prime}\dot{\psi}\right).\label{eq:I1 WZW 3}
\end{eqnarray}
Thus, if we use the expression \eqref{eq:Boundary terms in terms of H}
for the boundary term $B_{\infty}$, we finally obtain the following
reduced action at the boundary

\begin{eqnarray}
	I_{\left(k\right)} & = & \int dt\left[\int d\phi\left(\varphi^{\prime}\dot{\varphi}+\psi^{\prime}\dot{\psi}\right)-H_{\left(k\right)}\right].\label{eq:Total actions WZW 2-1-2-1}
\end{eqnarray}
This action describes the dynamics of the fields $\varphi$ and $\psi$,
whose interactions are described by the $k$-th Hamiltonian of the
mBoussinesq hierarchy. The members of the mBoussinesq hierarchy are
then recovered from the equations of motion derived from the action
\eqref{eq:Total actions WZW 2-1-2-1}, provided we identify the fields
according to \eqref{eq: Dynamical fields of Gauss decomposition-2}.
In this sense, the field equations coming from \eqref{eq:Total actions WZW 2-1-2-1}
define ``potential equations'' for the ones of the mBoussinesq hierarchy.

The action \eqref{eq:Total actions WZW 2-1-2-1} is invariant under
the following transformations
\[
\delta\varphi=\sqrt{\frac{\kappa}{8\pi}}\eta+f\left(t\right)\,,\qquad\qquad\delta\psi=\sqrt{\frac{\kappa}{8\pi}}\eta_{{\cal U}}+f_{{\cal U}}\left(t\right),
\]
where the parameter $\eta$ and $\eta_{{\cal U}}$, given by \eqref{eq:etas},
are associated to the infinite charges in involution of the integrable
system. Indeed, all the Hamiltonians of the hierarchy may be obtained
by a direct application of Noether theorem. On the other hand, the
arbitrary functions of the time $f\left(t\right)$ and $f_{{\cal U}}\left(t\right)$
define gauge symmetries of \eqref{eq:Total actions WZW 2-1-2-1} that
allow to gauge away the zero modes of these fields.

Furthermore, the action \eqref{eq:Total actions WZW 2-1-2-1} has
an additional Lifschitz scaling symmetry in the particular case when
only one of the two flows is considered. For the first flow, with
$\lambda_{1}=1$ and $\lambda_{2}=0$, the dynamical exponent is $z=3k-2$,
while for the second flow, with $\lambda_{1}=0$ and $\lambda_{2}=1$,
is $z=3k-1$, as expected from the invariance properties of the mBoussinesq
hierarchy described in section \ref{sec:Review-of-the}. For the $I$-th
flow of the $k$-th element of the hierarchy, the generators of the
Lifshitz algebra are given by

\[
H=H_{\left(k\right)}^{I}\,,\qquad\qquad P=H_{\left(1\right)}^{1}\,,\qquad\qquad D=-\frac{\kappa}{4\pi}\int d\phi\left(\frac{1}{2}\phi\left(\mathcal{J}^{2}+\mathcal{U}^{2}\right)\right)-ztH_{\left(k\right)}^{I}.
\]
Here, $H$ is the generators of translations in time, $P$ of translations
in space, and $D$ of anisotropic dilatations. Using the the Dirac
brackets \eqref{eq:Dirac} it is straightforward to show that they
close in the Lifshitz algebra
\[
\left\{ P,H\right\} ^{\star}=0,\qquad\left\{ D,P\right\} ^{\star}=P,\qquad\left\{ D,H\right\} ^{\star}=zH,
\]
where $z$ is the dynamical exponent associated to the corresponding
flow.

Let us consider as an example the case with $k=1$. The action then
takes the following form

\begin{eqnarray*}
	I_{\left(1\right)} & = & \int d\phi dt\left[\varphi^{\prime}\dot{\varphi}+\psi^{\prime}\dot{\psi}-\lambda_{1}\left(\varphi^{\prime2}+\psi^{\prime2}\right)+2\lambda_{2}\left(\varphi^{\prime}\psi^{\prime\prime}+\sqrt{\frac{8\pi}{\kappa}}\left(\varphi^{\prime2}\psi^{\prime}-\frac{1}{3}\psi^{\prime3}\right)\right)\right].
\end{eqnarray*}
For the flow with $\lambda_{1}=1$ and $\lambda_{2}=0$ we recover
the Floreanini-Jackiw action for two free chiral bosons \cite{Floreanini:1987as}.
On the other hand, for the flow with $\lambda_{1}=0$ and $\lambda_{2}=1$,
the action contains non-trivial interacting term and is invariant
under Lifshitz transformations with dynamical exponent $z=2$, as
expected.

\section{Generalized Gibbs ensemble\label{sec:Generalized Gibbs}}

The existence of an infinite number of commuting charges in the mBoussinesq
hierarchy opens the possibility of study more general thermodynamic
ensembles that generically might include all possible charges. They
are called ``Generalized Gibbs Ensemble'' (GGE).

In the case of a two-dimensional conformal field theory, a GGE is
constructed with the infinite set of charges in involution of the
KdV hierarchy, which are obtained as composite operators in terms
of the stress tensor \cite{Sasaki:1987mm,Eguchi:1989hs,Bazhanov:1994ft}
(see refs. \cite{Calabrese:2011vdk,Sotiriadis:2014uza,PhysRevLett.115.157201,Vidmar_2016,deBoer:2016bov,Perez:2016vqo,Pozsgay_2017,Dymarsky:2018lhf,Maloney:2018hdg,Maloney:2018yrz,Dymarsky:2018iwx,Brehm:2019fyy,Dymarsky:2019etq,Dymarsky:2020tjh}
for recent results on GGE).

In our context, the Hamiltonians of the mBoussinesq hierarchy can
be used to construct the GGE of a two-dimensional CFT with spin-3
currents. As discussed in subsection \ref{subsec:Relation-with-the},
the Miura transformation \eqref{eq:Miura} maps the Hamiltonians of
the mBoussinesq hierarchy into the Hamiltonians of the Boussinesq
one, which depend on the stress tensor $\mathcal{L}$ and the spin-3
current $\mathcal{W}$. These Hamiltonians belong to the enveloping
algebra of the $W_{3}$-algebra, and consequently define an infinite
set of commuting charges that are composite operators in terms of
$\mathcal{L}$ and $\mathcal{W}$. In this sense, the relation between
the mBoussinesq hierarchy and spin-3 gravity discussed in this article,
provides a natural holographic bulk dual description of this type
of GGE.

This may be implemented as follows. Instead of considering one particular
$H_{\left(k\right)}$ as the Hamiltonian of the dynamical system,
we deal with a linear combination of them, i.e.,
\begin{equation}
	H_{GGE}=\sum_{n=1}^{\infty}\sum_{I=1}^{2}\gamma_{n}\left(\lambda_{I}H_{\left(n\right)}^{I}\right).\label{eq:HGGE}
\end{equation}
If we want to interpret this general Hamiltonian as the one of a CFT$_{2}$
given by $H_{\left(1\right)}^{1}$, deformed by (multitrace) deformations
that include spin-3 currents, then we must identify the inverse (right)
temperature as $\beta_{+}=T_{+}^{-1}=\alpha_{1}\lambda_{1}$, and
the additional chemical potentials as $\mu_{n,I}:=T_{+}\alpha_{n}\lambda_{I}$.
However, this is not the only possibility. Any $H_{\left(k\right)}^{I}$
could be considered as the ``undeformed Hamiltonian,'' allowing
new branches that generically change the phase structure of the theory
\cite{Perezinprog}.

Black holes are described by static configurations of the dynamical
system with Hamiltonian \eqref{eq:HGGE}. The regularity condition
\eqref{eq:chemreg} then takes the form 
\[
\sum_{n=1}^{\infty}\sum_{I=1}^{2}\gamma_{n}\lambda_{I}\frac{\delta H_{\left(n\right)}^{I}}{\delta\mathcal{J}}=\frac{\kappa}{2}\,,\qquad\qquad\sum_{n=1}^{\infty}\sum_{I=1}^{2}\gamma_{n}\lambda_{I}\frac{\delta H_{\left(n\right)}^{I}}{\delta\mathcal{\mathcal{U}}}=0.
\]
These equations guarantee that the Euclidean action principle attains
an extremum, and hence they fully characterize the thermodynamics
of the GGE.

On the other hand, the boundary dynamics obtained by the Hamiltonian
reduction is easily generalized to the case when the Hamiltonian is
given by \eqref{eq:HGGE}. Indeed, the boundary action now becomes
\begin{eqnarray}
	I_{GGE} & = & \int dt\left[\int d\phi\left(\varphi^{\prime}\dot{\varphi}+\psi^{\prime}\dot{\psi}\right)-H_{GGE}\right].\label{eq:Total actionGGE}
\end{eqnarray}
spin-N gravity and modified Gelfand-Dickey hierarchies

\section{Higher spin gravity with gauge group $SL\left(N,\mathbb{R}\right)\times SL\left(N,\mathbb{R}\right)$
	and modified Gelfand-Dickey hierarchies\label{sec:Higher-spin-gravity SLN}}

Our results can be generalized to the case of three-dimensional higher
spin gravity with gauge group $SL\left(N,\mathbb{R}\right)\times SL\left(N,\mathbb{R}\right)$,
where the corresponding integrable systems are the called ``modified
Generalized KdV hierarchies,'' or ``modified Gelfand-Dickey (mGD)
hierarchies'' (see e.g. chapter 4 of ref. \cite{2003ASMP...26.....D}).
The link with higher spin gravity is based on the zero curvature formulation
of these integrable systems \cite{Drinfeld:1984qv}.

Let us consider asymptotic conditions for spin-$N$ gravity described
by the following auxiliary gauge connection valued on the $sl\left(N,\mathbb{R}\right)$
algebra
\begin{equation}
	a=\left({\cal J}d\phi+\zeta dt\right)L_{0}+\sum_{s=3}^{N}\sigma_{s}\left(\mathcal{U}^{\left(s\right)}d\phi+\zeta_{\mathcal{U}}^{\left(s\right)}dt\right)W_{0},\label{eq:Auxiliary Connections SLN}
\end{equation}
with $\sigma_{s}$ given by 
\[
\sigma_{s}=\left(\frac{\left(2s-1\right)!\left(2s-2\right)!}{48(s-1)!^{4}}\frac{1}{\prod_{i=2}^{s-1}\left(N^{2}-i^{2}\right)}\right)^{\frac{1}{2}}.
\]
The Chern-Simons action takes the same form as in \eqref{eq:Chern-Simons Action},
but replacing $\kappa_{3}\rightarrow\kappa_{N}=3l/(2N(N^{2}-1)G)$.
Hence, the variation of the boundary term of the canonical Chern-Simons
action becomes
\begin{equation}
	\delta B_{\infty}=-\frac{\kappa}{4\pi}\int dtd\phi\left(\zeta\delta\mathcal{J}+\sum_{s=3}^{N}\zeta_{\mathcal{U}}^{\left(s\right)}\delta\mathcal{U}^{\left(s\right)}\right).\label{eq:Variation of boundary terms-1}
\end{equation}
To make contact with the mGD hierarchies, we choose the Lagrange multipliers
as follows
\begin{equation}
	\zeta=\frac{4\pi}{\kappa}\frac{\delta H_{\left(k,N\right)}^{\text{mGD}}}{\delta\mathcal{J}},\qquad\qquad\zeta_{\mathcal{U}}^{\left(s\right)}=\frac{4\pi}{\kappa}\frac{\delta H_{\left(k,N\right)}^{\text{mGD}}}{\delta\mathcal{U}^{\left(s\right)}},\label{eq:chemical potentials-1}
\end{equation}
where $H_{\left(k,N\right)}^{\text{mGD}}$ corresponds to the $k$-th
Hamiltonian of the $N$-th hierarchy \cite{Drinfeld:1984qv}. With
this choice of boundary conditions, the boundary term of the canonical
Chern-Simons action integrates as 
\[
B_{\infty}=-\int dtH_{\left(k,N\right)}.
\]
As expected, the Hamiltonian of the higher spin theory coincides with
one of the hierarchy.

The Dirac brackets are described by $N-1$ $\hat{u}\left(1\right)$
current algebras
\begin{eqnarray}
	\left\{ \mathcal{J}\left(\phi\right),\mathcal{J}\left(\phi^{\prime}\right)\right\} ^{\star} & = & \frac{4\pi}{\kappa}\partial_{\phi}\delta\left(\phi-\phi^{\prime}\right),\nonumber \\
	\left\{ \mathcal{U}^{\left(s\right)}\left(\phi\right),\mathcal{U}^{\left(s'\right)}\left(\phi^{\prime}\right)\right\} ^{\star} & = & \frac{4\pi}{\kappa}\partial_{\phi}\delta\left(\phi-\phi^{\prime}\right)\delta^{s,s'},\label{eq:Dirac-1}
\end{eqnarray}
matching the first Poisson structure of the mGD hierarchies. Moreover,
their members are obtained from the equations of motion of the higher
spin theory with the boundary conditions \eqref{eq:Auxiliary Connections SLN},
\eqref{eq:chemical potentials-1}, and are given by
\[
\dot{\mathcal{J}}=\frac{4\pi}{\kappa}\partial_{\phi}\left(\frac{\delta H_{\left(k,N\right)}^{\text{mGD}}}{\delta\mathcal{J}}\right)\,,\qquad\qquad\dot{\mathcal{U}}^{\left(s\right)}=\frac{4\pi}{\kappa}\partial_{\phi}\left(\frac{\delta H_{\left(k,N\right)}^{\text{mGD}}}{\delta\mathcal{U}^{\left(s\right)}}\right).
\]

The mGD hierarchies are related to the called ``Generalized KdV hierarchies,''
or ``Gelfand-Dickey (GD) hierarchies'' by an appropriate generalization
of the Miura transformation. One of the two Poisson brackets of the
GD hierarchies is described by the $W_{N}$-algebra, whose generators
are composite in terms of the $\hat{u}\left(1\right)$ currents of
the mGD hierarchies. Hence, according to the Hamiltonian reduction
in \cite{Drinfeld:1984qv}, the generalized Miura transformation should
emerge geometrically from our boundary conditions once they are expressed
in the highest weight gauge, as in the case for $N=3$ described in
subsection \ref{subsec:Relation-with-the}. Note that since generically
the expression for the stress tensor of the $W_{N}$-algebra in terms
of $\hat{u}\left(1\right)$ currents is the one of a twisted Sugawara
construction, there is a particular flow in which the currents are
chiral. This case precisely corresponds to one of the proposals in
ref. \cite{Perez:2020klz}, for describing gravitational duals of
averaged CFT's on the Narain lattice \cite{Maloney:2020nni,Afkhami-Jeddi:2020ezh}
(see \cite{Cotler:2020ugk} for an alternative proposal for a possible
gravitational dual).

\chapter*{Conclusions \label{sec:5 Conclusions}}
\addcontentsline{toc}{chapter}{Conclusions}

In this thesis, a new set of boundary conditions for pure gravity on AdS$_3$ was proposed, with the very special property that the boundary dynamics of the gravitational theory is described by an integrable hierarchy of differential equations called the Gardner hierarchy \cite{Ojeda:2019xih}. These boundary conditions can be easily described using the diagonal gauge in the Chern-Simons formulation of General Relativity, and are closely related with the soft hairy ones in ref. \cite{Afshar:2016wfy}. The main difference is that now the chemical potentials are chosen to depend on the dynamical fields in very precise way, instead of being fixed at infinity. This different choice of chemical potentials produces a  drastic change in the boundary dynamics of the theory, and therefore, the whole integrable structure of the hierarchy can be recovered from the gravitational theory, including its infinite number of conserved charges in involution. Black hole solutions which fit within these boundary conditions were also found, and they were shown to be described by static configurations associated to the corresponding member of the Gardner hierarchy.

These results generalize along new directions the previous ones presented in ref. \cite{Perez:2016vqo}, reinforcing this new deep relationship between three-dimensional gravity with certain special boundary conditions and some two-dimensional integrable systems. Indeed, as was shown in ref. \cite{Ojeda:2020bgz} and explained in detail in chapter \ref{chapter:Gelfand-Dickey hierarchy Part}, it is also possible to further extend these results to the case of higher spin gravity on AdS$_3$ with gauge group $SL\left(N,\mathbb{R}\right)\times SL\left(N,\mathbb{R}\right)$, where the corresponding integrable system is now given by the modified Gelfand-Dickey hierarchy.

\newpage

Remarkably, this construction allows to establish a one-to-one map between three-dimensional (higher-spin) geometries that solve the bulk gravitational field equations, and solutions of the corresponding two-dimensional integrable system. This could allow, for example, to study unsolved problems in black hole physics using the powerful tools developed in the study of the integrable systems, Vice versa, one could try to understand different phenomena in the two-dimensional theory in terms of (higher-spin) geometry, which could be seen as a kind ``geometrization''  of the integrable system. In sum, this new link between three-dimensional gravitational theories in the bulk and two-dimensional integrable systems on the boundary provides a new kind of duality between these two seemingly different physical systems. This research field is still in an early stage, and one would expect that it should be able to provide new insights in the solution of some of the open problems in both physical theories.

\appendix

\chapter{Operator $P$ in KdV Lax pair\label{Appendix 1: Op P}}

The operator $P$ can be obtained directly from the square root of
the operator $L$ in eq. \eqref{eq:Lax L - KdV}. Let us propose the
following ansatz

\begin{equation}
	L^{\frac{1}{2}}=\partial_{\phi}+f_{0}+f_{1}\partial_{\phi}^{-1}+f_{2}\partial_{\phi}^{-2}+\ldots,\label{eq:L12 ansatz}
\end{equation}
which is a pseudo-differential operator, where $\partial_{\phi}^{-1}$
is an integration operator which satisfy $\partial_{\phi}\partial_{\phi}^{-1}=\partial_{\phi}^{-1}\partial_{\phi}=1$
and the general Leibniz rule

\[
\partial_{\phi}^{-1}h=h\partial_{\phi}^{-1}-h^{\prime}\partial_{\phi}^{-2}+h^{\prime\prime}\partial_{\phi}^{-3}-\ldots
\]
The $f_{n}$ in eq. \eqref{eq:L12 ansatz} are functions depending
on the field $\mathcal{L}$ and its spatial derivatives, they can
be obtained recursively through the condition

\[
L=L^{\frac{1}{2}}\cdot L^{\frac{1}{2}}=\partial_{\phi}^{2}+2f_{0}\partial_{\phi}+\left(f_{0}^{2}+2f_{1}+f_{0}^{\prime}\right)+\left(f_{1}^{\prime}+2f_{0}f_{1}+2f_{2}\right)\partial_{\phi}^{-1}+\ldots
\]
and using the definition of the operator $L$ in eq. \eqref{eq:Lax L - KdV},
we obtain

\begin{equation}
	L^{\frac{1}{2}}=\partial_{\phi}+\mathcal{L}\partial_{\phi}^{-1}-\frac{1}{2}\mathcal{L}^{\prime}\partial_{\phi}^{-2}+\ldots\label{eq:Root Operator L}
\end{equation}
where the lower orders in eq. \eqref{eq:Root Operator L} are needed
for calculate the odd powers of this root operator. The first one
is given by 

\begin{equation}
	L^{\frac{3}{2}}=L^{\frac{1}{2}}\cdot L=\partial_{\phi}^{3}+3\mathcal{L}\partial_{\phi}+\frac{3}{2}\mathcal{L}^{\prime}+\left(\frac{3}{2}\mathcal{L}^{2}+\frac{1}{4}\mathcal{L}^{\prime\prime}\right)\partial_{\phi}^{-1}+\left(\ldots\right)\partial_{\phi}^{-2}+\ldots,\label{eq:3/2 Operator L}
\end{equation}
Finally, the second operator $P$ is defined from the eq. \eqref{eq:3/2 Operator L}
as

\[
P=\left(L^{\frac{3}{2}}\right)_{\geq0}=\partial_{\phi}^{3}+\frac{3}{2}\mathcal{L}\partial_{\phi}+\frac{3}{4}\mathcal{L}^{\prime},
\]
where $\left(\ldots\right)_{\geq0}$ correspond to the purely differential
part of the operator.

\chapter{Gelfand-Dickey polynomials and Hamiltonians of the Gardner hierarchy\label{Appendix 2: Gelfand-Dickey pol Gardner}}

In this appendix, we explicitly display the first Hamiltonians $H_{\left(k\right)}$
associated to the Gardner integrable system, as well as their corresponding
generalized Gelfand-Dickey polynomials $R_{\left(k\right)}$. The first Hamiltonians $H_{\left(k\right)}$ are given by

\begin{eqnarray}
	\frac{4\pi}{\hat{\kappa}}H_{\left(-1\right)} & = & \int d\phi\left(a^{-1}\mathcal{J}\right),\nonumber \\
	\frac{4\pi}{\hat{\kappa}}H_{\left(0\right)} & = & \int d\phi\left(\frac{1}{2}\mathcal{J}^{2}\right),\nonumber \\
	\frac{4\pi}{\hat{\kappa}}H_{\left(1\right)} & = & \int d\phi\left(\frac{1}{2}a\mathcal{J}^{3}+\frac{1}{4}b\mathcal{J}^{4}+\mathcal{J}^{\prime2}\right),\label{eq:k-thH}\\
	\frac{4\pi}{\hat{\kappa}}H_{\left(2\right)} & = & \int d\phi\left(\frac{5}{8}a^{2}\mathcal{J}^{4}-\frac{5}{2}a\mathcal{J}^{2}\mathcal{J}^{\prime\prime}+\frac{1}{4}b^{2}\mathcal{J}^{6}-\frac{5}{3}b\mathcal{J}^{3}\mathcal{J}^{\prime\prime}+\frac{3}{4}ab\mathcal{J}^{5}+2\mathcal{J}^{\prime\prime2}\right),\nonumber \\
	\frac{4\pi}{\hat{\kappa}}H_{\left(3\right)} & = & \int d\phi\left(\frac{7}{8}a^{3}\mathcal{J}^{5}-\frac{35}{6}a^{2}\mathcal{J}^{3}\mathcal{J}^{\prime\prime}+7a\mathcal{J}^{2}\mathcal{J}^{\prime\prime\prime\prime}+\frac{5}{16}b^{3}\mathcal{J}^{8}+\frac{35}{2}b^{2}\mathcal{J}^{\prime2}\mathcal{J}^{4}\right.\nonumber \\
	&  & \left.+\frac{7}{3}b\left(2\mathcal{J}^{3}\mathcal{J}^{\prime\prime\prime\prime}+\mathcal{J}^{\prime4}\right)+\frac{7}{4}a^{2}b\mathcal{J}^{6}+\frac{5}{4}ab^{2}\mathcal{J}^{7}+35ab\mathcal{J}^{\prime2}\mathcal{J}^{3}-4\mathcal{J}\mathcal{J}^{\prime\prime\prime\prime\prime\prime}\right).\nonumber 
\end{eqnarray}
Note that the Hamiltonians of the Gardner hierarchy cannot be written
as the sum of the Hamiltonians of KdV $(b=0)$ and mKdV $\left(a=0\right)$,
because there are cross terms. The case $H_{\left(-1\right)}$, that
it is obtained by extending the hierarchy backwards, is special because
contains $a^{-1}$, and consequently it cannot be defined in the mKdV
hierarchy.

The generalized Gelfand-Dickey polynomials $R_{\left(k\right)}$ are
obtained using eq. \eqref{eq:G-RHJ}, and take the form

\begin{eqnarray}
	\frac{4\pi}{\hat{\kappa}}R_{\left(-1\right)} & = & \frac{1}{a},\nonumber \\
	\frac{4\pi}{\hat{\kappa}}R_{\left(0\right)} & = & \mathcal{J},\nonumber \\
	\frac{4\pi}{\hat{\kappa}}R_{\left(1\right)} & = & \left(\frac{3}{2}a\mathcal{J}^{2}+b\mathcal{J}^{3}-2\mathcal{J^{\prime\prime}}\right),\nonumber \\
	\frac{4\pi}{\hat{\kappa}}R_{\left(2\right)} & = & \left(\frac{5}{2}a^{2}\mathcal{J}^{3}-5a\left(\mathcal{J}^{\prime2}+2\mathcal{J}\mathcal{J}^{\prime\prime}\right)+\frac{3}{2}b^{2}\mathcal{J}^{5}\right.\label{eq:k-thR}\\
	&  & \left.-10b\left(\mathcal{J}\mathcal{J}^{\prime2}+\mathcal{J}^{2}\mathcal{\mathcal{J}^{\prime\prime}}\right)+\frac{15}{4}ab\mathcal{J}^{4}+4\mathcal{J}^{\prime\prime\prime\prime}\right),\nonumber \\
	\frac{4\pi}{\hat{\kappa}}R_{\left(3\right)} & = & \left(\frac{35}{8}a^{3}\mathcal{J}^{4}-35a^{2}\left(\mathcal{J}\mathcal{\mathcal{J}}^{\prime2}+\mathcal{J}^{2}\mathcal{\mathcal{J}^{\prime\prime}}\right)+7a\left(4\mathcal{J}\mathcal{J}^{\prime\prime\prime\prime}+6\mathcal{J}^{\prime\prime2}+8\mathcal{J}^{\prime}\mathcal{J}^{\prime\prime\prime}\right)\right.\nonumber \\
	&  & +\frac{5}{2}b^{3}\mathcal{J}^{7}+7b\left(4\mathcal{J}^{2}\mathcal{J}^{\prime\prime\prime\prime}+12\mathcal{J}\mathcal{J}^{\prime\prime2}+16\mathcal{J}\mathcal{J}^{\prime}\mathcal{J}^{\prime\prime\prime}+20\mathcal{J}^{\prime2}\mathcal{J}^{\prime\prime}\right)-8\mathcal{J}^{\prime\prime\prime\prime\prime\prime}\nonumber \\
	&  & \left.-35b^{2}\left(\mathcal{J}^{4}\mathcal{J}^{\prime\prime}+2\mathcal{J}^{3}\mathcal{J}^{\prime2}\right)+\frac{21}{2}a^{2}b\mathcal{J}^{5}+\frac{35}{4}ab^{2}\mathcal{J}^{6}-35ab\left(2\mathcal{J}^{3}\mathcal{J}^{\prime\prime}+3\mathcal{J}^{2}\mathcal{J}^{\prime2}\right)\right).\nonumber 
\end{eqnarray}

\chapter{Second Hamiltonian structure of the modified Boussinesq hierarchy\label{Appendix 3: Second Hamiltonian structure mBsq}}

The second Hamiltonian structure of the mBoussinesq hierarchy is defined
by the operator $\mathcal{D}_{\left(2\right)}$ in \eqref{eq:secondPstr},
whose explicit components are given by

\begin{eqnarray*}
	\frac{\hat{\kappa}}{4\pi}\mathcal{D}_{(2)}^{11} & = & -4\left(2\left({\cal U}^{\prime\prime}+2\left({\cal J}{\cal U}\right)^{\prime}\right)\partial_{\phi}^{-1}\left({\cal J}\partial_{\phi}+\partial_{\phi}^{2}\right)+{\cal J}^{\prime}\partial_{\phi}^{-1}\left(\left(4{\cal J}\mathcal{U}+3\mathcal{U}^{\prime}\right)\partial_{\phi}+\mathcal{U}\partial_{\phi}^{2}\right)\right.\\
	&  & \left.+\left(8{\cal J}^{2}\mathcal{U}-5{\cal U}\mathcal{J}^{\prime}-3{\cal U}^{\prime\prime}\right)\partial_{\phi}-3\mathcal{U}^{\prime}\partial_{\phi}^{2}-2{\cal U}\partial_{\phi}^{3}\right),\\
	\frac{\hat{\kappa}}{4\pi}\mathcal{D}_{(2)}^{12} & = & -4\left(2\left({\cal U}^{\prime\prime}+2\left({\cal J}{\cal U}\right)^{\prime}\right)\partial_{\phi}^{-1}\left({\cal U}\partial_{\phi}\right)+{\cal J}^{\prime}\partial_{\phi}^{-1}\left(\left(2\mathcal{J}^{2}-2\mathcal{U}^{2}+\mathcal{J}^{\prime}\right)\partial_{\phi}+3\mathcal{J}\partial_{\phi}^{2}+\partial_{\phi}^{3}\right)\right.\\
	&  & \left.+\left(2{\cal J}^{3}+2{\cal J}\mathcal{U}^{2}+2{\cal U}^{2\prime}-3{\cal J}{\cal J}^{\prime}-{\cal J}^{\prime\prime}\right)\partial_{\phi}+\left(\mathcal{J}^{2}+\mathcal{U}^{2}-4\mathcal{J}^{\prime}\right)\partial_{\phi}^{2}-2{\cal J}\partial_{\phi}^{3}-\partial_{\phi}^{4}\right),\\
	\frac{\hat{\kappa}}{4\pi}\mathcal{D}_{(2)}^{21} & = & -4\left(2\left({\cal J}^{2\prime}-{\cal U}^{2\prime}-{\cal J}^{\prime\prime}\right)\partial_{\phi}^{-1}\left({\cal J}\partial_{\phi}+\partial_{\phi}^{2}\right)+{\cal U}^{\prime}\partial_{\phi}^{-1}\left(\left(4{\cal J}\mathcal{U}+3\mathcal{U}^{\prime}\right)\partial_{\phi}+\mathcal{U}\partial_{\phi}^{2}\right)\right.\\
	&  & \left.+\left(2{\cal J}^{3}+2{\cal J}\mathcal{U}^{2}-4{\cal J}^{2\prime}+3{\cal U}{\cal U}^{\prime}+{\cal J}^{\prime\prime}\right)\partial_{\phi}-\left(3\mathcal{J}^{\prime}+\mathcal{J}^{2}+\mathcal{U}^{2}\right)\partial_{\phi}^{2}-2{\cal J}\partial_{\phi}^{3}+\partial_{\phi}^{4}\right),\\
	\frac{\hat{\kappa}}{4\pi}\mathcal{D}_{(2)}^{22} & = & -4\left(-2\left({\cal J}^{\prime\prime}+{\cal U}^{2\prime}-{\cal J}^{2\prime}\right)\partial_{\phi}^{-1}\left({\cal U}\partial_{\phi}\right)+{\cal U}^{\prime}\partial_{\phi}^{-1}\left(\left(2\mathcal{J}^{2}-2\mathcal{U}^{2}+\mathcal{J}^{\prime}\right)\partial_{\phi}+3\mathcal{J}\partial_{\phi}^{2}+\partial_{\phi}^{3}\right)\right.\\
	&  & \left.+\left(-4\mathcal{U}^{3}+4{\cal J}^{2}\mathcal{U}-3{\cal J}{\cal U}^{\prime}-4{\cal J}^{\prime}{\cal U}+{\cal U}^{\prime\prime}\right)\partial_{\phi}+2\mathcal{U}^{\prime}\partial_{\phi}^{2}+2{\cal U}\partial_{\phi}^{3}\right).
\end{eqnarray*}

\chapter{Gelfand-Dickey polynomials and Hamiltonians of the modified Boussinesq
	hierarchy\label{Appendix 4: Gelfand-Dikckey pol mBsq}}

In this appendix we exhibit the first Gelfand-Dickey polynomials and
Hamiltonians of the mBoussinesq hierarchy.

The Gelfand-Dickey polynomials can be explicitly constructed using
the recurrence relation \eqref{eq:rec}. The first of them are given
by

\begin{eqnarray*}
	\frac{4\pi}{\hat{\kappa}}R_{\mathcal{J}}^{\left(0\right)} & = & \lambda_{1},\\
	\frac{4\pi}{\hat{\kappa}}R_{\mathcal{J}}^{\left(1\right)} & = & \lambda_{1}\mathcal{J}-\lambda_{2}\left(\mathcal{U}^{\prime}+2\mathcal{J}\mathcal{U}\right),\\
	\frac{4\pi}{\hat{\kappa}}R_{\mathcal{J}}^{\left(2\right)} & = & 4\lambda_{1}\left(-4\mathcal{J}^{3}\mathcal{U}-\frac{4}{3}\mathcal{J}\mathcal{U}^{3}-2\mathcal{J}^{2}\mathcal{U}^{\prime}-2\mathcal{U}^{2}\mathcal{U}^{\prime}+2\mathcal{J}^{\prime}\mathcal{U}^{\prime}+2\mathcal{J}^{\prime\prime}\mathcal{U}+2\mathcal{J}\mathcal{U}^{\prime\prime}\right.\\
	{\color{red}} & {\color{red}} & \left.+\mathcal{U}^{\prime\prime\prime}\right)+4\lambda_{2}\left(\mathcal{J}^{5}+10\mathcal{J}^{3}\mathcal{U}^{2}-\frac{5}{3}\mathcal{J}\mathcal{U}^{4}-5\mathcal{J}\mathcal{J}^{\prime2}+5\mathcal{J}^{2}\mathcal{U}^{2\prime}-\frac{10}{12}\mathcal{U}^{4\prime}\right.\\
	&  & \left.-5\mathcal{J}^{\prime}\mathcal{U}^{2\prime}+5\mathcal{J}\mathcal{U}^{\prime2}-5\mathcal{J}^{2}\mathcal{J}^{\prime\prime}-5\mathcal{J}^{\prime\prime}\mathcal{U}^{2}-5\mathcal{J}^{\prime}\mathcal{J}^{\prime\prime}+5\mathcal{U}^{\prime}\mathcal{U}^{\prime\prime}+\mathcal{J}^{\prime\prime\prime\prime}\right),
\end{eqnarray*}
\begin{eqnarray*}
	\frac{4\pi}{\hat{\kappa}}R_{\mathcal{U}}^{\left(0\right)} & = & \lambda_{2},\\
	\frac{4\pi}{\hat{\kappa}}R_{\mathcal{U}}^{\left(1\right)} & = & \lambda_{1}\mathcal{U}+\lambda_{2}\left(\mathcal{J}^{\prime}-\mathcal{J}^{2}+\mathcal{U}^{2}\right),\\
	\frac{4\pi}{\hat{\kappa}}R_{\mathcal{U}}^{\left(2\right)} & = & 4\lambda_{1}\left(-\mathcal{J}^{4}-2\mathcal{J}^{2}\mathcal{U}^{2}+\frac{5}{3}\mathcal{U}^{4}+\frac{2}{3}\mathcal{J}^{3\prime}+2\mathcal{U}^{2}\mathcal{J}^{\prime}+\mathcal{J}^{\prime2}-\mathcal{U}^{\prime2}+2\mathcal{J}\mathcal{J}^{\prime\prime}\right.\\
	{\color{red}} & {\color{red}} & \left.-2\mathcal{U}\mathcal{U}^{\prime\prime}-\mathcal{J}^{\prime\prime\prime}\right)+4\lambda_{2}\left(5\mathcal{J}^{4}\mathcal{U}-\frac{10}{3}\mathcal{J}^{2}\mathcal{U}^{3}+\frac{7}{3}\mathcal{U}^{5}-\frac{10}{3}\mathcal{J}^{3\prime}\mathcal{U}+\frac{10}{3}\mathcal{J}^{\prime}\mathcal{U}^{3}\right.\\
	{\color{red}} & {\color{red}} & \left.+5\mathcal{U}\mathcal{J}^{\prime2}-5\mathcal{J}^{2\prime}\mathcal{U}^{\prime}-5\mathcal{U}\mathcal{U}^{\prime2}+5\mathcal{J}^{\prime\prime}\mathcal{U}^{\prime}-5\mathcal{J}^{2}\mathcal{U}^{\prime\prime}-5\mathcal{U}^{2}\mathcal{U}^{\prime\prime}+5\mathcal{J}^{\prime}\mathcal{U}^{\prime\prime}+\mathcal{U}^{\prime\prime\prime\prime}\right).
\end{eqnarray*}
The corresponding Hamiltonians can then be obtained using eq. \eqref{eq:GelfandDickey}.
Thus,

\begin{eqnarray*}
	\frac{4\pi}{\hat{\kappa}}H_{\left(0\right)} & = & \int d\phi\left(\lambda_{1}\mathcal{J}+\lambda_{2}\mathcal{U}\right),\\
	\frac{4\pi}{\hat{\kappa}}H_{\left(1\right)} & = & \int d\phi\left\{ \frac{\lambda_{1}}{2}\left(\mathcal{J}^{2}+\mathcal{U}^{2}\right)+\lambda_{2}\left(\frac{1}{3}\mathcal{U}^{3}-\mathcal{J}^{2}\mathcal{U}-\mathcal{J}\mathcal{U}^{\prime}\right)\right\} ,\\
	\frac{4\pi}{\hat{\kappa}}H_{\left(2\right)} & = & \int d\phi\left\{ \frac{4\lambda_{1}}{3}\left(\mathcal{U}^{5}-3\mathcal{J}^{4}\mathcal{U}-2\mathcal{J}^{2}\mathcal{U}^{3}-3\mathcal{J}^{\prime2}\mathcal{U}-2\mathcal{J}^{3}\mathcal{U}^{\prime}-2\mathcal{J}\mathcal{U}^{3\prime}\right.\right.\\
	{\color{red}} & {\color{red}} & \left.+3\mathcal{J}^{2}\mathcal{U}^{\prime\prime}-\frac{3}{2}\mathcal{U}^{2}\mathcal{U}^{\prime\prime}+3\mathcal{J}\mathcal{U}^{\prime\prime\prime}\right)+\frac{2\lambda_{2}}{3}\left(\mathcal{J}^{6}+15\mathcal{J}^{4}\mathcal{U}^{2}-5\mathcal{J}^{2}\mathcal{U}^{4}+\frac{7}{3}\mathcal{U}^{6}\right.\\
	&  & \left.+15\mathcal{J}^{2}\mathcal{J}^{\prime2}+15\mathcal{U}^{2}\mathcal{J}^{\prime2}+10\mathcal{J}^{3}\mathcal{U}^{2\prime}-5\mathcal{J}\mathcal{U}^{4\prime}+15\mathcal{J}^{2}\mathcal{U}^{\prime2}+15\mathcal{U}^{2}\mathcal{U}^{\prime2}\right.\\
	&  & \left.\left.-10\mathcal{J}\mathcal{J}^{\prime}\mathcal{J}^{\prime\prime}+30\mathcal{J}\mathcal{U}^{\prime}\mathcal{U}^{\prime\prime}+3\mathcal{J}\mathcal{J}^{\prime\prime\prime\prime}+3\mathcal{U}\mathcal{U}^{\prime\prime\prime\prime}\right)\right\} .
\end{eqnarray*}

\chapter{Boussinesq hierarchy\label{Appendix 5: Bsq and mBsq}}

The Boussinesq hierarchy is an integrable bi-Hamiltonian system which
possesses two different Poisson brackets defined by the following
operators
\begin{equation}
	\mathcal{D}_{\left(1\right)}^{\text{Bsq}}=\frac{\pi}{2\hat{\kappa}}\left(\begin{array}{cc}
		0 & \partial_{\phi}\\
		\partial_{\phi} & 0
	\end{array}\right),\label{eq:Bq Operators}
\end{equation}
\[
\mathcal{D}_{\left(2\right)}^{\text{Bsq}}=\frac{4\pi}{\hat{\kappa}}\begin{pmatrix}2{\cal L}\partial_{\phi}+{\cal L}^{\prime}-\partial_{\phi}^{3} & 3\mathcal{W}\partial_{\phi}+2\mathcal{W}^{\prime}\\
	3\mathcal{W}\partial_{\phi}+\mathcal{W}^{\prime} & -\frac{1}{2}\mathcal{L}^{\prime\prime\prime}+2\mathcal{L}^{2\prime}-\frac{9}{4}\left(\mathcal{L}^{\prime\prime}-\frac{16}{9}\mathcal{L}^{2}\right)\partial_{\phi}-\frac{15}{4}{\cal L}^{\prime}\partial_{\phi}^{2}-\frac{5}{2}{\cal L}\partial_{\phi}^{3}+\frac{1}{4}\partial_{\phi}^{5}
\end{pmatrix}.
\]
The Poisson bracket associated to the operator $\mathcal{D}_{\left(2\right)}^{\text{Bsq}}$
is given by the classical $W_{3}$-algebra.

The infinite Hamiltonians in involution can be obtained using the
following recursion relation
\[
\mathcal{D}_{(1)}^{\text{Bsq}}\left(\begin{array}{c}
	R_{\mathcal{L}}\\
	R_{\mathcal{W}}
\end{array}\right)_{\left(k+1\right)}=\mathcal{D}_{(2)}^{\text{Bsq}}\left(\begin{array}{c}
	R_{\mathcal{L}}\\
	R_{\mathcal{W}}
\end{array}\right)_{\left(k\right)}.
\]
Here the corresponding Gelfand-Dickey polynomials are defined through
\[
\left(\begin{array}{c}
	R_{\mathcal{L}}\\
	R_{\mathcal{W}}
\end{array}\right)_{\left(k\right)}=\begin{pmatrix}\frac{\delta H_{\left(k\right)}^{\text{Bsq}}}{\delta\mathcal{L}}\\
	\frac{\delta H_{\left(k\right)}^{\text{Bsq}}}{\delta\mathcal{\mathcal{W}}}
\end{pmatrix},
\]
where the first Hamiltonian is given by

\[
H_{\left(1\right)}^{\text{Bsq}}=\frac{\hat{\kappa}}{4\pi}\int d\phi\left(\lambda_{1}\mathcal{L}+\lambda_{2}\mathcal{W}\right).
\]
Therefore, the members of the hierarchy can be written as follows

\begin{equation}
	\left(\begin{array}{c}
		\dot{\mathcal{L}}\\
		\dot{\mathcal{W}}
	\end{array}\right)_{\left(k\right)}=\mathcal{D}_{(1)}^{\text{Bsq}}\left(\begin{array}{c}
		R_{\mathcal{L}}\\
		R_{\mathcal{W}}
	\end{array}\right)_{\left(k+1\right)}=\mathcal{D}_{(2)}^{\text{Bsq}}\left(\begin{array}{c}
		R_{\mathcal{L}}\\
		R_{\mathcal{W}}
	\end{array}\right)_{\left(k\right)}.\label{eq:Bq Hierarchy}
\end{equation}

As explained in section \ref{sec:Review-of-the}, the Boussinesq and
the mBoussinesq hierarchies are related by the Miura transformation
\eqref{eq:Miura}, that can be rewritten in the following vector form

\[
\left(\begin{array}{c}
	\mathcal{L}\\
	\mathcal{W}
\end{array}\right)=F\left[{\cal J},{\cal U}\right],
\]
for a functional $F$ defined through \eqref{eq:Miura}. Taking the
derivative with respect to the time one obtains

\begin{equation}
	\left(\begin{array}{c}
		\dot{\mathcal{L}}\\
		\dot{\mathcal{W}}
	\end{array}\right)=M\left(\begin{array}{c}
		\dot{\mathcal{J}}\\
		\dot{{\cal U}}
	\end{array}\right),\label{eq:Miura vector}
\end{equation}
where $M=M\left[{\cal J},{\cal U}\right]$ correspond to the Fréchet
derivative of $F$ with respect to ${\cal J}$ and ${\cal U}$ \cite{Mathieu:1991},
and is precisely given by the matrix $M$ in \eqref{eq:M}, i.e.,

\[
M=\left(\begin{array}{cc}
	{\cal J}+\partial_{\phi}\qquad & \mathcal{U}\\
	-2{\cal J}\mathcal{U}-\frac{1}{2}\mathcal{U}\partial_{\phi}-\frac{3}{2}{\cal U}^{\prime}\qquad & \mathcal{U}^{2}-{\cal J}^{2}-\frac{1}{2}{\cal J}^{\prime}-\frac{3}{2}{\cal J}\partial_{\phi}-\frac{1}{2}\partial_{\phi}^{2}
\end{array}\right).
\]
If one takes into account its formal adjoint

\[
M^{\dagger}=\left(\begin{array}{cc}
	\;\;{\cal J}-\partial_{\phi}\qquad & -2{\cal J}\mathcal{U}+\frac{1}{2}\mathcal{U}\partial_{\phi}-{\cal U}^{\prime}\\
	\;\;\mathcal{U}\qquad & \mathcal{U}^{2}-{\cal J}^{2}+{\cal J}^{\prime}+\frac{3}{2}{\cal J}\partial_{\phi}-\frac{1}{2}\partial_{\phi}^{2}
\end{array}\right),
\]
the Gelfand-Dickey polynomials of both hierarchies are then related
by

\begin{equation}
	\left(\begin{array}{c}
		R_{\mathcal{J}}\\
		R_{\mathcal{U}}
	\end{array}\right)=M^{\dagger}\left(\begin{array}{c}
		R_{\mathcal{L}}\\
		R_{\mathcal{W}}
	\end{array}\right).\label{eq:Miura Gelfand-Dickey}
\end{equation}
Taking into account eqs. \eqref{eq:Miura vector}, \eqref{eq:EOM}
and \eqref{eq:Miura Gelfand-Dickey} we can write

\[
\left(\begin{array}{c}
	\dot{\mathcal{L}}\\
	\dot{\mathcal{W}}
\end{array}\right)_{\left(k\right)}=M\mathcal{D}\left(\begin{array}{c}
	R_{\mathcal{J}}\\
	R_{\mathcal{U}}
\end{array}\right)_{\left(k\right)}=M\mathcal{D}M^{\dagger}\left(\begin{array}{c}
	R_{\mathcal{L}}\\
	R_{\mathcal{W}}
\end{array}\right)_{\left(k\right)}=\mathcal{D}_{(2)}^{\text{Bsq}}\left(\begin{array}{c}
	R_{\mathcal{L}}\\
	R_{\mathcal{W}}
\end{array}\right)_{\left(k\right)},
\]
which imply that the second Poisson structure for the Boussinesq hierarchy
can be expressed in terms of the first Poisson structure of the mBoussinesq
hierarchy according to
\[
\mathcal{D}_{(2)}^{\text{Bsq}}=M\mathcal{D}M^{\dagger}.
\]

\chapter{Fundamental representation of the principal embedding of $sl(2,\mathbb{R})$
	within $sl(N,\mathbb{R})$\label{Appendix 6: Fundamental rep}}

In the principal embedding of the $sl(2,\mathbb{R})$ algebra within
the $sl(N,\mathbb{R})$ algebra the generators can be written in the
basis \emph{$\left\{ L_{i},W_{m}^{(s)}\right\} $}, with $i=-1,0,1$,
$s=3,4,\dots$ and $m=-s+1,\dots,s-1$. In the fundamental representation
of $sl(N,\mathbb{R})$, the generators may be represented by the following
$N\times N$ matrices

\[
\begin{aligned}\left(L_{1}\right)_{jk} & =-\sqrt{j\left(N-j\right)}\delta_{j+1,k},\\
	\left(L_{-1}\right)_{jk} & =\sqrt{k\left(N-k\right)}\delta_{j,k+1},\\
	\left(L_{0}\right)_{jk} & =\frac{1}{2}\left(N+1-2j\right)\delta_{j,k},
\end{aligned}
\]
\[
\begin{aligned}W_{m}^{(s)} & =2\left(-1\right)^{s-m-1}\frac{\left(s+m-1\right)!}{\left(2s-2\right)!}\underbrace{\left[L_{-1},\left[L_{-1},\cdots\left[L_{-1},\left(L_{1}\right)^{s-1}\right]\cdots\right]\right]}_{s-m-1\text{ terms }},\\
	& =2\left(-1\right)^{s-m-1}\frac{\left(s+m-1\right)!}{\left(2s-2\right)!}\left(\mathrm{ad}_{L_{-1}}\right)^{s-m-1}\left(L_{1}\right)^{s-1}.
\end{aligned}
\]
with $j,k=1,\dots,N$, and where \emph{$\mathrm{ad}_{\mathrm{x}}\left(\mathrm{Y}\right):=\left[\mathrm{X},\mathrm{Y}\right]$.}
From the commutation relations
\[
\begin{aligned}\left[L_{i},L_{j}\right] & =\left(i-j\right)L_{i+j},\\
	\left[L_{i},W_{m}^{(s)}\right] & =\left(\left(s-1\right)i-m\right)W_{i+m}^{(s)},
\end{aligned}
\]
it can be seen that the $L_{i}$ generators close in a $sl(2,\mathbb{R})$
subalgebra, while the generators $W_{m}^{(s)}$ transform in a spin-$s$
representation under $sl(2,\mathbb{R})$.

\subsection*{$sl(3,\mathbb{R})$ generators}

The generators of $sl(3,\mathbb{R})$ algebra are given by the following
$3\times3$ matrices

\[
L_{-1}=\left(\begin{array}{ccc}
	0 & -\sqrt{2} & 0\\
	0 & 0 & -\sqrt{2}\\
	0 & 0 & 0
\end{array}\right),\quad\quad L_{0}=\left(\begin{array}{ccc}
	1 & 0 & 0\\
	0 & 0 & 0\\
	0 & 0 & -1
\end{array}\right),\quad\quad L_{1}=\left(\begin{array}{ccc}
	0 & 0 & 0\\
	\sqrt{2} & 0 & 0\\
	0 & \sqrt{2} & 0
\end{array}\right),
\]
and the spin--3 generators $W_{m}^{\left(3\right)}=W_{m}$

\[
W_{-2}=\left(\begin{array}{ccc}
	0 & 0 & 4\\
	0 & 0 & 0\\
	0 & 0 & 0
\end{array}\right),\quad\quad W_{-1}=\left(\begin{array}{ccc}
	0 & -\sqrt{2} & 0\\
	0 & 0 & \sqrt{2}\\
	0 & 0 & 0
\end{array}\right),\quad\quad W_{0}=\frac{2}{3}\left(\begin{array}{ccc}
	1 & 0 & 0\\
	0 & -2 & 0\\
	0 & 0 & 1
\end{array}\right),
\]

\[
W_{-1}=\left(\begin{array}{ccc}
	0 & 0 & 0\\
	\sqrt{2} & 0 & 0\\
	0 & -\sqrt{2} & 0
\end{array}\right),\qquad\qquad W_{2}=\left(\begin{array}{ccc}
	0 & 0 & 0\\
	0 & 0 & 0\\
	4 & 0 & 0
\end{array}\right).
\]

\chapter{Wess-Zumino term\label{Appendix 7: Wess-Zumino-term}}

Here we show that for our boundary conditions in \eqref{eq:A Connections},
\eqref{eq:Auxiliary Connections}, \eqref{eq:chemical potentials},
the Wess-Zumino term 
\begin{align*}
	I_{1} & =\frac{\kappa}{16\pi}\int dtdrd\phi\epsilon^{ij}\left\langle \partial_{t}\left(G^{-1}\right)\partial_{i}GG^{-1}\partial_{j}G\right\rangle ,
\end{align*}
in \eqref{eq: I1 WZW action} vanishes.

Let us perform the following Gauss decomposition of the group element

\begin{equation}
	G=e^{TL_{1}+MW_{1}+QW_{2}}e^{\Phi L_{0}+\Phi_{W}W_{0}}e^{XL_{-1}+YW_{-1}+ZW_{-2}},\label{eq:GWZW}
\end{equation}
where all the functions that appear in \eqref{eq:GWZW} generically
depend on $t$, $r$ and $\phi$. Then, if we replace \eqref{eq:GWZW}
in $I_{1}$, one can show that it reduces to a boundary term of the
form

\begin{eqnarray}
	I_{1} & = & \frac{\kappa}{16\pi}\int d\phi dt\left[2e^{\Phi+2\Phi_{W}}\left(\left(X^{\prime}+Y^{\prime}\right)\left(\dot{M}+\dot{T}\right)-\left(\dot{X}+\dot{Y}\right)\left(M^{\prime}+T^{\prime}\right)\right)\right.\nonumber \\
	&  & \left.-2e^{\Phi-2\Phi_{W}}\left(\left(X^{\prime}-Y^{\prime}\right)\left(\dot{M}-\dot{T}\right)-\left(\dot{X}-\dot{Y}\right)\left(M^{\prime}-T^{\prime}\right)\right)\right.\label{eq:I2ap}\\
	&  & +e^{2\Phi}\left(8\left(XY^{\prime}-YX^{\prime}\right)\dot{Q}-8\left(X\dot{Y}-Y\dot{X}\right)Q^{\prime}+8\left(T\dot{M}-M\dot{T}\right)Z^{\prime}-8\left(TM^{\prime}-MT^{\prime}\right)\dot{Z}\right.\nonumber \\
	&  & \left.\left.+4\left(X\dot{Y}-Y\dot{X}\right)\left(TM^{\prime}-MT^{\prime}\right)-4\left(XY^{\prime}-YX^{\prime}\right)\left(T\dot{M}-M\dot{T}\right)+16\left(Q^{\prime}\dot{Z}-Z^{\prime}\dot{Q}\right)\right)\right].\nonumber 
\end{eqnarray}
Now, it is useful to perform the following decomposition in the asymptotic
region which is compatible with \eqref{eq:A Connections}

\[
G=g\left(t,\phi\right)b\left(r\right),
\]
with

\[
g\left(t,\phi\right)=\exp\left[\sqrt{\frac{8\pi}{\kappa}}\varphi L_{0}+\sqrt{\frac{6\pi}{\kappa}}\psi W_{0}\right],
\]
and where $b\left(r\right)$ is an arbitrary gauge group element depending
on the radial coordinate that generically can be decomposed as 
\[
b\left(r\right)=b_{\left(+\right)}b_{\left(0\right)}b_{\left(-\right)},
\]
with

\begin{eqnarray*}
	b_{\left(+\right)} & = & e^{\left(b_{1}L_{1}+\bar{b}_{1}W_{1}+\bar{b}_{2}W_{2}\right)},\qquad b_{\left(0\right)}=e^{\left(b_{0}L_{0}+\bar{b}_{0}W_{0}\right)},\qquad b_{\left(-\right)}=e^{\left(b_{-1}L_{-1}+\bar{b}_{-1}W_{-1}+\bar{b}_{-2}W_{-2}\right)}.
\end{eqnarray*}
Consistency with \eqref{eq:GWZW} then implies the following conditions
\[
\Phi=b_{0}\left(r\right)+\sqrt{\frac{8\pi}{\kappa}}\varphi\left(t,\phi\right),\qquad\Phi_{W}=\bar{b}_{0}\left(r\right)+\sqrt{\frac{6\pi}{\kappa}}\psi\left(t,\phi\right),
\]
\[
X=b_{-1}\left(r\right),\qquad Y=\bar{b}_{-1}\left(r\right),\qquad Z=\bar{b}_{-2}\left(r\right),
\]
\[
Q=e^{-2\sqrt{\frac{8\pi}{\kappa}}\varphi}\bar{b}_{-2}\left(r\right),\qquad M=e^{-\sqrt{\frac{8\pi}{\kappa}}\varphi}\left(\bar{b}_{1}\left(r\right)\cosh\left(2\sqrt{\frac{6\pi}{\kappa}}\psi\right)-b_{1}\left(r\right)\sinh\left(2\sqrt{\frac{6\pi}{\kappa}}\psi\right)\right),
\]
\[
T=e^{-\sqrt{\frac{8\pi}{\kappa}}\varphi}\left(b_{1}\left(r\right)\cosh\left(2\sqrt{\frac{6\pi}{\kappa}}\psi\right)-\bar{b}_{1}\left(r\right)\sinh\left(2\sqrt{\frac{6\pi}{\kappa}}\psi\right)\right).
\]
Note that since, $X$, $Y$ and $Z$ depend only on the radial coordinate,
then the WZ term in eq. \eqref{eq:I2ap} identically vanishes.

\newpage

\singlespacing
\thispagestyle{empty}

\bibliographystyle{JHEP}
\bibliography{review}

\end{document}